\documentclass[prd,superscriptaddress,floatfix,notitlepage,nofootinbib,reprint]{revtex4-1} 
\pdfoutput=1
\usepackage{amssymb,epsfig}
\usepackage{amsmath}
\usepackage[]{empheq}
\usepackage{hyperref}
\usepackage{natbib,ifthen}

\usepackage{wasysym}
\usepackage{mathrsfs}
\usepackage[lofdepth,lotdepth,caption=false]{subfig}
\usepackage{color}
\usepackage{bbold}

\begin{document}

\newcommand{\jcap}{JCAP}
\newcommand{\apjl}{APJL~}
\newcommand{\aap}{Astronomy \& Astrophysics}
\newcommand{\mnras}{Mon.\ Not.\ R.\ Astron.\ Soc.}
\newcommand{\apjs}{Astrophys.\ J.\ Supp.}

\newcommand{\FFPT}{\textsf{FFT-PT}}
\def\Btheo{{B_\delta^{\textrm{theo}}}}
\newcommand{\todo}[1]{{\color{blue}{TODO: #1}}}
\newcommand{\comment}[1]{{\color{red}{#1}}}
\newcommand{\MS}[1]{{\color{red}{MS: #1}}}
\newcommand{\NEW}[1]{#1}
\newcommand{\US}[1]{{\color{red}{US: #1}}}
\newcommand{\lbar}[1]{\underline{l}_{#1}}
\newcommand{\drm}{\mathrm{d}}
\renewcommand{\d}{\mathrm{d}}
\renewcommand{\rm}[1]{\mathrm{#1}}
\newcommand{\gaensli}[1]{\lq #1\rq$ $}
\newcommand{\bartilde}[1]{\bar{\tilde #1}}
\newcommand{\barti}[1]{\bartilde{#1}}
\newcommand{\ti}{\tilde}
\newcommand{\oforder}[1]{\mathcal{O}(#1)}
\newcommand{\D}{\mathrm{D}}
\renewcommand{\(}{\left(}
\renewcommand{\)}{\right)}
\renewcommand{\[}{\left[}
\renewcommand{\]}{\right]}
\def\<{\left\langle}
\def\>{\right\rangle}
\newcommand{\mycaption}[1]{\caption{\footnotesize{#1}}}
\newcommand{\hattilde}[1]{\hat{\tilde #1}}
\newcommand{\mycite}[1]{[#1]}
\newcommand{\fnl}{{f_\mathrm{NL}}}
\newcommand{\fsky}{{f_\mathrm{sky}}}

\def\uk{{\bf \hat{k}}}
\def\un{{\bf \hat{n}}}
\def\ur{{\bf \hat{r}}}
\def\ux{{\bf \hat{x}}}
\def\bk{{\bf k}}
\def\bn{{\bf n}}
\def\br{{\bf r}}
\def\bx{{\bf x}}
\def\bK{{\bf K}}
\def\by{{\bf y}}
\def\bl{{\bf l}}
\def\bkp{{\bf k^\pr}}
\def\brp{{\bf r^\pr}}

\newcommand{\fixme}[1]{{\textbf{Fixme: #1}}}
\newcommand{\detD}{{\det\!\cld}}
\newcommand{\clh}{\mathcal{H}}
\newcommand{\ud}{{\rm d}}
\renewcommand{\eprint}[1]{\href{http://arxiv.org/abs/#1}{#1}}
\newcommand{\adsurl}[1]{\href{#1}{ADS}}
\newcommand{\ISBN}[1]{\href{http://cosmologist.info/ISBN/#1}{ISBN: #1}}
\newcommand{\vort}{\varpi}
\newcommand\ba{\begin{eqnarray}}
\newcommand\ea{\end{eqnarray}}
\newcommand\be{\begin{equation}}
\newcommand\ee{\end{equation}}
\newcommand\lagrange{{\cal L}}
\newcommand\cll{{\cal L}}
\newcommand\cln{{\cal N}}
\newcommand\clx{{\cal X}}
\newcommand\clz{{\cal Z}}
\newcommand\clv{{\cal V}}
\newcommand\cld{{\cal D}}
\newcommand\clt{{\cal T}}

\newcommand\clo{{\cal O}}
\newcommand{\cla}{{\cal A}}
\newcommand{\clp}{{\cal P}}
\newcommand{\clr}{{\cal R}}
\newcommand{\uD}{{\mathrm{D}}}
\newcommand{\calE}{{\cal E}}
\newcommand{\calB}{{\cal B}}
\newcommand{\curl}{\,\mbox{curl}\,}
\newcommand\del{\nabla}
\newcommand\Tr{{\rm Tr}}
\newcommand\half{{\frac{1}{2}}}
\newcommand\fourth{{1\over 8}}
\newcommand\bibi{\bibitem}
\newcommand{\kf}{\beta}
\newcommand{\kfprod}{\alpha}
\newcommand\calS{{\cal S}}
\renewcommand\H{{\cal H}}
\newcommand\K{{\rm K}}
\newcommand\mK{{\rm mK}}
\newcommand\synch{\text{syn}}
\newcommand\opacity{\tau_c^{-1}}

\newcommand{\Psil}{\Psi_l}
\newcommand{\bsigma}{{\bar{\sigma}}}
\newcommand{\bI}{\bar{I}}
\newcommand{\bq}{\bar{q}}
\newcommand{\bv}{\bar{v}}
\renewcommand\P{{\cal P}}
\newcommand{\numfrac}[2]{{\textstyle \frac{#1}{#2}}}

\newcommand{\la}{\langle}
\newcommand{\ra}{\rangle}
\newcommand{\lla}{\left\langle}
\newcommand{\rra}{\right\rangle}

\newcommand{\vnabla}{\ensuremath{\boldsymbol\nabla}}

\newcommand{\Omtot}{\Omega_{\mathrm{tot}}}
\newcommand\xx{\mbox{\boldmath $x$}}
\newcommand{\phpr} {\phi'}
\newcommand{\gam}{\gamma_{ij}}
\newcommand{\sqgam}{\sqrt{\gamma}}
\newcommand{\delk}{\Delta+3{\K}}
\newcommand{\dph}{\delta\phi}
\newcommand{\om} {\Omega}
\newcommand{\dom}{\delta^{(3)}\left(\Omega\right)}
\newcommand{\rar}{\rightarrow}
\newcommand{\Rar}{\Rightarrow}
\newcommand\gsim{ \lower .75ex \hbox{$\sim$} \llap{\raise .27ex \hbox{$>$}} }
\newcommand\lsim{ \lower .75ex \hbox{$\sim$} \llap{\raise .27ex \hbox{$<$}} }
\newcommand\bigdot[1] {\stackrel{\mbox{{\huge .}}}{#1}}
\newcommand\bigddot[1] {\stackrel{\mbox{{\huge ..}}}{#1}}
\newcommand{\Mpc}{\text{Mpc}}
\newcommand{\Al}{{A_l}}
\newcommand{\Bl}{{B_l}}
\newcommand{\eAl}{e^\Al}
\newcommand{\ix}{{(i)}}
\newcommand{\ixp}{{(i+1)}}
\renewcommand{\k}{\beta}
\newcommand{\HD}{\mathrm{D}}

\newcommand{\nonflat}[1]{#1}
\newcommand{\Cgl}{C_{\text{gl}}}
\newcommand{\Cgltwo}{C_{\text{gl},2}}
\newcommand{\He}{{\rm{He}}}
\newcommand{\Mhz}{{\rm MHz}}
\newcommand{\vx}{{\mathbf{x}}}
\newcommand{\ve}{{\mathbf{e}}}
\newcommand{\vv}{{\mathbf{v}}}
\newcommand{\vk}{{\mathbf{k}}}
\renewcommand{\vr}{{\mathbf{r}}}
\newcommand{\vn}{{\mathbf{n}}}
\newcommand{\vPsi}{{\mathbf{\Psi}}}
\newcommand{\vs}{{\mathbf{s}}}
\newcommand{\vH}{{\mathbf{H}}}
\newcommand{\theo}{\mathrm{th}}
\newcommand{\sgn}{\mathrm{sgn}}

\newcommand{\vnhat}{{\hat{\mathbf{n}}}}
\newcommand{\vkhat}{{\hat{\mathbf{k}}}}
\newcommand{\taueps}{{\tau_\epsilon}}
\newcommand{\vbeta}{\ensuremath{\boldsymbol\beta}}

\newcommand{\vgrad}{{\mathbf{\nabla}}}
\newcommand{\fbarln}{\bar{f}_{,\ln\epsilon}(\epsilon)}

\newcommand{\secref}[1]{Section~\ref{se:#1}}
\newcommand{\app}[1]{Appendix~\ref{app:#1}}
\newcommand{\expt}{\mathrm{expt}}
\newcommand{\eq}[1]{(\ref{eq:#1})} 
\newcommand{\eqq}[1]{Eq.~(\ref{eq:#1})} 
\newcommand{\fig}[1]{Fig.~\ref{fig:#1}} 
\renewcommand{\to}{\rightarrow}
\renewcommand{\(}{\left(}
\renewcommand{\)}{\right)}
\renewcommand{\[}{\left[}
\renewcommand{\]}{\right]}
\renewcommand{\vec}[1]{\mathbf{#1}}
\newcommand{\vy}{\vec{y}}
\newcommand{\vz}{\vec{z}}
\newcommand{\vq}{\vec{q}}
\newcommand{\vp}{\vec{p}}
\newcommand{\va}{\vec{a}}
\newcommand{\vb}{\vec{b}}
\newcommand{\VPsi}{\vec{\Psi}}
\newcommand{\vecv}{\vec{v}}
\newcommand{\vl}{{\boldsymbol \ell}}
\newcommand{\vell}{\boldsymbol{\ell}}
\newcommand{\VL}{\vec{L}}
\newcommand{\dl}{\d^2\vl}
\newcommand{\vd}{\vec{d}}
\newcommand{\vw}{\vec{w}}
\newcommand{\valpha}{\vec{\alpha}}
\newcommand{\vtheta}{\ensuremath{\boldsymbol\theta}}
\renewcommand{\L}{\mathscr{L}}

\newcommand{\abs}[1]{\lvert #1\rvert}

\newcommand{\ul}{\underline{l}}
\newcommand{\lin}{\mathrm{lin}}

\newcommand*{\df}  {\delta}
\newcommand*{\tf}  {\theta}
\renewcommand{\vec}{\textbf}
\newcommand*{\non}  {\nonumber}
\newcommand*{\lb}  {\left(}
\newcommand*{\rb}  {\right)}
\newcommand*{\ls}  {\left[}
\newcommand*{\rs}  {\right]}

\def\pvm#1{{\color{blue}[PM: {\it #1}] }}
\def\zv#1{{\color{red}[ZV: {\it #1}] }}


\thispagestyle{empty}

\title{Parameter constraints from cross-correlation of CMB lensing with galaxy clustering}

\author{Marcel Schmittfull}
\affiliation{Institute for Advanced Study, Einstein Drive, Princeton, NJ 08540, USA}

\author{Uro\v{s} Seljak}
\affiliation{Berkeley Center for Cosmological Physics, University of California,
Berkeley, CA 94720, USA}
\affiliation{Department of Astronomy and Department of Physics,
University of California, Berkeley, CA 94720, USA}
\affiliation{Lawrence Berkeley National Laboratory, 1 Cyclotron Road, Berkeley, CA
93720, USA}

\date{\today}

\begin{abstract}

The lensing convergence measurable with future CMB surveys like CMB-S4 will be highly correlated with the clustering observed by deep photometric large scale 
structure (LSS) surveys such as the LSST, with cross-correlation coefficient as high as 95\%. 
This will enable use of sample variance cancellation techniques to determine cosmological parameters, and use of cross-correlation measurements to break parameter degeneracies. 
Assuming large sky overlap between CMB-S4 and LSST,
we show that a joint analysis of CMB-S4 lensing and LSST clustering can yield very tight constraints on the matter amplitude $\sigma_8(z)$, halo bias, and $f_\mathrm{NL}$, competitive with the best stage IV experiment predictions, but using complementary methods, which may carry different and possibly lower systematics. 
Having no sky overlap between experiments degrades the precision of $\sigma_8(z)$ by a factor of 20, and that of $f_\mathrm{NL}$ by a factor of 1.5 to 2.
Without CMB lensing, the precision always degrades by an order of magnitude or more, showing that a joint analysis is critical.
Our results also suggest that CMB lensing in combination with 
LSS photometric surveys is a competitive probe 
of the evolution of structure in the redshift range $z\simeq 1-7$, probing a regime that is not well tested observationally.
We explore predictions against other surveys and experiment configurations, finding that wide patches with maximal sky overlap between CMB and LSS surveys are most powerful for $\sigma_8(z)$ and $f_\mathrm{NL}$.

\end{abstract}

\maketitle

\section{Introduction}

\label{se:intro}

Deep imaging surveys like the Large Synoptic Survey Telescope LSST \cite{LSSTwebsite,LSSTScienceBook} will transform the quality of large-scale structure (LSS) observations by cataloging positions and redshifts of billions of galaxies in the next decade.
With LSST we can hope to measure more than $10$ photometric redshifts per $\mathrm{arcmin}^{2}$ at redshift $0\le z\le 2$, and more than $0.1$ per $\mathrm{arcmin}^{2}$ at redshift $2\le z\le 4$ \cite{LSST1301}.
By reconstructing weak gravitational lensing of the cosmic microwave background (CMB) radiation, CMB experiments will also probe these large-scale structures in projection along the line of sight. 
Future LSST galaxy catalogs and CMB lensing maps are thus expected to be highly correlated.
The moderate accuracy of photometric redshifts is sufficient for this particular type of cross-correlation analysis because the CMB lensing kernel is very broad in redshift.

The science case of cross-correlations between clustering and CMB lensing can inform the design of
planned CMB experiments such as the Simons Observatory \cite{SimonsObservatory} and CMB-S4 \cite{CMBS4SciBook}, which can provide CMB lensing measurements that are signal dominated on scales $\ell\lesssim 1000$.
To take advantage of cross-correlations with imaging surveys, these CMB experiments need to maximize the overlap of their footprint with LSS surveys such as LSST.
This allows to cancel part of the sample variance that usually limits parameter constraints \cite{Uros0807}. 
LSST is particularly suited for this because it has a high number density out to high redshift, tracing the structures responsible for lensing of the CMB with relatively low stochasticity.
Using Fisher forecasts we will show that large sky overlap between LSS and CMB lensing observations can indeed improve sensitivity to certain parameters.

We will discuss three applications of CMB-lensing--clustering cross-correlations:
Measuring the amplitude of matter fluctuations $\sigma_8$ as a function of redshift, measuring local primordial non-Gaussianity $\fnl$ using
scale-dependent galaxy bias \cite{Dalal0710}, and measuring neutrino mass from a small scale-dependent bias effect due to a difference of transfer functions relevant for lensing and clustering \cite{Villaescusa-Navarro1311,Castorina1311,LoVerde1405,LoVerde1602}
 (see \app{FnlAndMnuBasicsAppdx} for a review of these scale-dependent bias effects).
Using cross-correlations for these applications does not only offer a way to cancel part of the cosmic variance, but it can also reduce parameter degeneracies and may be more robust to systematics than auto-correlation measurements.
An additional advantage of measuring neutrino mass with cross-correlations based on scale-dependent bias is that the transfer function difference is a pure low-redshift effect that is not limited by our knowledge of the optical depth $\tau$ to the CMB, which limits most other techniques to measure neutrino mass from LSS \cite{Allison1509}, although 
the effect we are after is very weak.
As one might expect, the success of the three applications depends on
the noise of the galaxy catalogs and CMB lensing, the overlap of galaxy samples with the CMB lensing redshift kernel, the scales that are probed, and the overlap and size of sky footprints.

Throughout this paper we take a rather optimistic point of view in terms of systematics and modeling, using only a simple linear bias model for the signal and its covariance. 
The reason for this is that we want to explore new opportunities with CMB-S4 and LSST and see how promising these opportunities can in principle be.
This can provide motivation for joint analyses of future CMB lensing and galaxy redshift surveys, and helps to understand what directions are useful to pursue further.
Given the optimistic nature of our analysis, it will be important to scrutinize and improve our forecasts by adding systematics and improving models.

Several groups have successfully measured the cross-correlation of CMB lensing and LSS clustering.
The first detections cross-correlated 
WMAP CMB lensing measurements with the NRAO VLA Sky Survey \cite{2007PhRvD..76d3510S}, and additionally with SDSS LRGs and quasars \cite{2008PhRvD..78d3520H}. 
These were also the first detections of the effect of CMB lensing itself.
More recent CMB-lensing--clustering cross-correlation measurements include Refs.~\cite{Bleem1203Xcorr,Sherwin1207QuasarsCMBlensing,Planck13Lensing,Giannantonio:2013kqa,Bianchini1410,Omori:2015qda,Giannantonio:2015ahz,Kuntz1510,Liu1601,Baxter:2016ziy,Singh1606,Nicola1612}.
Thanks to the large number of CMB and LSS surveys that are planned in the near future, the number of possible cross-correlation analyses will continue to grow rapidly. 
\NEW{Recent forecasts for cross-correlations between CMB-S4 and LSST showed promising results for calibrating multiplicative shear bias \cite{schaan1607}, as well as measuring the matter amplitude $\sigma_8$ \cite{Modi1706}, dark energy, and neutrino mass \cite{Banerjee1612}.}

Our paper is organized as follows. 
We first motivate why cross-correlation analyses are particularly useful for the three applications that we consider.
We then proceed in \secref{Expts} by specifying the assumptions we make about future CMB and LSS experiments, and discussing their redshift overlap and cross-correlation coefficient.
In \secref{PowerSpectra} we present power spectra and their expected signal-to-noise ratios, as well as the signals expected from scale-dependent bias.
\secref{AnalyticalSVCFisher} provides analytical estimates of the expected gain from sample variance cancellation.
In \secref{FisherSetup} we set up a more complete numerical Fisher analysis.
The resulting forecasts are presented in \secref{FisherResults}, where we also identify the main drivers and explore the impact of changing experimental configurations.
\NEW{
In \secref{zerrors} we study the impact of catastrophic redshift errors on the forecasts within a simple toy model. We conclude and discuss possible future directions in \secref{conclusions}.
In appendices we describe 3-D to 2-D projections, provide background on the scale-dependent bias effects from primordial non-Gaussianity and neutrino mass, discuss the sampling variance error of cross-spectra, and discuss how observations may be compressed to smaller data vectors to simplify analyses.
}

\section{Motivation for cross-correlation measurements}

\label{se:XCorrelMotivation}
Let us start by motivating in more detail why cross-correlating CMB-lensing and galaxy clustering is particularly well-suited to measure $\sigma_8(z)$ and constrain primordial non-Gaussianity and neutrino mass using their scale-dependent bias effect.

\subsection{Measuring \texorpdfstring{$\sigma_8(z)$}{sigma8(z)} and the distribution of dark matter in 3-D}
\label{se:S8Motivation}

Lensing observations are only sensitive to the cumulative matter distribution along the line of sight, collapsing the redshift dimension of the 3-D dark matter distribution.
Galaxy surveys, in contrast, measure that redshift dimension and are therefore 3-D, but they observe biased tracers of the dark matter and are therefore only sensitive to the parameter combination $b_1(z)\sigma_8(z)$, where $b_1$ is a bias factor that is typically not well known, and $\sigma_8(z)$ is the rms of the matter density in a sphere of radius $8\,h^{-1}\mathrm{Mpc}$ at redshift $z$.
Lensing observations or galaxy surveys alone can therefore not provide accurate measurements of $\sigma_8(z)$ or the 3-D matter distribution.

As is well known, cross-correlating lensing and clustering observations can break the above $b_1$-$\sigma_8$ degeneracy and determine the galaxy bias as a function of redshift, e.g.~using $b_1\simeq C^{gg}/C^{\kappa g}$, $b_1^2\simeq C^{gg}/C^{\kappa\kappa}$, or $b_1\simeq C^{\kappa g}/C^{\kappa\kappa}$.
We can then obtain the 3-D matter distribution by dividing the observed galaxy density by the estimated bias, $\delta_m(k,z)=\delta_g(k,z)/b_1(z)$.
From that we can compute the matter power spectrum as a function of redshift, and its amplitude, $\sigma_8(z)$.
Even if bias is treated as a scale-dependent function, $b_1(k,z)$, cross-correlating lensing and clustering can significantly improve the uncertainty of the matter power spectrum as a function of redshift if the cross-correlation coefficient between lensing and clustering is high \cite{Pen0402008}.
Maybe more futuristically, a better understanding of galaxy formation might predict the relation between dark matter and galaxies without requiring a general bias expansion.
In that case, lensing-clustering cross-correlations could help inform parameters of the galaxy formation models and thus improve the inferred 3-D dark matter maps.

Measuring the 3-D distribution of dark matter offers a direct way to test the growth of structure and expansion of the Universe as a function of time.
Both depend on the cosmological model, e.g.~on the time evolution of the dark energy equation of state or the sum of neutrino masses.
At low redshift,  $z\lesssim 0.5$, the motivation is to improve over current constraints.
At higher redshift, only little is known observationally about growth and expansion, so that entering this regime has significant discovery potential, especially if we can measure the matter amplitude $\sigma_8(z)$ with sub-percent-level precision.
\NEW{Such high-precision measurements of $\sigma_8(z)$ over a wide range of redshifts provide a promising tool to constrain the sum of neutrino masses through their imprint on the growth function, possibly even without calibrating against the amplitude of the CMB which is limited by the optical depth $\tau$ to the CMB \cite{ByeongheeInPrep}.
Mapping the cosmic growth history with such high precision also constrains a possible time dependence of the equation of state of dark energy.
}

Sample variance cancellation can help to improve constraints on galaxy bias parameters, because they enter only the galaxy density but not the CMB lensing convergence, which are both due to the same underlying 3-D Fourier modes at redshifts where they overlap.
Improved bias constraints can then improve the precision of the 3-D matter distribution.

For simplicity we will only quote the precision of $\sigma_8$ and $b_1$ assuming all other cosmological parameters are fixed.
If other cosmological parameters are allowed to be free, the cross-correlation measurements constrain certain combinations of them, for example roughly $\sigma_8\Omega_m$ at low redshift \cite{Jain:1996st}.
Our forecasts should therefore be interpreted as constraints on such parameter combinations.

\subsection{Motivation for \texorpdfstring{$\fnl$}{fNL} from cross-correlations}

\begin{figure}[tbp]
\includegraphics[width=0.4\textwidth]{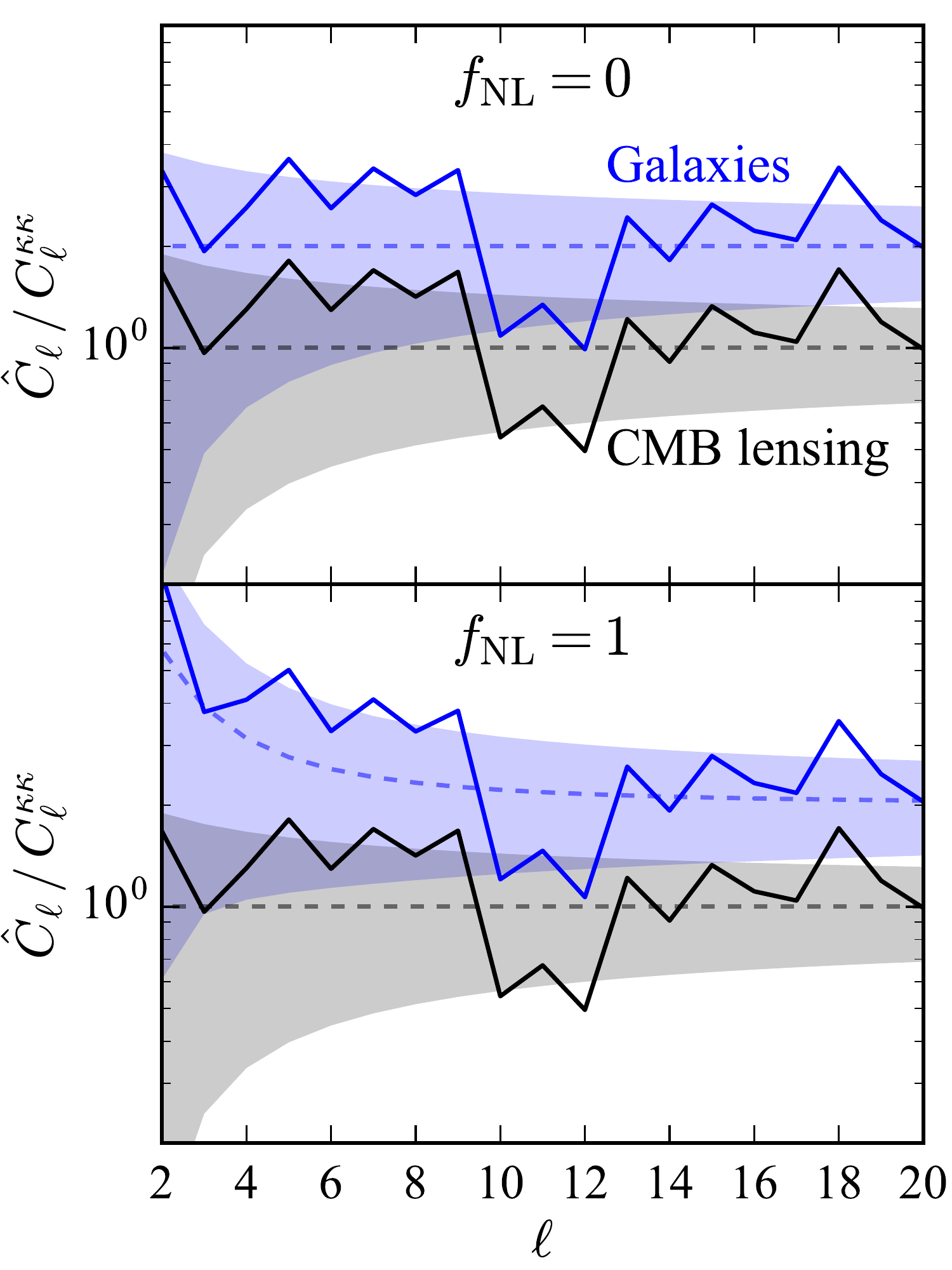}
\caption{Illustration of the primordial non-Gaussianity signal from scale-dependent galaxy bias \cite{Dalal0710}, in an idealized toy example where galaxies (blue) perfectly trace the matter fluctuations observed by CMB lensing (black), $\hat C_\ell^{gg}= b^2 \hat C_\ell^{\kappa\kappa}$.
The signal for $\fnl=1$ is smaller than cosmic variance (shaded regions), but the ratio of the observed galaxy and lensing power spectrum realizations has no cosmic variance, so that the non-Gaussianity amplitude $f_\mathrm{NL}$ can be measured with infinite precision from a single Fourier mode  \cite{Uros0807}.
In practice, this is limited by nonzero stochasticity between the observed CMB lensing convergence and galaxy density.
}
\label{fig:fnlcartoon}
\end{figure}

Primordial non-Gaussianity of the local type, parameterized by the amplitude $\fnl$, induces a scale-dependent galaxy bias that scales as $\fnl k^{-2}$ on large scales \cite{Dalal0710}.
We review this effect and the motivation to measure it in \app{fnlBasicsAppdx}.
Since the effect is largest on large scales, the precision of $\fnl$
is limited by the number of large-scale Fourier modes in the volume of the galaxy survey.
This cosmic variance noise can be partially cancelled by observing unbiased and biased tracers of LSS and searching for a scale-dependent difference in their power spectra \cite{Uros0807}. 
\fig{fnlcartoon} illustrates this idea for an idealized toy model where CMB lensing (an unbiased tracer) and galaxy number counts (a biased tracer) are assumed to originate from the exact same Fourier modes.
The prospect of sample variance cancellation is an important motivation for searching for $\fnl$ in CMB-lensing--galaxy-clustering cross-correlations rather than in galaxy auto-spectra.

The second motivation for measuring $\fnl$ from cross-correlations is its potential superiority over galaxy auto-spectra in terms of systematics (e.g., \cite{Giannantonio:2013kqa,Rhodes:2013fyq}).
On the large scales where the $k^{-2}$ scale-dependent bias is largest, systematics like stellar contamination can add galaxy auto-power and thus mimic an $\fnl$ signal.
This has been a major concern for recent $\fnl$ analyses, e.g.~\cite{Leistedt1404,Leistedt1405}, although not for one QSO sample 
in the earliest data analysis of this effect \cite{Slosar0805}. 
\NEW{Another important large-scale systematic is depth variation.}
While known systematics can of course be subtracted, it is difficult to establish ahead of time that there are no unknown or poorly understood systematics that could lead to an enhanced galaxy auto power spectrum on large scales (it is easier to establish an upper limit, since absence of power on very large scales 
can only be explained by absence of both systematics and primordial 
non-Gaussianity).
Cross-correlations can be helpful in this regard because they are unaffected by many (additive) observational systematics:  Only systematics that contribute in a correlated way to CMB lensing and galaxy counts can change the cross-correlation power spectrum.

\NEW{The large $\fnl$ signal in cross-correlations between CMB lensing and high-redshift galaxies has already been pointed out in  \cite{Jeong0910} as a promising route to search for $\fnl$.}

\subsection{Motivation for neutrino mass from cross-correlations}

\begin{figure}[tbp]
\includegraphics[width=0.5\textwidth]{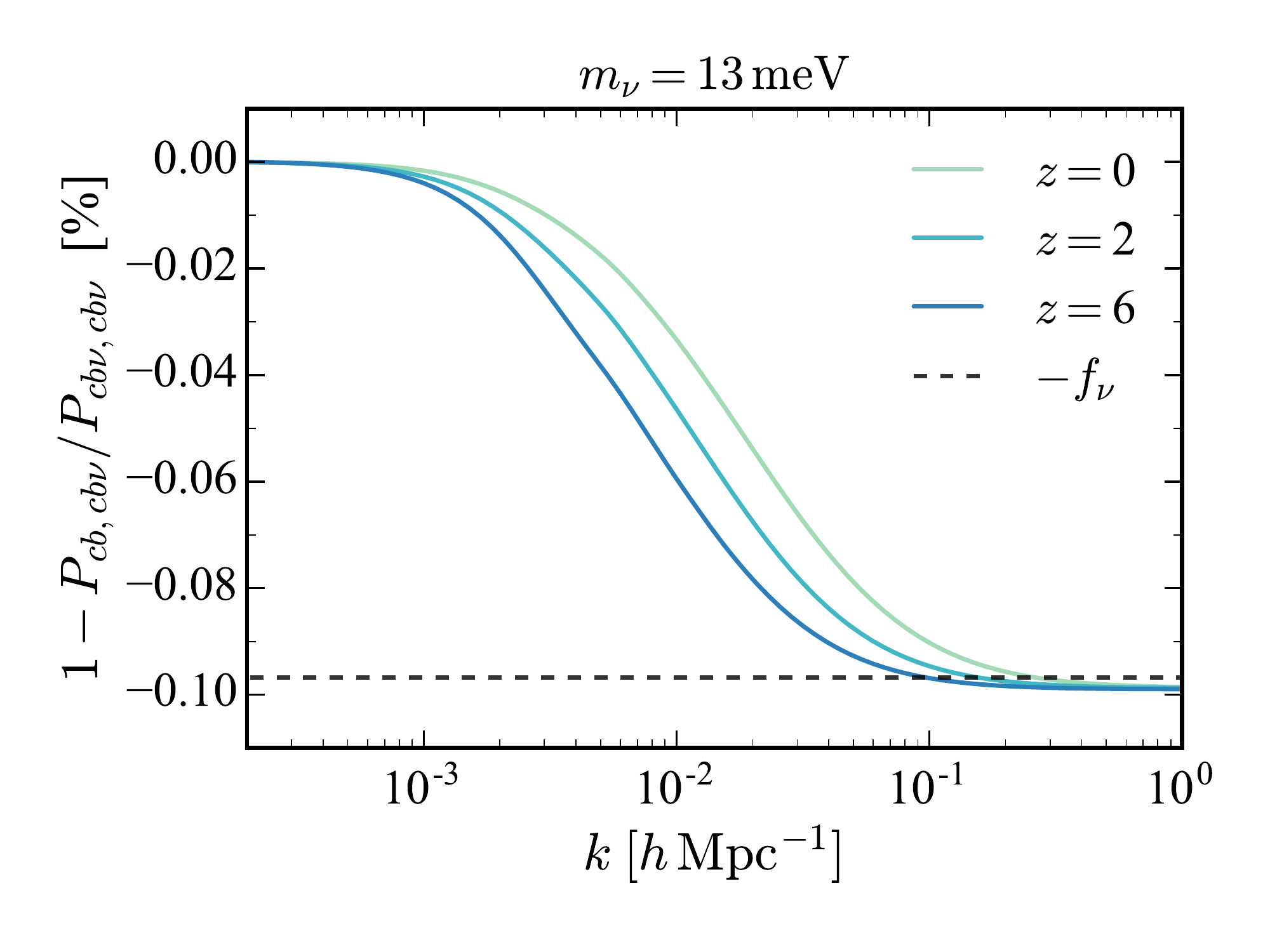}
\caption{On scales smaller than the neutrino free-streaming scale, massive neutrinos suppress the auto-power spectrum of the total matter density (`$cb\nu$') relative to its cross-spectrum with the cold-dark-matter--baryon density (`$cb$'). 
Since lensing is sensitive to the total matter density while galaxies form at peaks of the cold-dark-matter--baryon density, these spectra determine the observable lensing-lensing auto-power and the lensing-galaxy cross-power, respectively.
The suppression is $0.1\%$ for $m_\nu=13\,\mathrm{meV}$, and scales roughly linearly with $m_\nu$.
}
\label{fig:NeutrinoTransfer}
\end{figure}

The third application of cross-correlations that we consider is measuring the sum of neutrino masses $m_\nu$ using a subtle scale-dependent bias effect between lensing and clustering \cite{Villaescusa-Navarro1311,Castorina1311,LoVerde1405,LoVerde1602}.
Gravitational lensing is sensitive to all matter, so it is computed using the total matter transfer function $T_{cb\nu}$, involving cold dark matter $c$, baryons $c$, and neutrinos $\nu$. 
Galaxies, however, form in regions where cold dark matter and baryons have gravitationally collapsed, so their overdensity is computed using the transfer function $T_{cb}$, without being sensitive to the neutrino overdensity. 
As shown in \fig{NeutrinoTransfer}, the transfer functions, $T_{cb\nu}(k)$ for lensing and $T_{cb}(k)$ for clustering, are slightly different, because neutrinos free-stream on small scales and thus suppress small-scale clustering while still contributing to the energy budget responsible for the expansion of the Universe \cite{PanKnox1506}.  The different transfer functions lead to a small scale-dependent bias between lensing and clustering.
That can be used to measure neutrino mass without making any assumption about the shape of the underlying total matter power spectrum, thus providing a clean probe of neutrino mass that relies only on linear physics.
We describe this scale-dependent bias effect more quantitatively in \app{NeutrinoBasicsAppdx}.

Unfortunately, the effect is very small, leading a relative power spectrum suppression of at most $f_\nu\equiv\Omega_\nu/(\Omega_m+\Omega_\nu)$; for example, the power spectrum is suppressed by only $0.1\%$ for $m_\nu=13\,\mathrm{meV}$.
This is so small that forecasts \NEW{for determining neutrino mass using only this effect} with galaxy lensing shear and galaxy clustering do not seem promising for realistic shear shape noise and galaxy number densities \cite{LoVerde1602}.
\NEW{We study here whether the same conclusion holds for CMB-S4 CMB lensing cross-correlated with LSST galaxy clustering.}

A neutrino mass constraint from scale-dependent bias would be independent from the conventional measurement of neutrino mass that measures the suppression of small-scale low-redshift power (e.g.~in CMB lensing or galaxy clustering) relative to that expected from extrapolating the CMB to low redshift. 
In particular, neutrino mass from scale-dependent bias is not limited by the precision of the optical depth $\tau$ to the CMB, which is a major limitation for the precision of the primordial scalar amplitude and conventional cosmological neutrino mass measurements \cite{Allison1509,CMBS4SciBook}.
\NEW{Another promising avenue to measure neutrino mass without $\tau$ information, which will be explored in \cite{ByeongheeInPrep}, is to search for a small change in the cosmic growth history caused by nonzero neutrino mass using the measurements of $\sigma_8(z)$ described above.}

\section{Experiments}
\label{se:Expts}

To determine how well the above effects can be measured in the future, we study a combination of CMB and LSS experiments that we describe in this section.
We will focus on CMB-S4 CMB lensing and LSST clustering because of their low noise and significant redshift overlap, but we will also include some additional LSS samples.
At the end of the section we compute the expected cross-correlation coefficient between the CMB lensing and clustering measurements.

\subsection{CMB-S4 CMB lensing specifications}

For CMB lensing, we work with a possible CMB-S4 configuration assuming a 1 arcmin beam and $\Delta_{T}=1\,\mu$K arcmin noise \cite{CMBS4SciBook}.
We assume that the lensing reconstruction can be performed with CMB polarization modes up to $\ell_\mathrm{max}^{E,B}=5000$, and with CMB temperature modes up to $\ell_\mathrm{max}^{T}=3000$, reflecting the difficulty to clean temperature foregrounds at $\ell>3000$ using groundbased experiments.

\begin{figure}[tbp]
\includegraphics[width=0.5\textwidth]{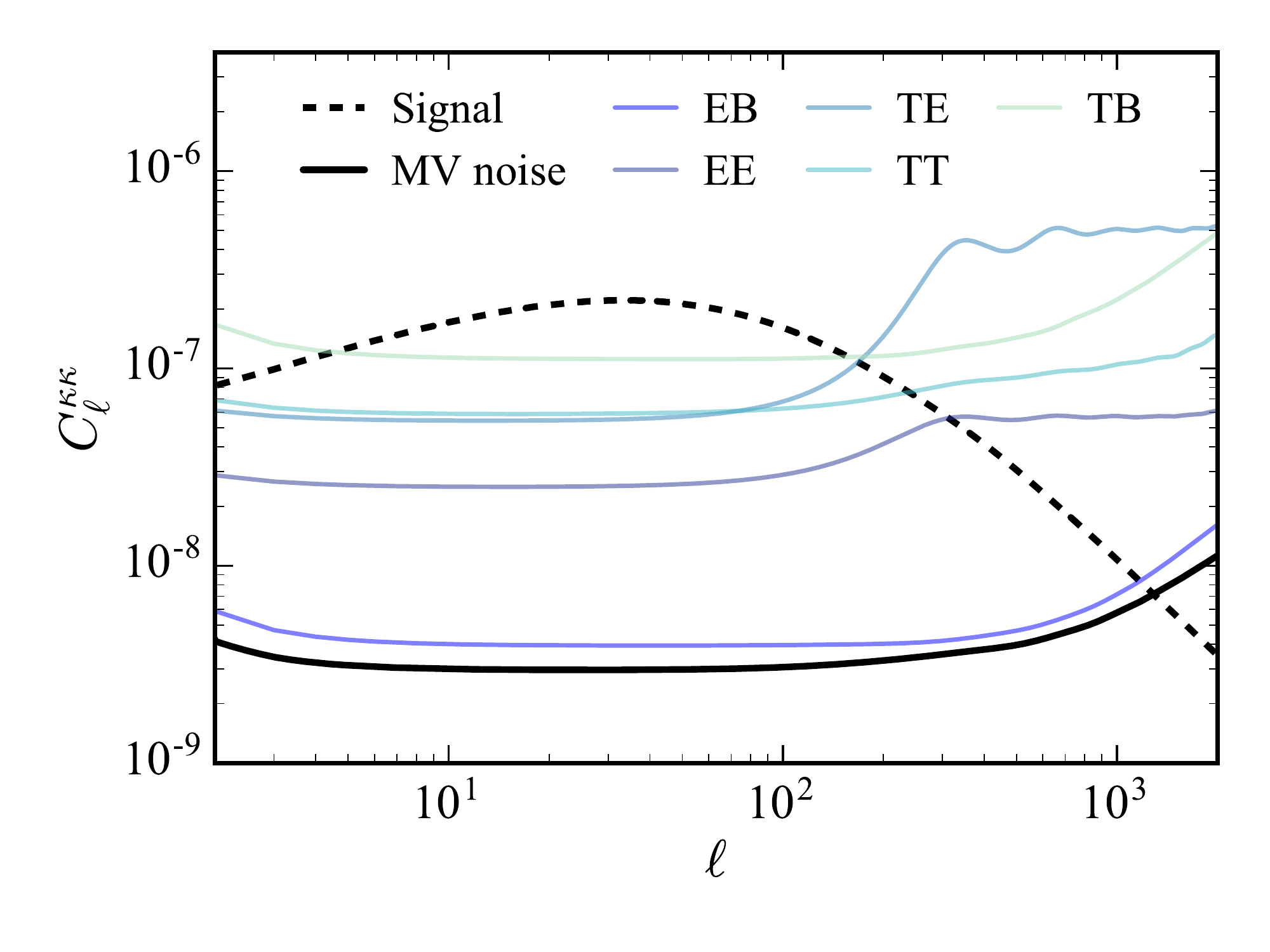}
\caption{CMB lensing power spectrum signal $C_\ell^{\kappa\kappa}$ (dashed) and expected Gaussian noise, $N^{(0),\kappa\kappa}_\ell$, from the minimum variance combination (solid black) of five individual CMB lensing estimators (solid colored). 
This assumes a possible CMB-S4 experiment with 1 arcmin beam, $\Delta_{T}=1\,\mu$K arcmin noise, $\ell_\mathrm{max}^{T}=3000$ and $\ell_\mathrm{max}^{E,B}=5000$. All noise curves were computed with \texttt{quicklens} \cite{quicklens,PlanckLensing2015} assuming quadratic estimator lens reconstruction on the full sky. The $EB$ noise is divided by a factor of $2.5$ to approximately match the expected improvement from iterative lens reconstruction \cite{HirataSeljakIterativeRec,SmithHanson1010} for CMB-S4 (this factor was obtained by matching Fig.~46 of \cite{CMBS4SciBook}). 
The signal is computed exactly at $\ell\le 50$ and using the Limber approximation at $\ell>50$.
}
\label{fig:N0}
\end{figure}

\fig{N0} shows the expected CMB lensing signal and Gaussian noise, which is essentially given by the number of signal-dominated background CMB modes.
The CMB lensing measurement is signal-dominated for $\ell\lesssim 1000$. 
On large scales, $\ell\sim 30$, the signal-to-noise per mode reaches more than 70.
This high signal-to-noise measurement of individual CMB lensing modes is important to reduce stochasticity between the CMB lensing map and maps of biased tracers.
The minimum-variance combination of the lensing estimators is dominated by the $EB$ lens reconstruction, especially after including the factor $2.5$ improvement due to iterative reconstruction over quadratic reconstruction expected for CMB-S4 (see caption of \fig{N0}).

\subsection{LSST clustering specifications}
\begin{figure}[tbp]
\includegraphics[width=0.45\textwidth]{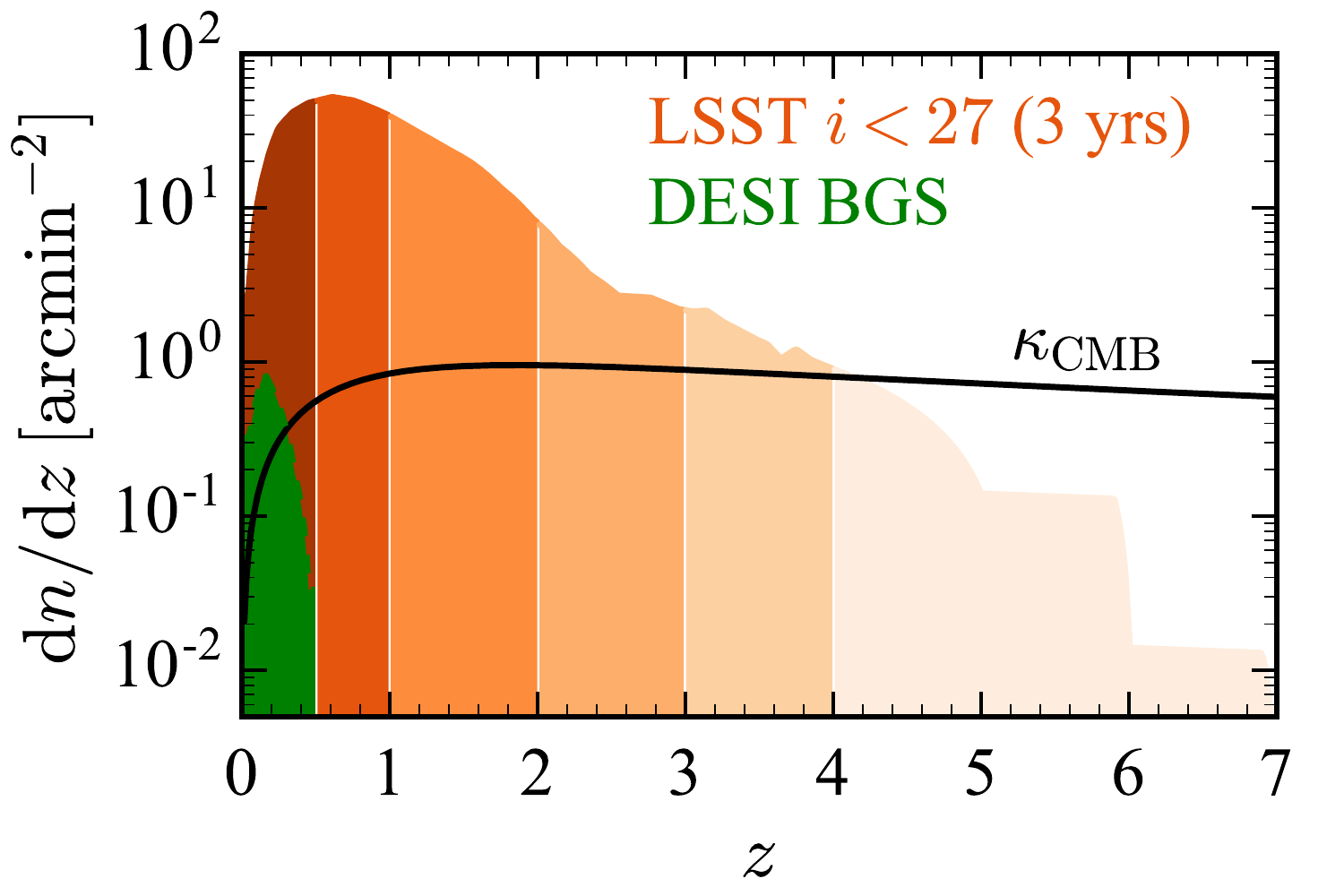}
\caption{Number density of tomographic LSST redshift bins, and one low-redshift bin from DESI.
For comparison, we also show the CMB lensing kernel (solid black), corresponding to the number density $\mathrm{d}n/\mathrm{d}z$ that would yield $C_\ell^{\kappa\kappa}$ if integrated over, with arbitrary normalization. It peaks around $z= 2$ and drops at lower and higher redshift (this is somewhat difficult to see because of the logarithmic vertical axis).
}
\label{fig:dndz_z7}
\end{figure}

For LSST clustering measurements, we assume a number density of galaxies that can be achieved by selecting with an $i<27$ magnitude cut with $S/N>5$ in at least the $i$ band for three years of observations, based on Fig.~7 in Ref.~\cite{LSST1301}.
The resulting galaxy number density is shown in \fig{dndz_z7}. 
It peaks at about $50\,\mathrm{arcmin}^{-2}$ at $z\simeq 0.6$. 
\NEW{This is approximately two times larger than the maximum number density of the LSST gold sample after 3 years of observation.
Although these galaxies might not be suitable for measuring galaxy lensing shear, their angular positions can still be used for measuring clustering and cross-correlation with the broad-kernel CMB lensing convergence, which is all we use here.
For some applications it may also be sufficient to use the measured cross-correlation coefficient between galaxies and CMB lensing convergence even if the underlying redshift distribution or other properties of the galaxies are unknown (similarly to delensing the CMB with the cosmic infrared background CIB \cite{SherwinSchmittfull1502}, or combining clustering and galaxy-galaxy-lensing to reconstruct the dark matter correlation function \cite{Baldauf0911,2013MNRAS.432.1544M}).}

\NEW{At higher redshift, $z>4$, imaging surveys with broad bands in the optical/near infrared can identify Lyman break galaxies (LBGs) using the dropout technique; see \cite{Dunlop1205} for a review. 
This technique identifies the Lyman break in galaxy spectra caused by neutral hydrogen absorption of rest-frame UV continuum emission, by looking for galaxies that are visible in short-wavelength bands but disappear in long-wavelength bands.
Recently, the Great Optically Luminous Dropout Research Using Subaru HSC (GOLDRUSH) program used this technique to identify $579,\!565$ dropout candidates at $z\simeq 4-7$ using $100\,\mathrm{deg}^2$ of Hyper Suprime-Cam observations \cite{Goldrush1}.
The sample was split into $540,\!011$, $38,\!944$, and $537$ LBGs at $z\sim 4$, 5, and 6, respectively, to measure angular clustering of these galaxies \cite{Goldrush2}.
Repeating these HSC observations on $18,\!000\,\mathrm{deg}^2$ would yield 180 times more galaxies, leading to 100 million dropout galaxies $z=4-7$.
To the extent that LSST filters and magnitude limits are comparable to those of HSC, which is approximately the case, we therefore expect roughly 100 million dropout galaxies at $z=4-7$ from LSST.
To include such a possible high-redshift LSST dropout sample in our forecast, we extrapolate the LSST redshift distribution from $z\le 4$ to the redshift range $4\le z \le 5$ as shown in \fig{dndz_z7}.
At higher redshift, we assume $\mathrm{d}n/\mathrm{d}z=0.14\,\mathrm{arcmin}^{-2}$ at $5 \le z \le 6$, and $\mathrm{d}n/\mathrm{d}z=0.014\,\mathrm{arcmin}^{-2}$ at $6 \le z \le 7$.
This corresponds to 43 million dropout galaxies at $z=4-7$, which is about two times less than the 100 million galaxies estimated above.
}

We split the LSST galaxies into six broad tomographic redshift bins, $z=0-0.5, 0.5-1, 1-2, 2-3, 3-4$, and $z=4-7$.
There is no need for finer redshift bins to study the effects we are after, which all have rather broad redshift kernels, but it is important to use more than one redshift bin to be able to weight the redshift bins to match the CMB lensing kernel, \NEW{and we find that parameter contraints improve somewhat when choosing finer redshifts bins, likely because the kernels are not perfectly constant within the broad redshift bins above.}
Should one be looking at features that 
are narrow in redshift, a finer redshift binning would become more 
important.
In the main forecasts, we do not account for photometric redshift uncertainties, noting that redshifts should be sufficiently accurate to correctly assign them to the broad redshift bins defined above; \NEW{we will revisit this assumption in \secref{zerrors}}.
For a survey area of $18,000\,\mathrm{deg}^2$, the total number of galaxies in each redshift bin is $N_\mathrm{tot}=9.3\times 10^8, 1.55\times 10^9, 1.40\times 10^9, 2.4\times 10^8, 9.4\times 10^7$, and $4.3\times 10^7$, corresponding to about one billion objects in each of the low-redshift bins, and more than 40 million objects at $z=4-7$ (see discussion above). 
We assume a fiducial linear galaxy bias of $b(z)=1+z$ \cite{LSSTScienceBook}.

\subsection{Other LSS surveys}
In our default forecasts, we also include number counts from SDSS \cite{SDSSwebsite}, BOSS  \cite{BOSSwebsite} and DESI \cite{DESIwebsite}.

For SDSS, we assume the number density of $r<22$ photometric redshifts obtained in Ref.~\cite{Rahman1512} using the clustering redshift technique \cite{2008ApJ...684...88N,McQuinn:2013ib,Menard:2013aaa}. We split the sample in two tomographic redshift bins, one at $0\le z\le 0.5$ and one at $0.5\le z\le 0.8$.
For a survey area of $4,800\,\mathrm{deg}^2$, this gives $N_\mathrm{tot}=1.1\times 10^8$ objects in each bin.
We assume the bias to be $b(z)=1$ for $z<0.1$ and $b(z)=1+(z-0.1)$ for $z\ge 0.1$.

For BOSS, we use spectroscopic redshifts of luminous red galaxies (LRGs) with the same number density as in Table II of Ref.~\cite{FontRibera1308}. 
We use a single redshift bin $0\le z\le 0.9$.
On a sky area of $9,329\,\mathrm{deg}^2$ this would give $1.3\times 10^6$ galaxies. Splitting the sample into multiple redshift bins does not improve our forecasts because we cross-correlate against CMB lensing, so that redshift accuracy is much less important than number density.
We assume a bias of $b(z)=1.7\bar D^{-1}(z)$ where $\bar D(z=0)=1$.

For DESI, we use five redshift samples, with number densities from Table 2.3 in Ref.~\cite{DESIFDRDoc}: The low-redshift BGS sample at $0\le z\le 0.5$ with $9.6\times 10^6$ objects and bias $b(z)=1.34\bar D^{-1}(z)$, the LRG sample at $0.6\le z \le 1.2$ with $3.9\times 10^6$ objects and bias $b(z)=1.7\bar D^{-1}(z)$, one ELG sample at $0.6\le z\le 0.8$ with $3.5\times 10^6$ objects and bias $b(z)=0.84\bar D^{-1}(z)$, a second ELG sample at $0.8\le z\le 1.7$ with $1.3\times 10^7$ objects and the same bias, and a QSO sample at $0.6\le z\le 1.9$ with $1.4\times 10^6$ objects and bias $1.2\bar D^{-1}(z)$.
In each case, the number of objects refers to a survey area of $14,000\,\mathrm{deg}^2$.

\subsection{CMB lensing--LSS correlation coefficient}

\begin{figure}[tbp]
\includegraphics[width=0.45\textwidth]{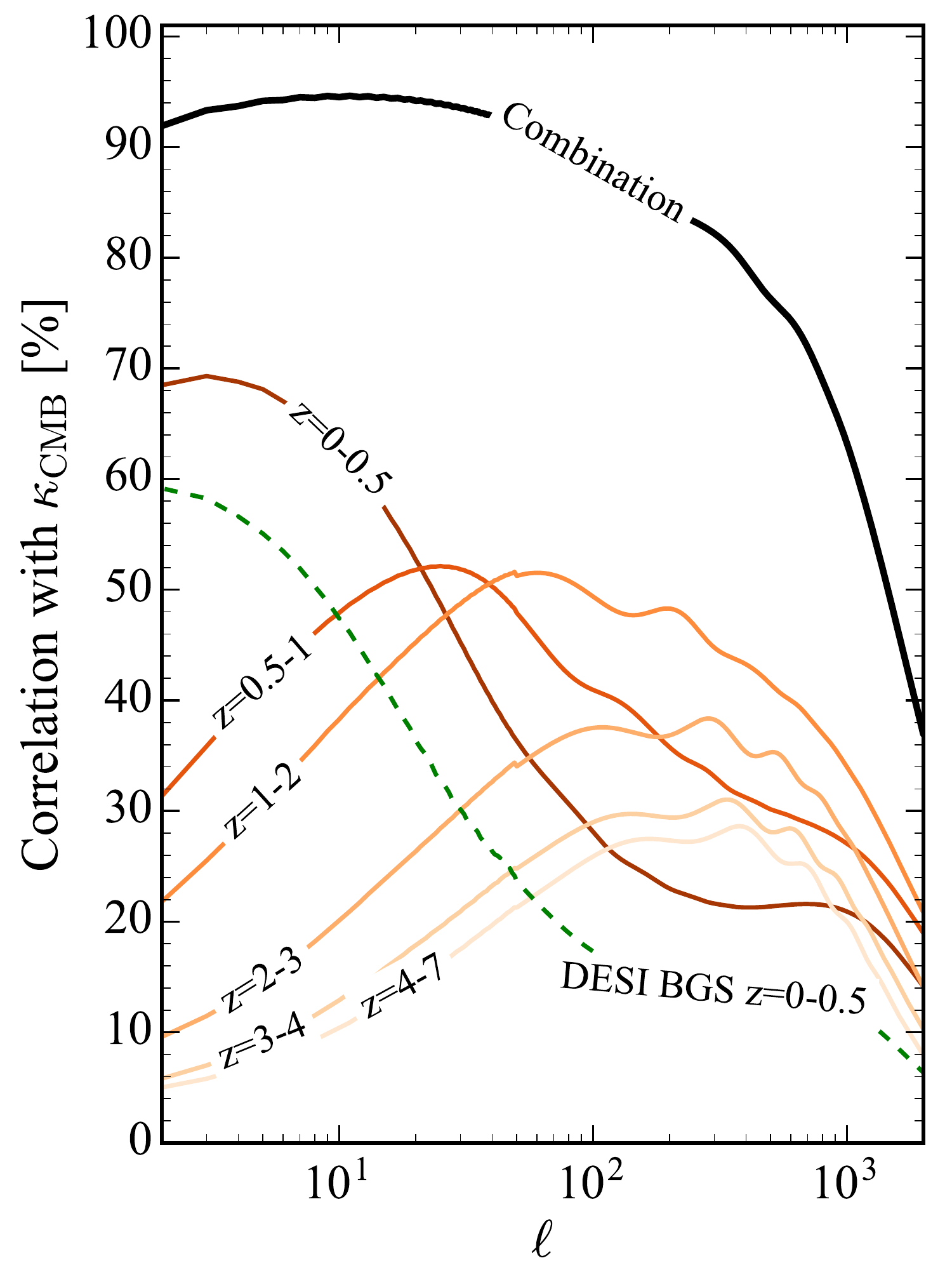}
\caption{Expected correlation coefficient $r_\ell=C^{\kappa g}_\ell(\hat C^{\kappa\kappa}_\ell\hat C^{gg}_\ell)^{-1/2}$ of CMB-S4 lensing measurements with six tomographic LSST samples (orange), with the low-redshift DESI BGS sample (green dashed), and with the optimal combination of these LSS tracers (black), as a function of wavenumber $\ell$.
The level of correlation is determined by the redshift overlap between CMB lensing and LSS samples and by their noise levels.
The plot includes CMB-S4 lensing noise and galaxy shot noise given by the number density in \fig{dndz_z7}.
The Limber approximation would wrongly predict the low-$\ell$ correlation of individual redshift bins to be $5$ to $10\%$ higher than the exact result shown here.
}
\label{fig:rho}
\end{figure}

The performance of the cross-correlation analyses depends on the cross-correlation coefficient
\begin{align}
  \label{eq:rcc}
  r_\ell = \frac{C_\ell^{\kappa g}}{\sqrt{\hat C_\ell^{\kappa\kappa}\hat C_\ell^{gg}}}
\end{align}
between the measured CMB lensing convergence $\kappa$ and the observed galaxy density $\delta_g$, where the power spectra $\hat C$ include lensing reconstruction noise and shot noise.
\fig{rho} shows the correlation coefficient of tomographic LSST redshift bins with lensing measurements expected from CMB-S4.

The correlation of the low-redshift LSST bin at $z=0-0.5$ with CMB lensing peaks at $70\%$ on very large scales, $\ell=3$, and drops on smaller scales.
The LSST samples at higher redshift reach their maximum correlation with CMB lensing at higher $\ell$, corresponding approximately to the peak of the 3-D power spectrum at $k_\mathrm{peak}\sim 2\times 10^{-2}\,h\mathrm{Mpc}^{-1}$, which is mapped to higher $\ell$ for higher redshift ($\ell_\mathrm{peak}=k_\mathrm{peak}\chi(z)$ where $\chi$ ranges from $\chi(z=0.1)\sim 400\,h^{-1}\mathrm{Mpc}$ to $\chi(z=7)\sim 9\,h^{-1}\mathrm{Gpc}$).
The low-redshift DESI BGS sample also has a substantial correlation with CMB lensing, reaching up to $60\%$ at low $\ell$.

The tomographic redshift bins can be combined into a single joint LSS sample, with redshift bins weighted to match the CMB lensing kernel.
Choosing these weights such that they maximize the correlation coefficient between the joint LSS sample and CMB lensing \cite{BlakeMarcel1502} gives the correlation coefficient shown in black in \fig{rho}.
The combined LSS sample is more than $92\%$ correlated with the CMB-S4 lensing measurement at $\ell\lesssim 40$, reaching a maximal correlation of $r=94.6\%$ at $\ell\simeq 10$.
This is combining all LSST redshift bins and the DESI BGS sample.
Additionally including SDSS and all other DESI samples described above increases the maximal correlation only mildly, to $r=94.8\%$.
The high correlation coefficient motivates exploring sample variance cancellation techniques for these experiments.
On smaller scales, the cross-correlation drops, but is still $60\%$ for the combined LSS sample at $\ell=1000$.

It may be surprising that the cross-correlation coefficient of the combined LSS sample can be as high as 
95\% despite the CMB lensing kernel being very broad (\fig{dndz_z7}) and extending all the way 
to $ z\sim 1100$. The reason is that at low $\ell$ the scales at 
cosmological distances $\chi$ (typically a few $h^{-1}\mathrm{Gpc}$) correspond to a 
very low $k$ ($k=\ell/\chi$, so for $\ell=10$ typically $k \sim 10^{-2}\,h\mathrm{Mpc}^{-1}$). Since this $k$ is lower than the peak of the power spectrum at $k_\mathrm{peak} \sim 2 \times 10^{-2}\,h\mathrm{Mpc}^{-1}$,
 the power spectrum has more power on 
smaller scales, so the projection integral picks most of the power from low 
values of $\chi$ and thus from low $z$. 
At higher $\ell$ we move to scales smaller than the peak 
of the power spectrum and the 
contribution from $z>4$ LSS becomes more and more important. Moreover, 
even though LSST has some sources at $z>4$ they are sparse and the 
corresponding shot noise reduces the cross-correlation coefficient.

One can also rephrase the above sample variance cancellation argument using delensing:
The more the tracers are correlated with the true CMB lensing, the better they delens the CMB modes; the delensed $B$ mode power follows by replacing $C^{\kappa\kappa}\rightarrow C^{\kappa\kappa}(1-\rho^2)$, where $\rho$ is the cross-correlation coefficient of optimally combined tracers with the true CMB lensing convergence without lens reconstruction noise \cite{BlakeMarcel1502}.
Thus, the more one can delens by combining multiple tracers, the more noise one removes from the cross-correlation of those tracers with CMB lensing. It is important to emphasize that the tracers need to cover as much redshift range as possible.

\section{Power spectra}
\label{se:PowerSpectra}

Assuming the above experiment specifications, we can compute angular power spectra, their expected statistical uncertainties, and how they change in presence of scale-dependent bias caused by $\fnl$ or neutrino mass.

\subsection{Angular power spectra and noise}

\begin{figure*}[tbp]
\includegraphics[width=0.95\textwidth]{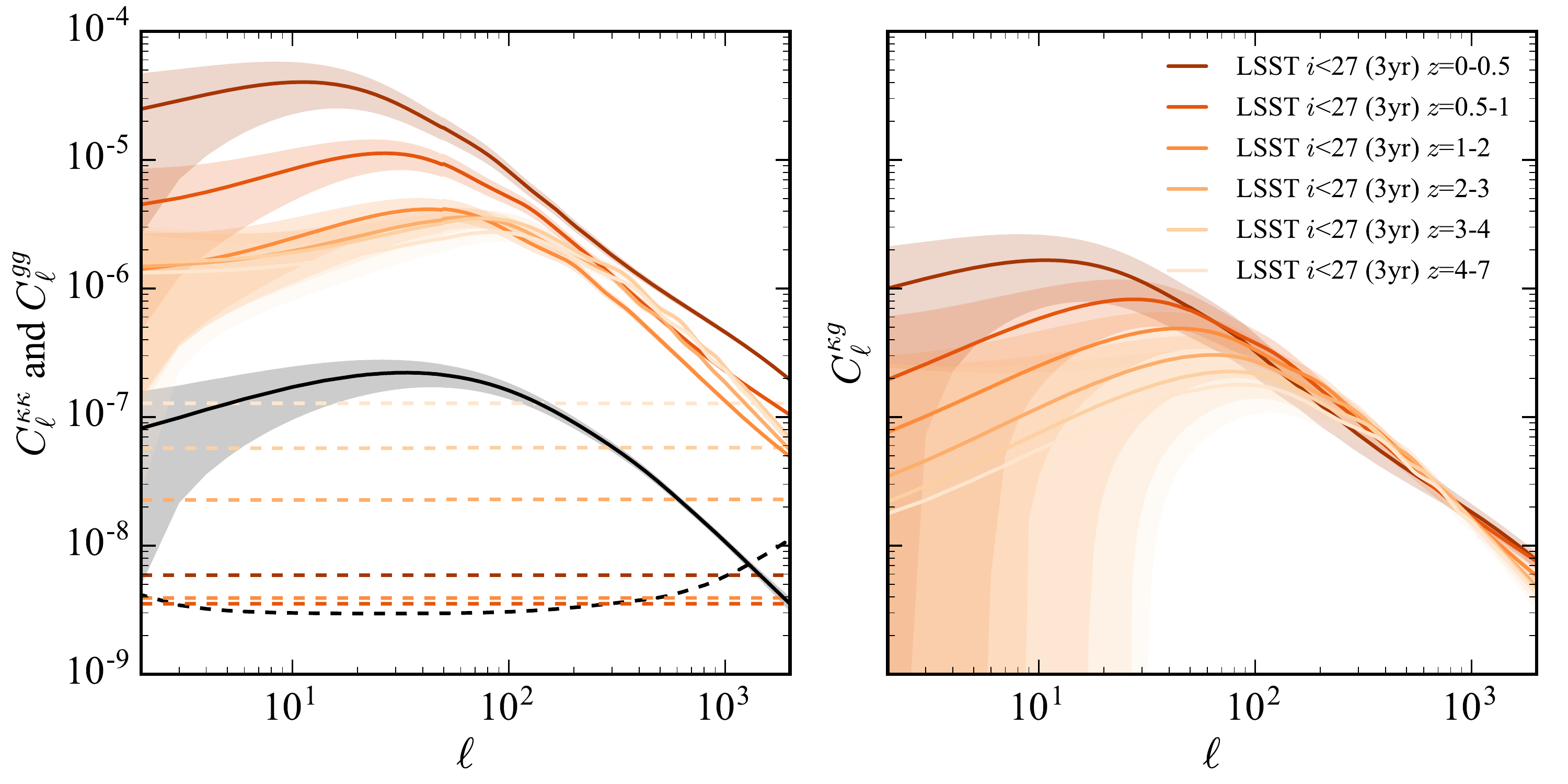}
\caption{\emph{Left panel:} Angular auto-power spectra of CMB-S4 lensing convergence $\kappa$ (black) and LSST galaxy density (colored). 
Solid lines show the signal power (not including lensing noise or shot noise), and shaded regions show 1$\sigma$ error bars assuming the Gaussian covariance \eq{GaussianCov}, $\fsky=0.5$, minimum variance lensing noise expected for CMB-S4, and LSST number density shown in \fig{dndz_z7}.
Dashed lines show lensing reconstruction noise (black) and shot noise (colored).
\emph{Right panel:} Angular cross-spectra between CMB lensing and LSST galaxy density.
}
\label{fig:ClsWithNoise}
\end{figure*}

In the left panel of \fig{ClsWithNoise} we show angular auto-power spectra of CMB-S4 lensing and LSST clustering. The shaded regions show the expected uncertainty
\begin{align}
  \label{eq:34}
  \sigma(C^{XX}_\ell) = \left[\frac{2}{\fsky (2\ell+1)}\left(\hat C^{XX}_\ell\right)^2\right]^{1/2}
\end{align}
due to sampling variance, CMB lensing reconstruction noise, and shot noise (included in $\hat C^{XX}$).
The spectra are signal-dominated up to at least $\ell=1000$ thanks to the low CMB lensing noise and high LSST number density. 
The overall shape of the angular power spectra is similar to the 3-D matter power spectrum, with the peak at the physical scale $k_\mathrm{peak}^{-1}$ mapped to smaller angular scales (higher $\ell$) for increasing redshift.

\begin{table}[tbp]
\centering
\renewcommand{\arraystretch}{1.0}
\begin{tabular}{@{}p{4.0cm}lllllll@{}}
\toprule
 & \phantom{} & \multicolumn{5}{l}{$\ell_\mathrm{max}$}  \\ 
SNR of $C^{XX}$ &&  500  &\phantom{$\,$}&  1000  &\phantom{$\,$}&  2000 \\ 
\colrule 
$\kappa_\mathrm{CMB}$  && 233 && 406 && 539 \\
BOSS LRG $z$=0-0.9  && 140 && 187 && 230 \\
SDSS $r<22$ $z$=0-0.5  && 247 && 487 && 936 \\
SDSS $r<22$ $z$=0.5-0.8  && 247 && 487 && 936 \\
DESI BGS $z$=0-0.5  && 230 && 417 && 665 \\
DESI ELG $z$=0.6-0.8  && 158 && 210 && 256 \\
DESI ELG $z$=0.8-1.7  && 150 && 194 && 225 \\
DESI LRG $z$=0.6-1.2  && 184 && 267 && 349 \\
DESI QSO $z$=0.6-1.9  && 44.8 && 48.8 && 50.8 \\
LSST $i<27$ (3yr) $z$=0-0.5  && 250 && 496 && 982 \\
LSST $i<27$ (3yr) $z$=0.5-1  && 250 && 496 && 979 \\
LSST $i<27$ (3yr) $z$=1-2  && 249 && 492 && 956 \\
LSST $i<27$ (3yr) $z$=2-3  && 245 && 469 && 830 \\
LSST $i<27$ (3yr) $z$=3-4  && 239 && 444 && 724 \\
LSST $i<27$ (3yr) $z$=4-7  && 224 && 387 && 555 \\
\botrule
\end{tabular}
\caption{Total signal-to-noise 
 of auto-power spectra $C^{XX}_\ell$ of CMB lensing convergence and galaxy density in tomographic redshift bins.
We assume $\fsky=0.5$, $\ell_\mathrm{min}=2$, and $\ell_\mathrm{max}\in\{500,1000,2000\}$ in different columns.
The noise includes CMB-S4 lensing reconstruction noise and shot noise.
}
\label{tab:AutoSNR}
\end{table}

In Table~\ref{tab:AutoSNR} we show the total signal-to-noise ratio 
\begin{align}
  \label{eq:10}
  \mathrm{SNR} = 
\left[\sum_{\ell=\ell_\mathrm{min}}^{\ell_\mathrm{max}} \left(\frac{C^{XX}_\ell}{\sigma(C^{XX}_\ell)}\right)^2\right]^{1/2}
\end{align}
of these auto-power spectra.
The CMB-S4 lensing auto-power spectrum has a signal-to-noise of $406$ for $\ell_\mathrm{max}=1000$. For $\ell_\mathrm{max}=2000$ this improves only moderately to a signal-to-noise of $539$ because CMB lensing noise becomes relevant at lensing scales $\ell>1000$.
The tomographic LSS redshift bins have comparable signal-to-noise for $\ell_\mathrm{max}=1000$; for example the photometric redshift samples of SDSS, the DESI BGS low-redshift sample, and each of the 6 LSST redshift bins have a total signal-to-noise of $\sim 400$.
Going to $\ell_\mathrm{max}=2000$ improves the signal-to-noise of most of these samples to $\sim 900$.

\begin{table}[tbp]
\centering
\renewcommand{\arraystretch}{1.0}
\begin{tabular}{@{}p{4.0cm}lllllll@{}}
\toprule
 & \phantom{} & \multicolumn{5}{l}{$\ell_\mathrm{max}$}  \\ 
SNR of $C^{\kappa_\mathrm{CMB}X}$ &&  500  &\phantom{$\,$}&  1000  &\phantom{$\,$}&  2000 \\ 
\colrule 
BOSS LRG $z$=0-0.9  && 77.3 && 117 && 159 \\
SDSS $r<22$ $z$=0-0.5  && 88.3 && 167 && 284 \\
SDSS $r<22$ $z$=0.5-0.8  && 88.3 && 167 && 284 \\
DESI BGS $z$=0-0.5  && 50.1 && 93.5 && 144 \\
DESI ELG $z$=0.6-0.8  && 50.7 && 73.5 && 97 \\
DESI ELG $z$=0.8-1.7  && 103 && 148 && 185 \\
DESI LRG $z$=0.6-1.2  && 86.7 && 133 && 182 \\
DESI QSO $z$=0.6-1.9  && 74.9 && 94.5 && 108 \\
LSST $i<27$ (3yr) $z$=0-0.5  && 78.1 && 150 && 258 \\
LSST $i<27$ (3yr) $z$=0.5-1  && 112 && 202 && 338 \\
LSST $i<27$ (3yr) $z$=1-2  && 144 && 259 && 406 \\
LSST $i<27$ (3yr) $z$=2-3  && 121 && 219 && 324 \\
LSST $i<27$ (3yr) $z$=3-4  && 101 && 182 && 261 \\
LSST $i<27$ (3yr) $z$=4-7  && 94 && 167 && 229 \\
\botrule
\end{tabular}
\caption{Like Table~\ref{tab:AutoSNR} but for CMB-lensing--clustering cross-spectra $C^{\kappa g}_\ell$.}
\label{tab:CrossSNR}
\end{table}

The cross-spectra between CMB lensing and galaxy clustering can also be measured very accurately.
This is shown in the right panel of \fig{ClsWithNoise} and in Table~\ref{tab:CrossSNR}.
The total signal-to-noise of those cross-spectra reaches more than $200$ for $\ell_\mathrm{max}=1000$, and up to $400$ for $\ell_\mathrm{max}=2000$ in the case of the LSST redshift bin at $z=1-2$ where the CMB lensing kernel peaks.
Even though the overall error is larger than for auto-power, at low $\ell$ 
the two errors share the sampling variance term, and upon taking the 
ratio of the two measurements this error cancels out. This is the 
basis of the sampling variance cancellation method.

\subsection{\texorpdfstring{$\fnl$}{fNL} signal and signal-to-noise}

\begin{figure}[tbp]
\includegraphics[width=0.45\textwidth]{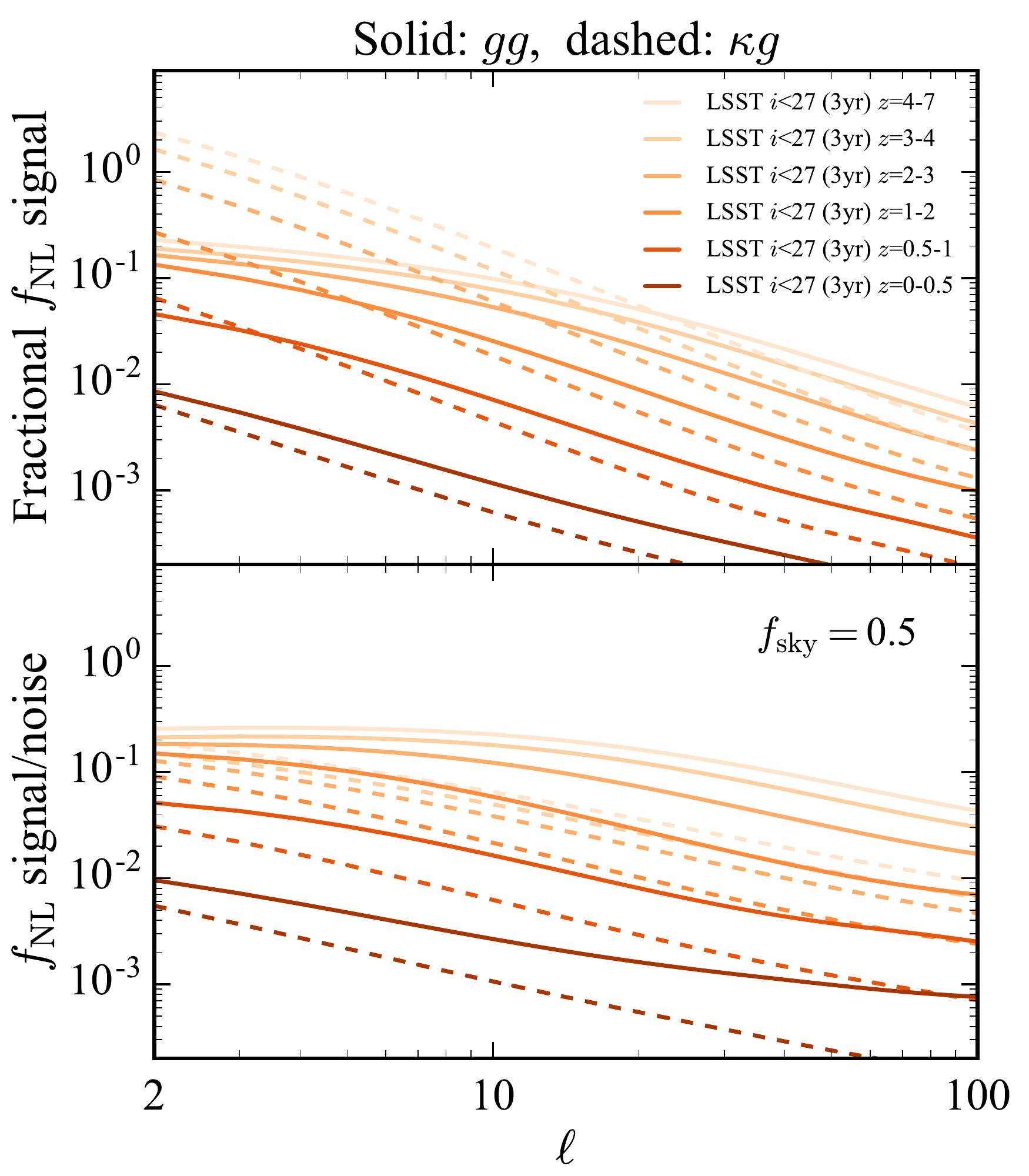}
\caption{\emph{Upper panel:} Fractional $f_\mathrm{NL}$ signal $(\partial C_\ell/\partial \fnl)/\hat C_\ell$ from scale-dependent bias as described in \app{fnlBasicsAppdx}, for $C^{gg}_\ell$ (solid) and $C^{\kappa g}_\ell$ (dashed).
\emph{Lower panel:} $\fnl$ signal divided by cosmic variance noise and shot noise, $(\partial C_\ell/\partial \fnl)/\sigma(\hat C_\ell)$.
Most of the $\fnl$ signal-to-noise comes from large scales and high redshift, as expected.
By cancelling part of the cosmic variance, a joint analysis can yield tighter constraints than naively expected by adding up the shown signal-to-noise in quadrature.
}
\label{fig:fnlSignalNoise}
\end{figure}

\fig{fnlSignalNoise} shows the fractional $\fnl$ signal from scale-dependent bias for galaxy auto-spectra $C^{gg}$ (solid), and for CMB lensing--galaxy clustering cross-spectra $C^{\kappa g}$ (dashed).

For galaxy auto-spectra, $\fnl=1$ can change the signal by more than 10\% on large scales at high redshift ($\ell\lesssim 5$, $z\gtrsim 2$).
At $\ell=20$ the signal is still $5\%$ at high redshift, but less than a percent at low redshift. 
The lower panel of \fig{fnlSignalNoise} compares these signals against the cosmic variance of each spectrum, without combining any measurements or exploiting sample variance cancellation.
This shows that the $\fnl$ signal-to-noise can be larger than $0.1\sigma$ per mode at $\ell\lesssim 30$ for high-redshift tracers.
For lower-redshift tracers this is significantly smaller; for example, the $\fnl$ signal-to-noise of the $z=0.5-1$ bin is $0.05\sigma$ per mode at $\ell=2$, and $0.01\sigma$ per mode at $\ell=20$.

\NEW{For $\kappa g$ cross-spectra, $\fnl=1$ changes the signal at $\ell=10$ by up to $20\%$ for the highest-redshift galaxies, and by several percent for galaxies at lower redshifts.  This is comparable to \cite{Jeong0910} who found a $4\%$ change for $\fnl=1$ at $\ell=10$.}
The $\kappa g$ cross-spectra have a larger fractional $\fnl$ signal than $gg$ auto-spectra on large scales (if beyond-Limber corrections are included).
However, the $\fnl$ signal-to-noise of each $\kappa g$ is always less than the corresponding $gg$ auto-spectrum.
The reason for this is that the cross-correlation coefficient $r_\ell$ between CMB lensing and each individual tomographic redshift bin, shown in \fig{rho} above,
is relatively small, which adds noise to the $\kappa g$ cross-spectrum that is not present in $gg$ spectra.
This can be seen explicitly by writing the fractional uncertainty of $C_\ell^{\kappa g}$ in terms of the correlation coefficient $r_\ell$ (also see \eqq{CrossFractError} below):
\begin{align}
  \label{eq:47}
\frac{\sigma(C_\ell^{\kappa g})}{C_\ell^{\kappa g}}
=
\left[\frac{1+r_\ell^{-2}}{(2\ell+1)\fsky}\right]^{1/2}.
\end{align}

The trends of the $\fnl$ signal and signal-to-noise with redshift and wavenumber are easily understood. 
Both signal and signal-to-noise are larger at lower $\ell$ and higher redshift (brighter colors), because the signal scales as $1+\fnl\beta$, where $\beta\propto(b-1)/b/k^2$ is the fractional bias change for $\fnl=1$ (defined in \eqq{fnlbias}), and the Gaussian bias $b$ increases with redshift.
In the regime where the Limber approximation holds, $\ell\gtrsim 30$, the fractional $\fnl$ signal of $gg$ spectra is about twice that of $\kappa g$ spectra,
because the former scale as 
$(1+\fnl\beta)^2\approx 1+2\fnl\beta$
whereas the latter scale as $1+\fnl\beta$.

If all experiments observe mutually independent patches of the sky, the total signal-to-noise would be given by adding the individual significances in quadrature.
If all experiments observe the same patch, however, the spectra can be correlated so that the total uncertainty can be smaller than the  uncertainty expected from the naive estimates of the signal relative to the cosmic variance of each spectrum \cite{Uros0807}.
We will discuss this improvement from sample variance cancellation more quantitatively later using Fisher forecasts.

\subsection{Neutrino mass signal and signal-to-noise}

\begin{figure}[tbp]
\includegraphics[width=0.45\textwidth]{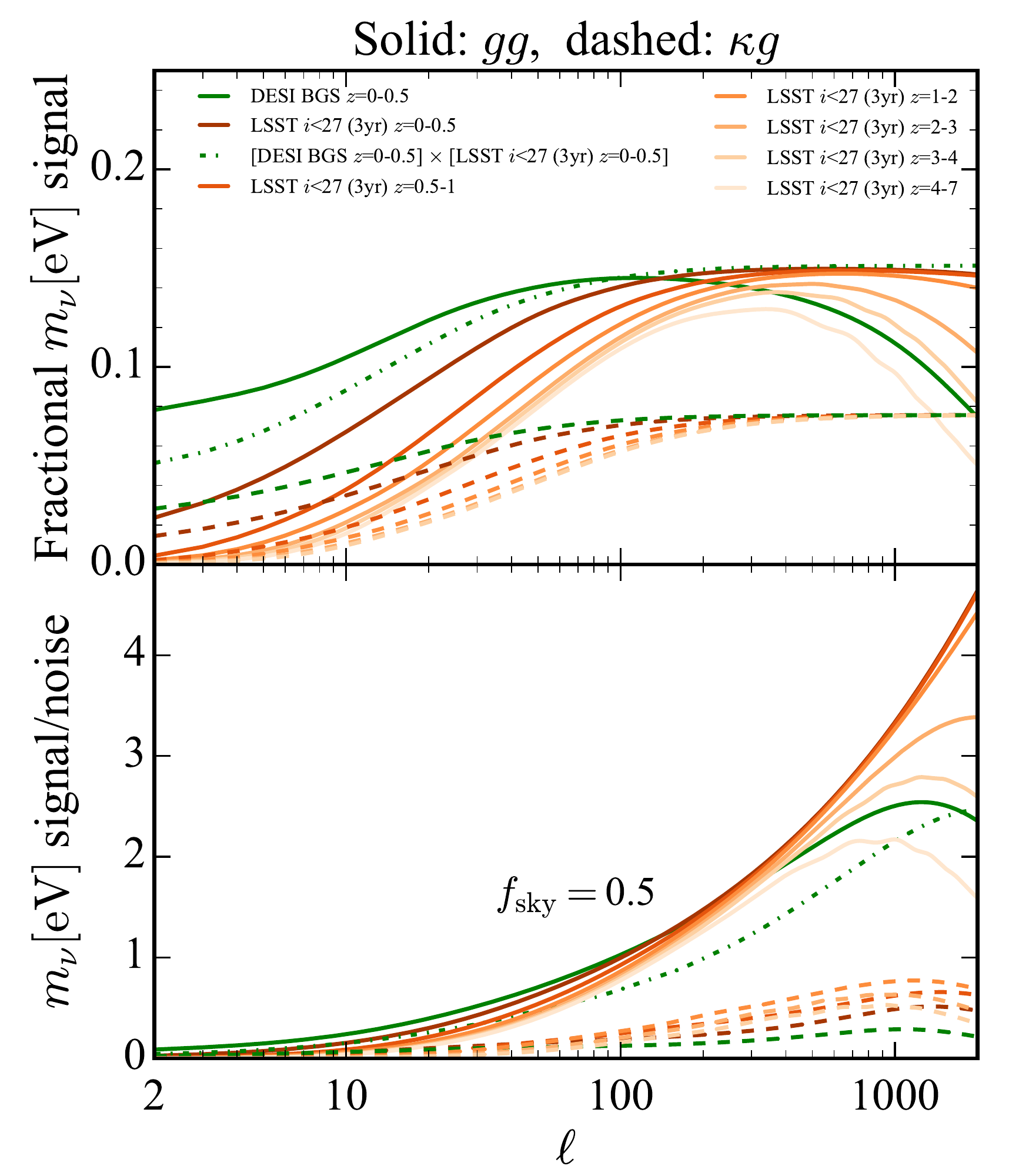}
\caption{\emph{Upper panel:} Fractional neutrino mass signal $(\partial C_\ell/\partial m_\nu)/\hat C_\ell$ from scale-dependent bias as described in \app{NeutrinoBasicsAppdx} for some $C^{gg}$ (solid) and $C^{\kappa g}$ (dashed) power spectra.
\emph{Lower panel:} Neutrino mass signal divided by cosmic variance and shot noise, $(\partial C_\ell/\partial m_\nu)/\sigma(\hat C_\ell)$, for $m_\nu=1\,\mathrm{eV}$.
In both panels, $gg$ spectra drop at high $\ell$ because we include shot noise in all $C_\ell$'s.
To include only the signal from differences in lensing and clustering, we will marginalize over galaxy bias and a fake parameter $m_\nu^\mathrm{fake}$ that rescales $\kappa\kappa$, $\kappa g$ and $gg$ spectra in the same way (with a shape matched to the scale-dependent bias).
In the upper panel, the signal of high-redshift samples starts to rise at higher $\ell$ than for low-redshift samples; this is because high-redshift samples have more power at higher $\ell$ because of the 3-D to 2-D mapping.
}
\label{fig:mnuSignalNoise}
\end{figure}

In \fig{mnuSignalNoise} we show the fractional signal of the scale-dependent neutrino mass bias caused by the different transfer functions relevant for angular $gg$ and $\kappa g$ power spectra, assuming a very large neutrino mass of $m_\nu=1\,\mathrm{eV}$.
The scale-dependent transition shown for the 3-D power spectrum in \fig{NeutrinoTransfer} is mapped to angular wavenumbers $10\lesssim \ell\lesssim 300$ in 2-D.
The $gg$ spectra have about twice the signal than $\kappa g$ spectra, and even more signal-to-noise.
The signal-to-noise rises steeply with wavenumber $\ell$, although we effectively exclude the signal at very high $\ell$ by marginalizing over galaxy bias parameters, so that most of the constraining power comes from $\ell\sim $ few hundred.
These scales are sufficiently small that we will assume the Limber approximation on all scales for neutrino mass forecasts.

\section{Analytical estimate}

\label{se:AnalyticalSVCFisher}

Before presenting detailed numerical forecasts based on the above scale-dependent bias signals, we estimate analytically what precision we might expect for the fractional error of a generic scale-dependent bias amplitude $\alpha$, which can be $\alpha=\fnl$ or $\alpha=m_\nu$ depending on the application.
For simplicity we will not marginalize over any parameters in this section.

Let us assume that all LSS tracers are optimally combined to a single tracer $g=\sum_i \delta_{g_i}$.
We then compute the Fisher information of $\alpha$ if the data vector is given by the CMB lensing convergence map and the combined tracer map, $(\kappa, g)$.
The Fisher information at the field level is given by
\begin{align}
  \label{eq:FisherFieldLevel}
F_{\alpha\alpha}=\sum_\ell (2\ell+1)(F_{\alpha\alpha})_\ell,
\end{align}
where the Fisher matrix per $\ell$ is
\begin{align}
  \label{eq:FisherPerMode}
  (F_{\alpha\alpha})_\ell = \frac{1}{2}\sum_{abcd\in\{\kappa,g\}}
\frac{\partial C^{ab}_\ell}{\partial\alpha}
(\hat C^{-1})^{bc}_\ell
\frac{\partial C^{cd}_\ell}{\partial\alpha}
(\hat C^{-1})^{da}_\ell.
\end{align}
Here, 
\begin{align}
  \label{eq:9}
  \hat C^{-1} = \frac{1}{\hat C^{\kappa\kappa}\hat C^{gg}(1-r_\ell^2)}
  \begin{pmatrix}
    \hat C^{gg} & - C^{\kappa g} \\
    -C^{\kappa g} & \hat C^{\kappa\kappa}
  \end{pmatrix}.
\end{align}
is the inverse of the covariance matrix
\begin{align}
  \label{eq:9}
  \hat C = 
  \begin{pmatrix}
    \hat C^{\kappa\kappa} & C^{\kappa g} \\
    C^{\kappa g} & \hat C^{gg}
  \end{pmatrix}
\end{align}
of the data vector $(\kappa,g)$.
Recall that $\hat C^{\kappa\kappa}$ includes lensing reconstruction noise and $\hat C^{gg}$ includes shot noise, while $\hat C^{\kappa g}=C^{\kappa g}$ is just the signal.
Using the notation $\partial C_\ell/\partial \alpha=C_{\ell,\alpha}$, and noting that $C^{\kappa\kappa}_{\ell,\alpha}=0$, a lengthy but straightforward calculation gives
\begin{align}
  (F_{\alpha\alpha})_\ell = \,&
\frac{1}{2\left(1-r_\ell^2\right)^2}\Bigg[ 
\left(\frac{C^{gg}_{\ell,\alpha}}{\hat C_\ell^{gg}}\right)^2
-4r_\ell^2 
\frac{C^{gg}_{\ell,\alpha}}{\hat C^{gg}_\ell}
\frac{C^{\kappa g}_{\ell,\alpha}}{C^{\kappa g}_\ell}
\non\\
&\;\;+2r_l^2(1+r_\ell^2)
\left(\frac{C^{\kappa g}_{\ell,\alpha}}{C^{\kappa g}_\ell}\right)^2\,
\Bigg].
  \label{eq:FisherAlphaFinal}
\end{align}
The result depends only on the signal-to-noise ratio of the $\alpha$ signal in $\kappa g$ and $gg$ power spectra, and the cross-correlation coefficient $r_\ell=C^{\kappa g}(\hat C^{\kappa\kappa}\hat C^{gg})^{-1/2}$ between $\kappa$ and $g$.
Completing the square,
\begin{align}
  \label{eq:FisherAlphaFinal2}
    (F_{\alpha\alpha})_\ell =\, & \frac{1}{2\left(1-r_\ell^2\right)^2}\Bigg[ 
\left(\frac{C^{gg}_{\ell,\alpha}}{\hat C^{gg}_\ell}-2r^2_\ell\frac{C^{\kappa g}_{\ell,\alpha}}{C^{\kappa g}_\ell}\right)^2\non\\
&\;\; +2r_\ell^2(1-r_\ell^2)\left(\frac{C^{\kappa g}_{\ell,\alpha}}{C^{\kappa g}_\ell}\right)^2
\,\Bigg].
\end{align}
This result for the Fisher information of a generic scale-dependent bias amplitude $\alpha$ is exact if $\kappa$ and $g$ are Gaussian fields.

We can simplify \eqq{FisherAlphaFinal2} by making some approximations.
We expect the $gg$ 3-D power to be roughly twice as sensitive to scale-dependent bias than the 3-D $\kappa g$ power, because bias enters quadratically in $P_{gg}\propto (1+\alpha\beta)^2\approx 1+2\alpha\beta$ but only linearly in $P_{\kappa g}\propto 1+\alpha\beta$.
Projecting on the 2-D sky, this is still true if the redshift kernels of $g$ and $\kappa$ match, i.e.~$W_g(z)\approx W_\kappa(z)$, and if the fiducial galaxy bias is independent of redshift.
In this idealized limit we thus have
\begin{align}
  \label{eq:DerivApprox}
  \frac{C^{gg}_{\ell,\alpha}}{\hat C^{gg}_\ell} \approx 2 \frac{C^{\kappa g}_{\ell,\alpha}}{C^{\kappa g}_\ell}.
\end{align}
The Fisher information then simplifies to
\begin{align}
  \label{eq:FalphalphaRCC}
  (F_{\alpha\alpha})_\ell = \frac{2-r_\ell^2}{1-r_\ell^2} \left(\frac{C^{\kappa g}_{\ell,\alpha}}{C^{\kappa g}_\ell}\right)^2.
\end{align}

\eqq{FalphalphaRCC} can be interpreted as the signal-to-noise-squared for detecting $\alpha=1$ from a single $\ell$. 
If the correlation approaches $r_\ell\rightarrow 1$, the signal-to-noise-squared becomes arbitrarily large, scaling as $(1-r_\ell^2)^{-1}$.
This is precisely the scaling expected from sampling variance cancellation \cite{Uros0807,McDonaldSeljak0810}.
For example, if $r_\ell=(0.9,0.95,0.99,0.999)$, the improvement factor of the signal-to-noise is $(1-r_\ell^2)^{-1/2}=(2.3, 3.2, 7.1, 22)$.

We can use \eqq{FalphalphaRCC} to compute a rough estimate for the uncertainty of $\alpha=f_\mathrm{NL}$ based on the cross-correlation coefficient between tracers and CMB lensing.
Noting that there are $f_\mathrm{sky}(2\ell+1)$ modes per multipole $\ell$, the error per $\ell$ is
\begin{align}
  \label{eq:SigmaFNLSimple}
  (\sigma_{f_\mathrm{NL}})_\ell = 
\left[f_\mathrm{sky}(2\ell+1)
\frac{2-r_\ell^2}{1-r_\ell^2}
\left(\frac{C^{\kappa g}_{\ell,\alpha}}{C^{\kappa g}_\ell}\right)^2 \right]^{-1/2}.
\end{align}
The $\fnl$ signal-to-noise $C^{\kappa g}_{\ell,\alpha}/C^{\kappa g}_\ell$ dominates at low $\ell$ as shown in \fig{fnlSignalNoise} above.
For example, assuming optimally combined DESI and LSST tracers correlated with CMB-S4 lensing and using only power spectra at $\ell=2$, we have $C^{\kappa g}_{\ell,\alpha}/C^{\kappa g}_{\ell}|_{\ell=2}\simeq 0.2$ from \fig{fnlSignalNoise} and $r_{\ell=2}\simeq 0.92$ from \fig{rho}, so that the simple expression in \eqq{SigmaFNLSimple} gives $(\sigma_{f_\mathrm{NL}})_{\ell=2}\simeq 1.1$ for $\fsky=0.5$.
At higher $\ell$, the signal-to-noise per $\ell$ from \eqq{SigmaFNLSimple} is $(\sigma_{f_\mathrm{NL}})_{\ell=5}\simeq 1.4$, $(\sigma_{f_\mathrm{NL}})_{\ell=10}\simeq 1.6$ and $(\sigma_{f_\mathrm{NL}})_{\ell=20}\simeq 2.4$.
This shows that combining these low multipoles can yield $\sigma(\fnl)\sim 1$.
This includes the improvement factor from sample variance cancellation, which is approximately $(1-r_\ell^2)^{-1/2}\simeq 3.1$ for the maximum correlation of $r_{\ell=10}=0.946$, and $(1-r_\ell^2)^{-1/2}\simeq 2.6$ for $r_\ell=0.92$ which holds at $\ell\lesssim 40$.

The simple analytical estimates above suggest that sample variance cancellation can improve $\fnl$ constraints by a factor of 2 to 3 for CMB-S4 and LSST, thus achieving $\sigma(\fnl)\sim 1$. 
In the fully numerical Fisher analysis presented in the next sections we will find comparable improvement factors, although the final $\fnl$ precision will turn out somewhat better than suggested by the analytical estimates here. 
One possible reason for this (other than possible inaccuracies of the analytical estimates) is that the full Fisher analysis takes into account all possible power spectra and their scale- and redshift dependence rather than combining LSS tracers before measuring spectra as assumed for the analytical estimates above.

For Gaussian initial conditions $\fnl=0$, the sample variance cancellation technique can improve measurements of galaxy bias, because the bias enters linearly in $\kappa g$ and quadratically in $gg$ power spectra.
Indeed, we can just replace $\fnl\rightarrow b$ in all equations above to get the precision of bias measurements, with the same improvement factor of $(1-r_\ell^2)^{-1/2}$, if we assume that $\sigma_8$ is perfectly known.
If we marginalize over $\sigma_8$, the sample variance cancellation for bias still works in the low-noise limit but not in general \cite{McDonaldSeljak0810}. 
We confirmed this using the full Fisher analysis described in the next section, finding that in absence of noise (setting lensing noise and shot noise to zero and adding a biased tracer with number density matched to the CMB lensing kernel) the bias error becomes extremely small even when we marginalize over $\sigma_8$. 
However for realisitic noise levels the marginalization over $\sigma_8$ does matter, which makes it difficult in practice to exploit sample variance cancellation for bias when marginalizing over $\sigma_8$.
Note that while the sample variance cancellation technique can in principle improve bias it cannot directly improve $\sigma_8$, which enters $\kappa g$ and $gg$ power spectra in the same way.

\section{Fisher analysis setup}
\label{se:FisherSetup}

The above analytical calculations are only rough estimates because we did not include all power spectra and we did not marginalize over parameters that could be degenerate with the effects we are looking for.  
We improve this using a numerical Fisher analysis that we describe in this section. 
The results will be discussed in \secref{FisherResults}.

In the baseline analysis, we include all auto- and cross-spectra of the CMB-S4 lensing convergence and the 14 tomographic LSS redshift bins defined in \secref{Expts}.
With these $N=15$ fields, we have $15$ auto-spectra and $N(N-1)/2=105$ cross-spectra, obtaining 120 power spectra in total.\footnote{While the large number of power spectra is not a problem for Fisher forecasts with Gaussian covariances, this may be more challenging for actual data analyses.  
In that case one may want to compress the observations before forming power spectra (see \app{CombiningObs}).
\NEW{Although many of the cross-spectra have zero signal in the Limber approximation because they correlate non-overlapping redshift bins, the Fisher matrix can benefit from including them, because they can be correlated with other measured spectra, for example with $\kappa g$ cross-spectra ($\mathrm{cov}(C^{ij},C^{\kappa i})\propto C^{ii}C^{j\kappa}\ne 0$).}
}
Some of the most relevant spectra are listed in Tables~\ref{tab:AutoSNR} and \ref{tab:CrossSNR}.   
The power spectra would capture all cosmological information if the observed lensing convergence and galaxy density were Gaussian random fields. For simplicity we will assume this throughout, ignoring information from higher-order statistics.

\NEW{We compute 3D power spectra assuming a nonlinear halofit \cite{Takahashi:2012em,Mead:2015yca,Mead:2016zqy,Smith:2002dz} matter power spectrum with linear galaxy bias and project it to 2D using the Limber approximation at $\ell>50$ and including beyond-Limber corrections at $\ell\leq 50$ (see \app{3dto2d}).
The linear galaxy bias follows a fiducial redshift evolution within each tomographic redshift bin, and we marginalize over its amplitude in each bin.
}

We assume Gaussian covariances for all power spectra,
\begin{align}
  \label{eq:GaussianCov}
  \mathrm{cov}(\hat C^{ij}_{\ell},\hat C^{i'j'}_{\ell'})=
\frac{\delta_{\ell\ell'}}{f_\mathrm{sky}(2\ell+1)}\left(
\hat C^{ii'}_{\ell}\hat C^{jj'}_{\ell}
+\hat C^{ij'}_{\ell}\hat C^{ji'}_{\ell}
\right).
\end{align}
This ignores non-Gaussian corrections to the CMB lensing covariance \cite{Marcel1308,Peloton1611,Motloch1612}, the LSS clustering covariance (e.g., \cite{Li:2014sga,Krause:2016jvl,Barreira:2017kxd} and references therein), and their cross-covariance.
$\hat C$ are power spectra that would be observed without any noise bias subtraction, i.e.~they are the sum of signal and noise.
The upper indices $i,j,i',j'\in\{\kappa_\mathrm{CMB},\delta^\mathrm{LSST}_{z=0-0.5}, \delta^\mathrm{LSST}_{z=0.5-1}, \dots\}$ label the observable fields.

To speed up covariance inversion, we define a large one-dimensional data vector that starts with all spectra at $\ell_\mathrm{min}$, continues with all spectra at $\ell_\mathrm{min}+1$, etc:
\begin{align}
  \label{eq:36}
  \vec d = \big(\vec d_{\ell_\mathrm{min}}, \vec d_{\ell_\mathrm{min}+1}, \dots,
\vec d_{\ell_\mathrm{max}}\big).
\end{align}
At each $\ell$,
\begin{align}
  \label{eq:39}
  \vec d_\ell = \big(
C^{11}_\ell, C^{12}_\ell, \dots, C^{NN}_\ell
\big)
\end{align}
contains $N(N+1)/2$ spectra $C_\ell^{ij}$ with $j\ge i$.
Assuming \eqq{GaussianCov}, the covariance $\mathrm{cov}(\vec d,\vec d)$ is then a block-diagonal matrix with $\ell_\mathrm{max}-\ell_\mathrm{min}+1$ blocks of size $N(N+1)/2$ $\times$ $N(N+1)/2$, which is easily inverted if the number of fields is $N\lesssim 100$.
The Fisher matrix at the power spectrum level is then
\begin{align}
  \label{eq:FisherFast}
  F_{ab} = \sum_{\ell=\ell_\mathrm{min}}^{\ell_\mathrm{max}} \frac{\partial\vec d_\ell}{\partial\theta_a}
[\mathrm{cov}(\vec d_\ell,\vec d_\ell)]^{-1}
\frac{\partial\vec d_\ell}{\partial\theta_b}.
\end{align}
We evaluate this without binning in $\ell$.\footnote{Binning is less accurate and does not speed up our implementation because binning the covariance is slow.}

The above analysis assumes that all experiments observe the same patch of sky, because it includes cross-spectra between all observed fields in the data vector and covariance. 
The forecast therefore includes (a) sample variance cancellation from observing the same modes multiple times, and (b) breaking of parameter degeneracies using $\kappa g$ cross-spectra.
Both effects can only be exploited in a joint analysis of CMB-S4 CMB lensing and LSS clustering on the same patch of sky.  
To determine how much these effects contribute to the forecasted parameter precisions, we will compare against a modified forecast, where we assume that each observed field (CMB lensing map or galaxy redshift bin) is on an independent patch of the sky.
In that case there is no sky overlap between any two observed fields so that all fields are independent from each other and neither (a) nor (b) are used.
We implement this by dropping all cross-spectra between two different fields from the data vector and setting all cross-spectra to zero in covariances, i.e.~
\begin{align}
  \label{eq:29}
  \text{No sky overlap: } & 
\textbf{d}_\ell=\left(
C^{11}_\ell, C^{22}_\ell, \dots, C^{NN}_\ell
\right),
\nonumber\\
& \mathrm{cov}\left(C^{ii}_\ell,C^{jj}_\ell\right) = \delta_{ij}\frac{2}{2\ell+1}(C^{ii}_\ell)^2,
\end{align}
excluding, e.g., $\la g^\mathrm{LSST}_{z=0-0.5},\kappa_\mathrm{CMB}\ra$ and $\la g^\mathrm{LSST}_{z=0-0.5},g^\mathrm{DESI}_\mathrm{BGS}\ra$ from the data vector and covariance.\footnote{A potential concern of the forecasts with no sky overlap is that the independent patches probe a larger total volume, increasing the number of independent Fourier modes that are measured.
For example, for two samples, working on two independent patches increases the total number of Fourier modes by a factor two, which should reduce sample variance error bars by a factor $\sqrt{2}$.
This can unintentionally improve parameter precisions, for example when constraining $\sigma_8$ assuming fixed bias parameters.
The analysis with no sky overlap might therefore be better than it should be, so that we might underestimate the true improvement factors caused by sample variance cancellation and breaking of parameter degeneracies using $\kappa g$ spectra.
A practical argument for comparing analyses with and without sky overlap is that this can inform observing strategies of experiments, quantifying how much gain there is if the surveys are on the same  rather than independent patches.
}

To marginalize over linear galaxy bias,
we will rescale the fiducial galaxy bias $b(z)$ of each tomographic redshift bin by a redshift-independent amplitude, $b(z)\rightarrow B_ib(z)$, and marginalize over the bias amplitudes $B_i$ of all tomographic redshift bins.
This implicitly assumes that the redshift-dependence within each tomographic redshift bin is known.

As a basic validation of our implementation of the Fisher matrix in \eqq{FisherFast} we checked that if all power spectra are included it numerically agrees with the Fisher matrix at the field level given by Eqs.~\eq{FisherFieldLevel} and \eq{FisherPerMode} above.
Below we will only use the Fisher analysis at the power spectrum level \eqq{FisherFast}  because it allows to exclude individual power spectra from the analysis and analyze their importance.

\section{Fisher analysis results}
\label{se:FisherResults}

Based on the experiments, signals, and Fisher analysis setup described above, we now present forecasts for $\sigma_8(z)$, primordial non-Gaussianity $\fnl$, and neutrino mass.
The forecasts use all power spectra $(C^{\kappa\kappa},C^{\kappa g_i},C^{g_ig_j})$ of CMB-S4 lensing convergence and tomographic LSS redshift bins  of SDSS, DESI and LSST clustering as described in the previous section.

\subsection{Amplitude of matter fluctuations \texorpdfstring{$\sigma_8(z)$}{sigma8(z)}}
\label{se:sigma8Results}

\subsubsection{Setup}

As motivated in \secref{S8Motivation}, the amplitude of matter fluctuations as a function of redshift $\sigma_8(z)$ carries important information about the growth of structure and the expansion of the Universe.
To forecast the expected precision of $\sigma_8(z)$, we rescale the fiducial amplitude of matter fluctuations in broad redshift bins,
\begin{align}
  \label{eq:sigma8Rescaling}
  P_\mathrm{mm}(k,z) \,\rightarrow \, \sum_i (1+s_i)^2 V_i(z) P_\mathrm{mm}(k,z)
\end{align}
where $s_i\equiv \sigma_{8,i}/\sigma_{8,\mathrm{fid}}-1$ is the fractional change of $\sigma_8$ in the $i$th redshift bin.
We work with seven broad redshift bins for $\sigma_8$, defined by $z=0-0.5,0.5-1,1-2,2-3,3-4,4-7,7-100$, and treat the amplitude $s_i$ in each bin as a parameter in the Fisher analysis.
The redshift binning function is $V_i(z)=1$ for $z_{i,\mathrm{min}}\le z< z_{i,\mathrm{max}}$ and $V_i(z)=0$ otherwise.
We marginalize over linear galaxy bias amplitude parameters $B_i$ as described at the end of \secref{FisherSetup} (also see, e.g., \eqq{Cggfnl} below).

\subsubsection{Baseline results}

\begin{figure}[tbp]
\includegraphics[width=0.5\textwidth]{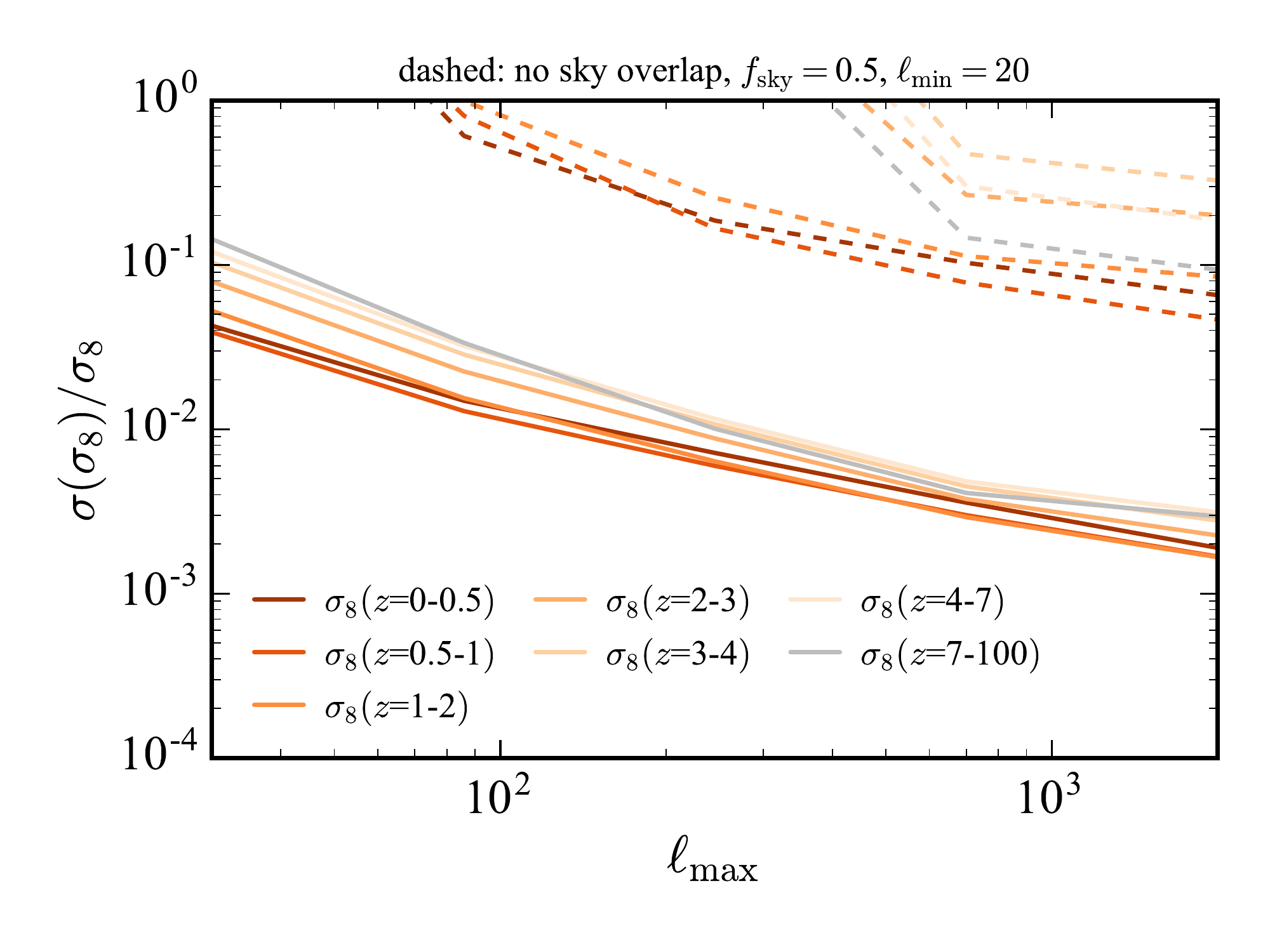}
\caption{Fractional statistical uncertainty of the amplitude of matter fluctuations, $\sigma_8$, defined in broad redshift bins, $z=0-0.5,0.5-1,1-2,2-3,3-4,4-7,7-100$, as a function of $\ell_\mathrm{max}$. The forecast uses all power spectra of  CMB-S4 lensing and SDSS, DESI and LSST ($i<27$, 3yr, $z<7$) clustering, and assumes $\ell_\mathrm{min}=20$ and $\fsky=0.5$.
Solid lines assume all observations are on the same patch of sky, while dashed lines assume  all fields are observed on independent non-overlapping patches (see end of \secref{FisherSetup}).
As in all other $\sigma_8$ forecasts we marginalize over one linear galaxy bias parameter in each redshift bin, but ignore nonlinear galaxy bias that can degrade the precision for high $\ell_\mathrm{max}$ (see \secref{Sigma8Caveats} for discussion).
}
\label{fig:s8_lmax_skyoverlap}
\end{figure}

\fig{s8_lmax_skyoverlap} shows the forecasted precision of $\sigma_8(z)$ bins as a function of the highest wavenumber $\ell_\mathrm{max}$ included in the analysis.
Using modes $20\le \ell\le 200$ on half the sky, $\sigma_8(z)$ can be determined to $\sim 1\%$ for all redshift bins. 
Including smaller scales, $20\le\ell\le 1000$, improves the precision to $0.2\%$ to $0.3\%$ in each redshift bin.
This subpercent-level precision on $\sigma_8(z)$ can lead to impressive constraints on dark energy and neutrino mass, which should be quantified in more detail in the future.

\subsubsection{Driving factors}

What drives the $\sigma_8$ forecast?
One key driver is to include small scales, because the precision of $\sigma_8$ in \fig{s8_lmax_skyoverlap} roughly scales as $\sigma(\sigma_8)\propto \ell_\mathrm{max}^{-1}$.
This is as expected: Since $\sigma_8$ affects power spectra at all $\ell$, its precision is  determined by the number of modes: $\sigma(\sigma_8)\propto N_\mathrm{modes}^{-1/2}\propto f_\mathrm{sky}^{-1/2}\ell_\mathrm{max}^{-1}$.
At high $\ell_\mathrm{max}$, the scaling becomes somewhat weaker because lensing noise and shot noise become relevant.
In practice, the maximum $\ell_\mathrm{max}$ should be set by the smallest scale where we can still model the observations.

To achieve subpercent-level $\sigma_8(z)$ precision it is also critical that CMB lensing and LSS clustering are observed on the same patch of sky so that cross-spectra can be measured:
Without sky overlap (dashed in \fig{s8_lmax_skyoverlap}) the $\sigma_8$ precision degrades by more than a factor of 20.
This is caused by the galaxy bias-$\sigma_8$ degeneracy that can only be broken with $\kappa g$ cross-spectra on the same patch.
Restricting the data to CMB lensing alone or galaxy clustering alone yields even lower precision than the dashed curves in \fig{s8_lmax_skyoverlap} (which combine $\kappa\kappa$ and $gg$), emphasizing even more the importance of a joint analysis of CMB lensing and galaxy clustering.

\begin{figure}[tbp]
\includegraphics[width=0.5\textwidth]{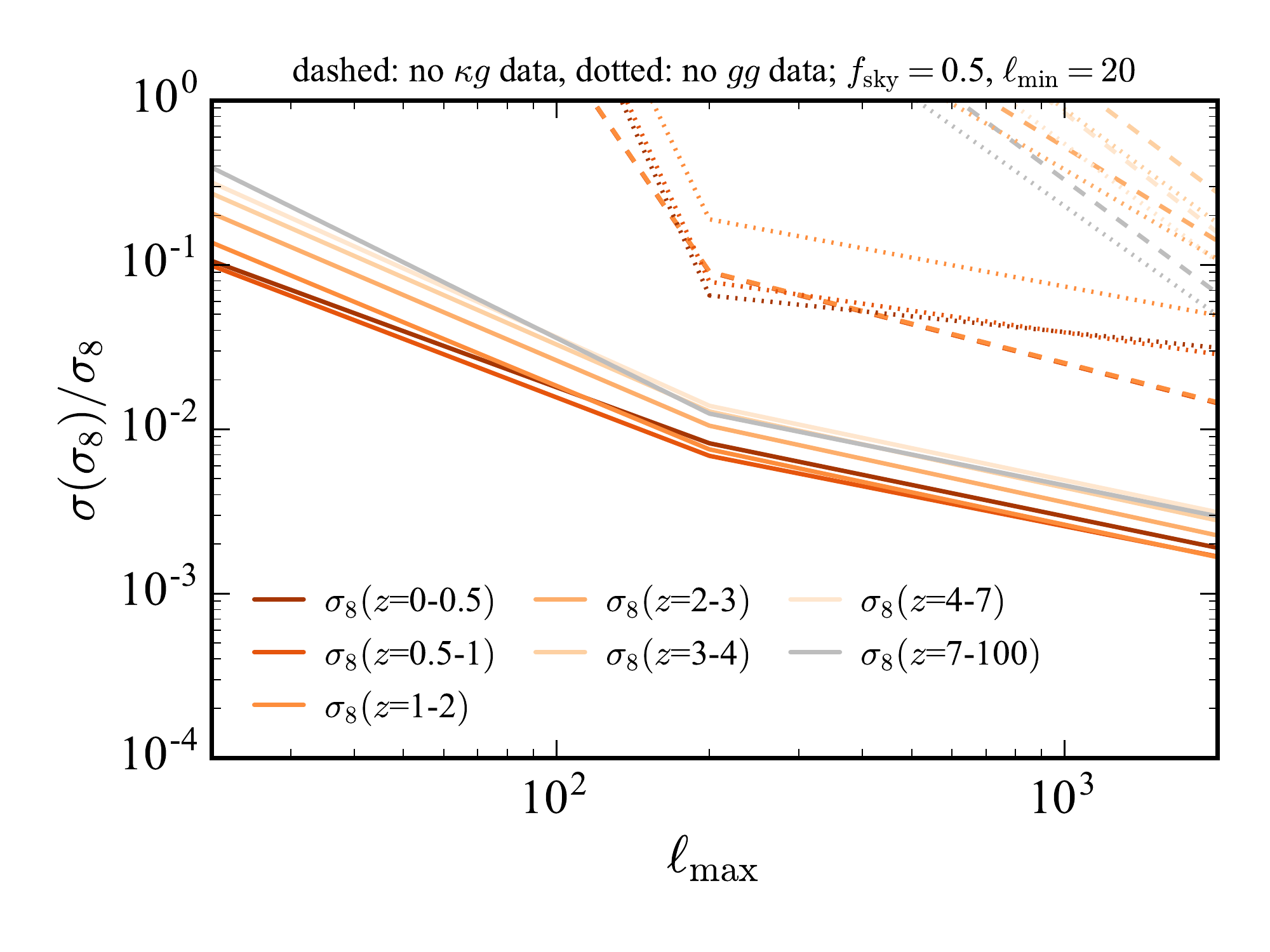}
\caption{Impact of different measured spectra on $\sigma_8$ precision.  Solid assumes we observe all spectra, $\kappa\kappa$, $\kappa g$ and $gg$; dashed assumes we cannot measure $\kappa g$; dotted assumes we cannot measure $gg$.
If we cannot measure $\kappa\kappa$, the precision of $\sigma_8$ bins at $z<7$ is the same as the solid lines, but $\sigma_8$ at $z>7$  cannot be constrained at all.
}
\label{fig:s8_lmax_spectraImpact}
\end{figure}

To check in more detail where most of the constraining power comes from, \fig{s8_lmax_spectraImpact} shows the impact of different measured power spectra on the $\sigma_8$ precision.
The precision degrades by a factor of 10 ore more if $\kappa g$ spectra are dropped (dashed in \fig{s8_lmax_spectraImpact}) or if $gg$ spectra are dropped (dotted in \fig{s8_lmax_spectraImpact}). 
In contrast, dropping $\kappa\kappa$ has no impact on $\sigma_8$ constraints at $z<7$ where LSS tracers are available and $\kappa g$ and $gg$ spectra have nonzero signal.
This shows that the combination of $\kappa g$ and $gg$ spectra determines $\sigma_8$ at all redshifts where we observe tracers.  These spectra also determine the galaxy bias parameters that we marginalize over.

The $\kappa\kappa$ auto-power spectrum is only useful to measure $\sigma_8$ at $z>7$, because it is the only spectrum sensitive to such high redshift in our forecasts, which assume vanishing galaxy number density at $z>7$ for all surveys.\footnote{One might wonder why $\sigma_8(z$=7-100$)$ is determined with similar precision as the $\sigma_8$ bins at lower redshift, although we do not incude any LSS tracers at $z>7$.
We have checked that high-redshift amplitudes in narrower bins, for example $\sigma_8(z$=7-8$)$ or $\sigma_8(z$=8-9$)$, cannot be constrained nearly as well.
The tight constraint of $\sigma_8(z$=7-100$)$ thus comes from an integral constraint on the $\kappa\kappa$ spectrum over many $\ell$, with $z<7$ contributions calibrated using $\sigma_8$ at $z<7$ and bias measured from the cross-correlations with tracers at those redshifts $z<7$.
The precision of that calibration then also limits the precision of $\sigma_8(z$=7-100$)$   measured from $\kappa\kappa$.
}

\subsubsection{Optimizing experiments}

\begin{figure}[tbp]
\includegraphics[width=0.5\textwidth]{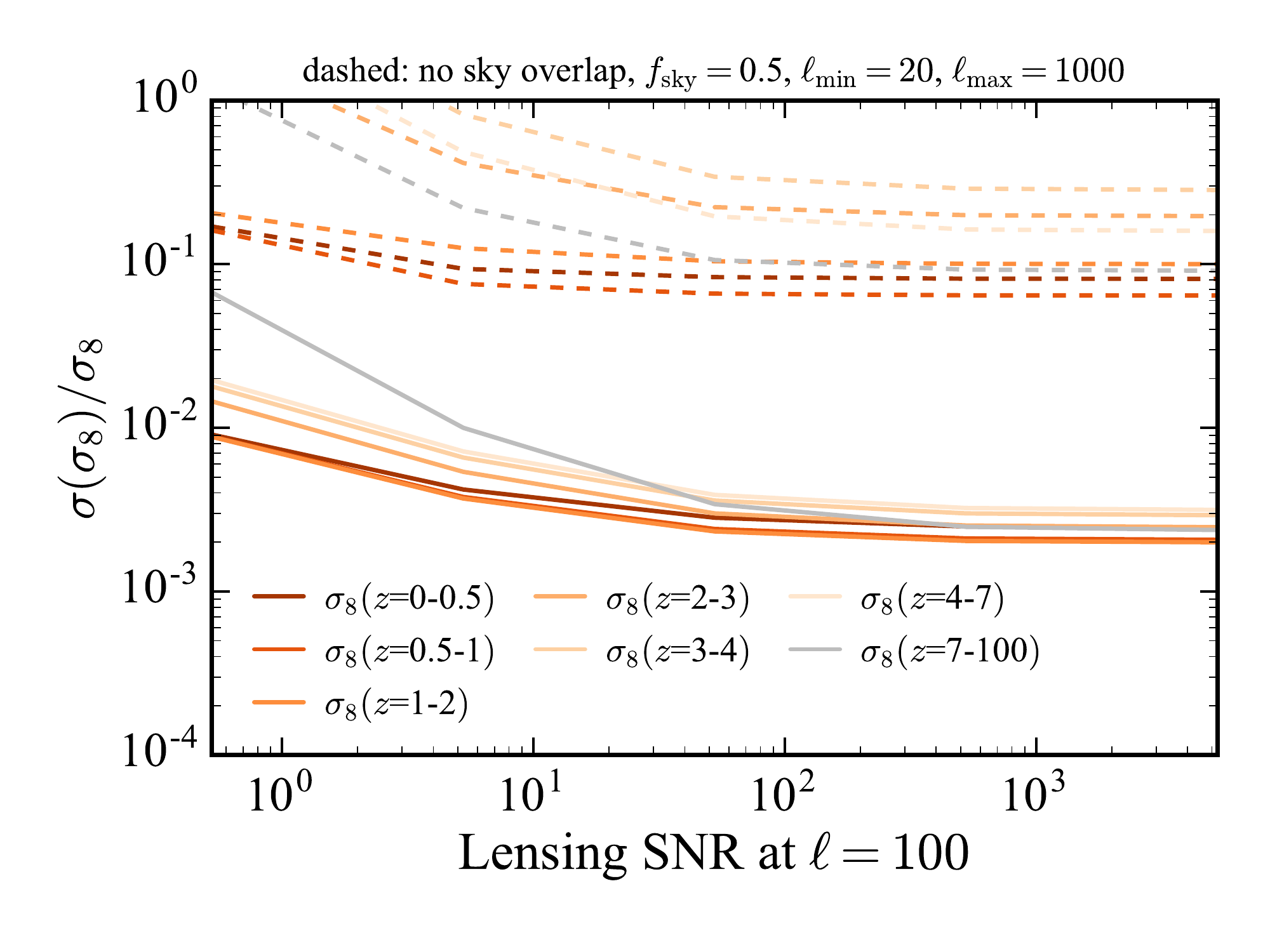}
\caption{Fractional statistical uncertainty of $\sigma_8$ as a function of CMB lensing signal-to-noise.
We rescale the lensing noise by an $\ell$-independent factor and quote on the horizontal axis the CMB lensing signal-to-noise ratio at $\ell=100$, defined as $\mathrm{SNR}=C^{\kappa\kappa}_{\ell=100}/N^{(0)}_{\ell=100}$.
CMB-S4 with iterative lensing reconstruction corresponds to $\mathrm{SNR}=50$ on the horizontal axis; see \cite{CMBS4SciBook} and \fig{N0}.
Planck corresponds roughly to $\mathrm{SNR}\sim 0.5$ \cite{PlanckLensing2015}.
The forecast uses modes $20\le \ell\le 1000$ of all power spectra that can be formed between CMB lensing and SDSS, DESI and LSST ($i<27$, 3yr, $z<7$) galaxy clustering. 
}
\label{fig:s8_lensingnoise}
\end{figure}

\begin{figure}[tbp]
\includegraphics[width=0.5\textwidth]{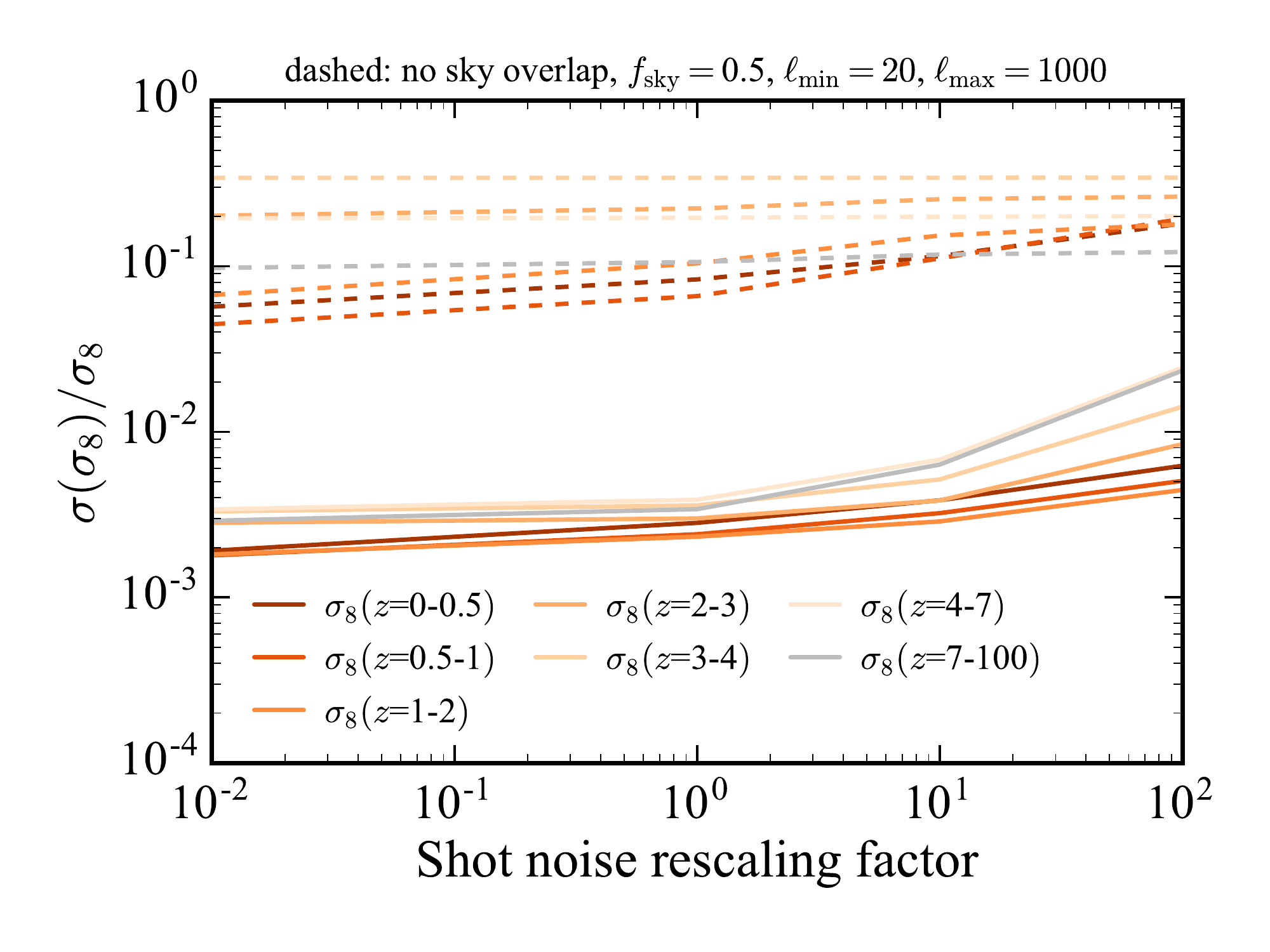}
\caption{Fractional statistical uncertainty of $\sigma_8$ as a function of the shot noise level of galaxy surveys.
The fiducial SDSS, DESI and LSST ($i<27$, 3yr, $z<7$) number densities correspond to a rescaling factor of 1 on the horizontal axis.
The forecast uses $20\le\ell\le 1000$ and marginalizes over one bias amplitude parameter in each tomographic redshift bin.
Lower shot noise relative to the fiducial case gives only small improvements because the uncertainty is dominated by cosmic variance at the smallest scale, $\ell=1000$, rather than shot noise.
Increasing the shot noise by factors of 10 or more relative to the fiducial case makes shot noise important, so that constraints degrade.
Overall the dependence on shot noise is rather mild.
}
\label{fig:s8_shotnoise}
\end{figure}

To see how the CMB lensing experiment can be optimized to measure $\sigma_8(z)$, \fig{s8_lensingnoise} shows $\sigma_8(z)$ forecasts as a function of the signal-to-noise-ratio (SNR) of the reconstructed CMB lensing power spectrum at $\ell=100$, with $\mathrm{SNR}\sim 0.5$ corresponding to Planck and $\mathrm{SNR}=50$ corresponding to CMB-S4.
While the $\sigma_8$ precision improves from Planck to CMB-S4 by a factor of 2 or more, there is not much improvement beyond CMB-S4, at least for our assumed version of LSST.

\fig{s8_shotnoise} shows how the shot noise level of the LSS surveys affects the
 $\sigma_8$ precision. 
It degrades somewhat if shot noise is higher than for LSST, but it does not improve much if the shot noise falls below LSST levels, at least assuming CMB-S4 like CMB lensing measurements and $\ell_\mathrm{max}=1000$ as in \fig{s8_shotnoise}.
If we can push models to higher $\ell_\mathrm{max}$, lower shot noise levels will likely be more useful.

These results suggest that CMB-S4 and LSST lie at a sweet spot for constraining $\sigma_8(z)$, and one would have to improve both experiments rather than any one of them to improve $\sigma_8(z)$, at least assuming $\ell_\mathrm{max}=1000$. 
This can be understood from Table~\ref{tab:AutoSNR} above which showed that CMB-S4 lensing and LSST power spectra have roughly the same total signal-to-noise for $\ell_\mathrm{max}=1000$. 
If we only improve CMB-S4 but not LSST, the noise in LSST will limit $\sigma_8$, and vice versa.

\begin{figure}[tbp]
\includegraphics[width=0.5\textwidth]{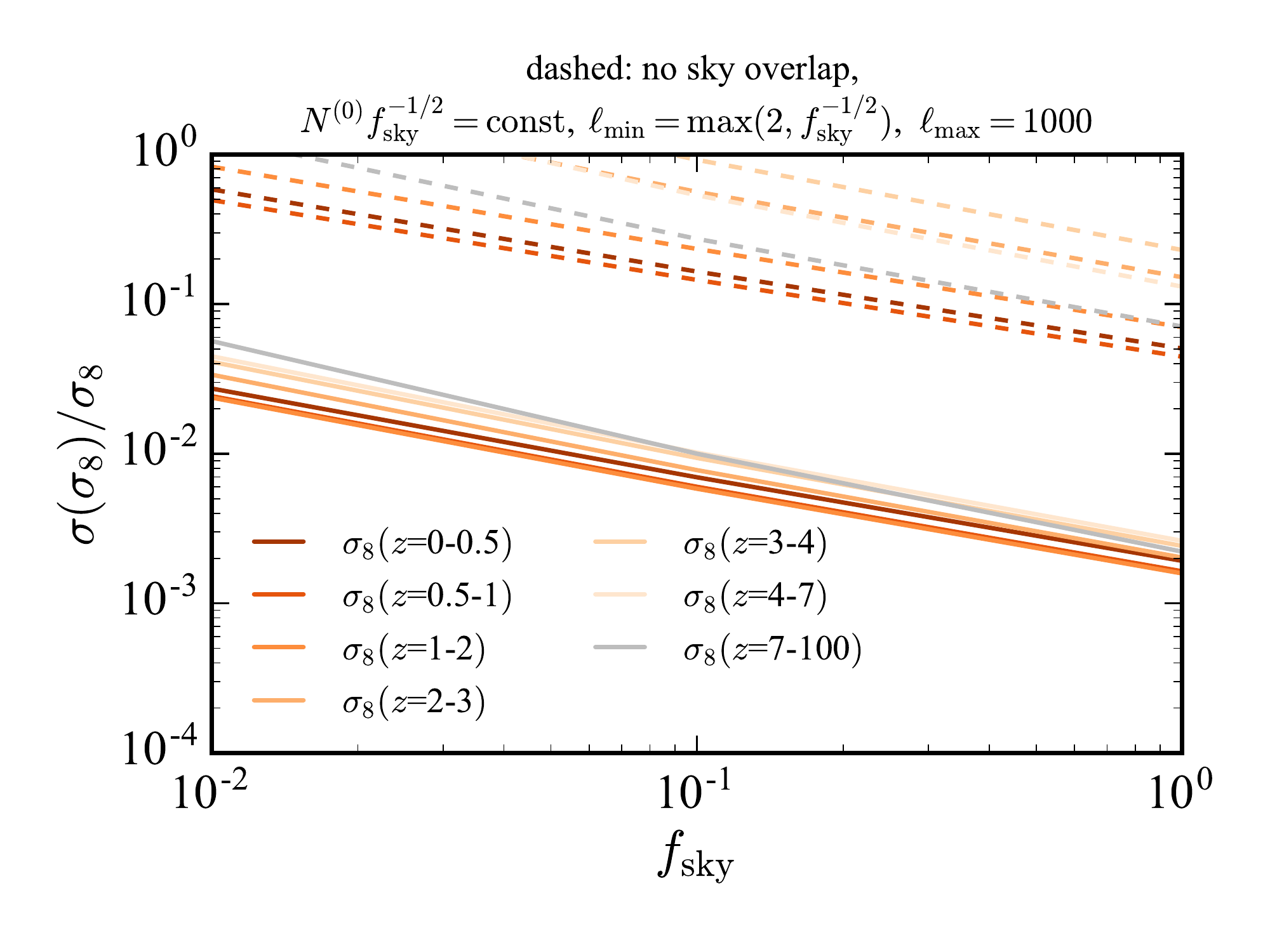}
\caption{$\sigma_8$ precision as a function of the sky fraction of all experiments.
We assume approximately fixed CMB observation time by lowering the lensing noise for smaller $\fsky$, keeping $N^{(0)}f_\mathrm{sky}^{-1/2}=N^{(0)}_\mathrm{S4}0.5^{-1/2}$ constant.
Results are marginalized over bias parameters, and assume CMB-S4 lensing and clustering from SDSS, DESI and LSST. 
As in the previous plots, solid curves assume all observations are on the same patch, while dashed curves assume mutually independent patches.
}
\label{fig:s8_fsky}
\end{figure}

\fig{s8_fsky} shows the $\sigma_8$ precision as a function of sky fraction $\fsky$.
While varying the sky fraction, we keep the CMB observation time approximately constant by lowering the lensing noise when decreasing $\fsky$.
Small $\fsky$ thus corresponds to a small deep patch, whereas large $\fsky$ corresponds to a wide shallow patch.
Specifically, we keep $N^{(0)}f_\mathrm{sky}^{-1/2}=N^{(0)}_\mathrm{S4}0.5^{-1/2}$ constant while varying $\fsky$.
This corresponds to constant CMB observation time if the lensing noise decreases linearly with observation time and if observing a larger sky fraction is quadratic in observation time.
As shown in \fig{s8_fsky}, the $\sigma_8$ precision strongly improves with larger sky fraction, scaling like $\sigma(\sigma_8)\propto f_\mathrm{sky}^{-1/2}$.
As mentioned before, this is precisely the scaling expected just from increasing the number of modes, which means that there is no degradation from the larger lensing noise that we assume for wider patches (assuming CMB-S4 lensing noise at $\fsky=0.5$).
For CMB-S4 this means that we should aim for a wide shallow patch rather than a small deep patch when trying to measure $\sigma_8$.

\subsubsection{Caveats and discussion}
\label{se:Sigma8Caveats}
An important caveat of our forecast is that we ignore nonlinear corrections to the bias relation between galaxies and dark matter.
These corrections become important on scales corresponding to the Lagrangian size of halos, which is independent of redshift.
At high redshift, $z\gtrsim 2$, nonlinear corrections to the bias relation can thus be more relevant than nonlinear corrections to the DM density \cite{Modi1706}. 
The cross-correlation of high-redshift galaxies with CMB lensing, and the auto power spectra of those galaxies, are thus more affected by nonlinear biasing than one might naively expect given the nonlinear scale of the DM density at those redshifts.
Indeed, as shown recently \cite{Modi1706}, marginalizing over nonlinear bias parameters can degrade  $\sigma_8$ forecasts like ours by a factor of up to 5.\footnote{Even assuming only linear bias, the constraints of \cite{Modi1706} are  about 2-3 times weaker than the ones we find.
Reasons for this may be that Ref.~\cite{Modi1706} assumes lower number density for LSST (they use the LSST $i<25$ gold sample whereas we use the $i<27$ 3-year sample), they use no SDSS or DESI observations, and their CMB-S4 lensing noise does not include improvements from the iterative $EB$ estimator, making their lensing noise a few times higher than ours.
We also choose broader tomographic redshift bins at $z>1$ as described in \secref{Expts}, which decreases the uncertainty of $\sigma_8$ in those bins.
Additional differences may be due to different models of the $\kappa g$ signal, noting that  \cite{Modi1706} employs a more accurate model than the linearly biased halofit used here.
}
Our forecasts assuming linear bias may thus be overly optimistic, especially for high $\ell_\mathrm{max}$.

On the other hand, one might argue that treating all nonlinear galaxy bias parameters as  completely free parameters may be overly pessimistic, because by the time we get data from CMB-S4 and LSST we might be able to describe the galaxy-matter relation with more restrictive models than now.
For example, it may be feasible to parameterize nonlinear halo bias parameters as functions of linear bias or halo mass (e.g., \cite{Lazeyras1511,Baldauf1511,Castorina1611}).
That would reduce the number of free parameters in the model and thus lead to tighter $\sigma_8$ constraints than if all nonlinear bias parameters are free and marginalized over.
Realistically, such relationships between bias parameters may never be perfect, but even broad priors on nonlinear bias parameters may help; for example $1\%$-$10\%$ priors on nonlinear bias parameters may be sufficient \cite{Baldauf1602}.

Additionally to theoretical progress, it is possible to obtain observational priors on bias parameters by measuring the anisotropic power spectrum in redshift space or higher-order N-point functions. 
For example, measurements of the bispectrum \cite{Gil-Marin:2014sta} and 3-point correlation function \cite{Slepian:2015hca} of spectroscopic SDSS BOSS galaxies constrained the allowed value of the quadratic bias of these galaxies (also see, e.g., \cite{SongTaruyaOka1502} for DESI and \cite{Spherex1412} for SPHEREx forecasts).
These nonlinear bias constraints could be used as a prior when modeling cross-correlations with CMB lensing.
Achieving our $\sigma_8$ forecasts that are based on just a single degree of freedom to model the galaxy-matter connection might still be optimistic, but there is hope that they could come within reach if nonlinear galaxy bias can be better modeled or observationally constrained in the future.

\NEW{Another potential caveat is super-sample variance (e.g., \cite{2013PhRvD..87l3504T}), which we ignored.
This should be added to the error bars of $\sigma_8$, especially at low redshift where the observed volume is relatively small.
It would be interesting to calculate the impact of super-sample variance on our type of forecasts, but this is beyond the scope of this paper.
}

\subsubsection{Measuring galaxy bias}

\begin{figure}[tbp]
\includegraphics[width=0.5\textwidth]{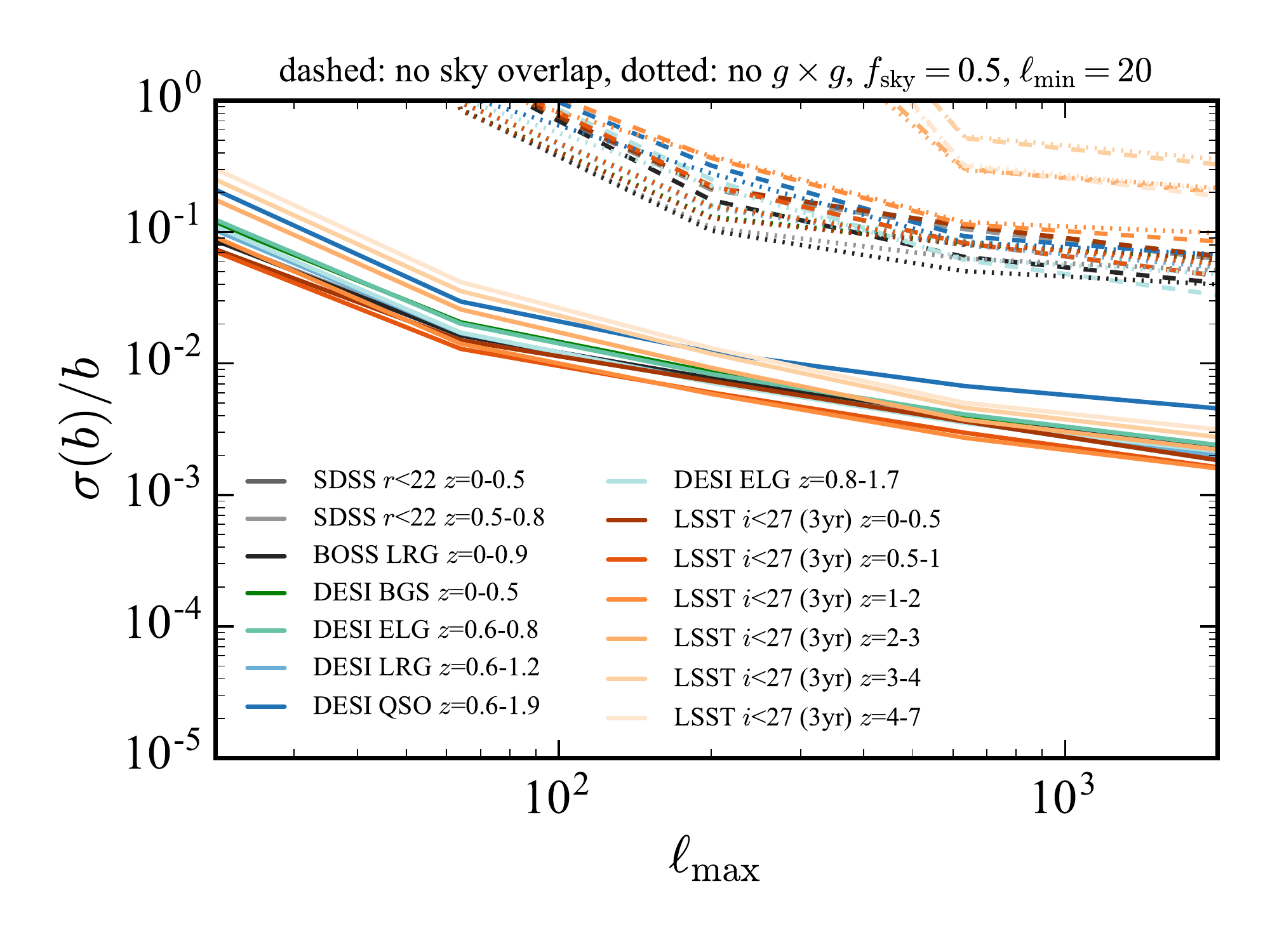}
\caption{Fractional precision of linear bias parameters marginalized over the matter amplitude $\sigma_8$ in broad redshift bins, $z=0-0.5,0.5-1,1-2,2-3,3-4,4-7$, and $7-100$. 
We assume $\ell_\mathrm{min}=20$ and $\fsky=0.5$.
Without sky overlap (dashed), or without $gg$ spectra (dotted), the constraints degrade by an order of magnitude.
If $\sigma_8$ was perfectly known (not shown), most constraints for full sky overlap would improve by a factor of $\sim 4-5$ for high $\ell_\mathrm{max}$.
}
\label{fig:biasparams_lmax_marginalize_s8}
\end{figure}

Rather than marginalizing over galaxy bias and determining the matter amplitude $\sigma_8$ as above, we can use lensing-clustering cross-correlation measurements to determine galaxy bias parameters while marginalizing over $\sigma_8$.
We show the expected precision of  linear bias parameters for marginalized $\sigma_8$ in \fig{biasparams_lmax_marginalize_s8}, finding that the bias can be measured rather accurately.
For example, the modes $20\le\ell\le 200$ can determine linear LSST bias parameters to about $1\%$ precision,  assuming $\fsky=0.5$ and ignoring nonlinear bias.
If smaller scales are included, $20\le \ell\le 1000$, the bias amplitudes can be measured with $0.2$ to $0.4\%$ precision (except DESI QSO which is slightly worse).

If we instead assume $\sigma_8$ to be perfectly known, the bias constraints improve by up to a factor of 5 for high $\ell_\mathrm{max}$, showing that the degeneracy between $\sigma_8$ and bias plays an important role for our noise levels.
This also explains why the precision of $\sigma_8$ and bias are similar when we measure one and marginalize over the other. 
If we decrease noise sufficiently, and cover the CMB lensing kernel with enough galaxies out to high redshift, this situation will change at some point and bias can benefit from sampling variance cancellation, so in principle it could be determined much more accurately than $\sigma_8$ which is always limited by sampling variance \cite{McDonaldSeljak0810}
(also see end of \secref{AnalyticalSVCFisher} above).
Our forecasts suggest that this may require experiments beyond CMB-S4 and LSST, but we leave a more detailed investigation  for future work.

\subsection{Local primordial non-Gaussianity}

\subsubsection{Setup}

\label{se:FnlBiasMarg}

We continue with forecasts for local primordial non-Gaussianity from its scale-dependent bias effect.
To allow some freedom of the shape of the matter power spectrum, 
we marginalize
over a `fake' parameter $f_\mathrm{NL}^\mathrm{fake}$ that rescales the matter power spectrum using the same scale- and redshift-dependence as the scale-dependent bias:
\begin{align}
  C^{\kappa\kappa}_\ell =&\, \int_z W_\kappa^2(z)\,[1+f_\mathrm{NL}^\mathrm{fake}\beta(k,z=1)]^2\, P_{\mathrm{mm}}(k,z),
\label{eq:Ckkfnl}
\end{align}
\begin{align}
  C^{\kappa g_i}_\ell =&\, \int_z W_\kappa(z)W_{g_i}(z) B_i b_i(z)\,[1+f_\mathrm{NL}\beta(k,z)]\non\\
&\quad\;\,\times[1+f_\mathrm{NL}^\mathrm{fake}\beta(k,z=1)]^2\,P_{\mathrm{mm}}(k,z),
\label{eq:Ckgfnl}
\end{align}
and
\begin{align}
  C^{g_ig_j}_\ell =&\, 
\int_z W_{g_i}(z)W_{g_j}(z) B_ib_i(z) B_jb_j(z)P_{\mathrm{mm}}(k,z)\non\\
&\quad\;\,\times[1+f_\mathrm{NL}\beta(k,z)]^2\,[1+f_\mathrm{NL}^\mathrm{fake}\beta(k,z=1)]^2
\non\\
& + \delta_{ij}^K N^{g_ig_i}_\ell.
\label{eq:Cggfnl}
\end{align}
Here, $\beta(k,z)\propto k^{-2}$ is the fractional change of the bias for $\fnl=1$ as defined in \eqq{fnlbias}.
The equations assume the Limber approximation with $k=\ell/\chi(z)$, but we include beyond-Limber corrections on large scales $\ell\le 50$ as described in \app{3dto2d}, where we also define the redshift kernels $W$ and the shot noise $N^{gg}$.
The redshift integrals include a conversion factor given by \eqq{RedshiftMeasure}.
A more complete analysis would marginalize over all changes of the matter power spectrum due to changes in cosmological parameters within some priors, but we expect that marginalizing over $f_\mathrm{NL}^\mathrm{fake}$ captures the worst possible case because its shape is perfectly degenerate with that of the true $\fnl$.

We also marginalize over linear galaxy bias by marginalizing over the bias amplitude parameters $B_i$ of each tomographic redshift bin, assuming that the redshift evolution of the bias within each redshift bin is known, and assuming no priors for the amplitudes $B_i$.

\subsubsection{Baseline results}

\begin{figure}[tbp]
\includegraphics[width=0.5\textwidth]{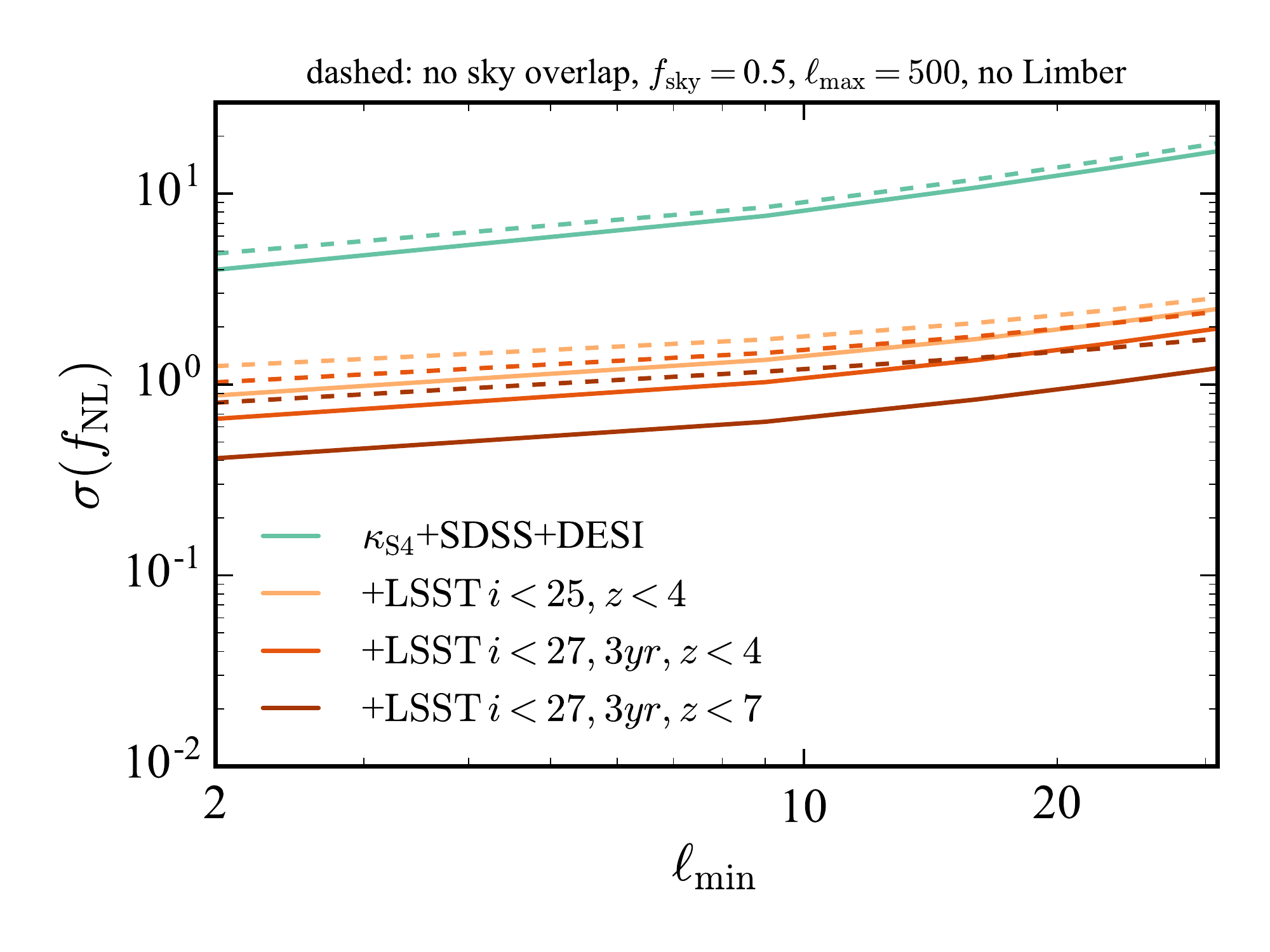}
\caption{Forecasted precision of the amplitude of local primordial non-Gaussianity $\fnl$ as a function of minimum wavenumber $\ell_\mathrm{min}$ of CMB lensing $\kappa$ and galaxy overdensities, for different LSS surveys (colors), assuming $\ell_\mathrm{max}=500$ and $\fsky=0.5$.
Solid curves assume all experiments observe the same patch of sky, whereas dashed curves assume mutually independent patches with no sky overlap.
We marginalize over galaxy bias and over $f_\mathrm{NL}^\mathrm{fake}$ defined in Eqs.~\eq{Ckkfnl}-\eq{Cggfnl} to marginalize over changes in the matter power spectrum that mimic the effect of $\fnl$.
Integrations along the line of sight are computed exactly for $\ell\le 50$ and with the Limber approximation for $\ell>50$. 
}
\label{fig:fnl_LSST_S4}
\end{figure}

In \fig{fnl_LSST_S4} we show the expected $\fnl$ precision as a function of the largest scale or minimum wavenumber $\ell_\mathrm{min}$ included in the analysis.
A joint analysis of CMB-S4 lensing with 3-year $i<27$ LSST clustering measurements at $z=0-7$ is able to reach $\sigma(\fnl)=0.4$ for $\ell_\mathrm{min}=2$, $\sigma(\fnl)=0.7$ for $\ell_\mathrm{min}=10$, and $\sigma(\fnl)=1$ for $\ell_\mathrm{min}=20$.
This is twelve to five times stronger than the best current constraint, $\sigma(\fnl)=5.0$ \cite{Planck15fnl}.
Combining CMB lensing with LSS clustering on large scales thus offers an intriguing method to test if $\fnl$ is larger or smaller than one, which is very exciting because a detection of $\fnl>\mathcal{O}(1)$ would rule out  single-field inflation in a model-independent way (see \app{fnlBasicsAppdx}).

\subsubsection{Driving factors}

The baseline $\fnl$ forecast is driven by several factors that we discuss next.

First, as already indicated above and shown in  \fig{fnl_LSST_S4}, the $\fnl$ precision improves rather strongly with the largest scale (lowest $\ell$) of CMB lensing and galaxy clustering included in the analysis. This is of course expected because the $k^{-2}$ bias is largest on large scales.
At very low $\ell$, however, the improvement is somewhat less strong than naively expected; for example the improvement from $\ell_\mathrm{min}=2$ relative to $\ell_\mathrm{min}=10$ is less than a factor of 2.
This is a consequence of computing line-of-sight integrals exactly rather than using the Limber approximation (we will get back to this in \secref{FnlLimberDiscussion} below).

A second important aspect is that CMB-S4 and LSST should observe the same patch of sky:
For $\ell_\mathrm{min}=2$, perfect sky overlap improves the $\fnl$ precision by about a factor of 2 relative to observing on independent patches (solid vs dashed in \fig{fnl_LSST_S4}).
For $\ell_\mathrm{min}=20$ the improvement due to sky overlap is somewhat smaller but still a factor of $1.5$.
The improvements are due to sample variance cancellation and the breaking of degeneracies between $\fnl$ and $f_\mathrm{NL}^\mathrm{fake}$ using $\kappa g$ spectra that are only available on overlapping patches.
The improvement due to sky overlap is larger for better versions of LSST because they have a higher cross-correlation coefficient with CMB lensing.  

Third, it is important to include galaxies at very high redshift:
Including the high-redshift LSST dropout galaxies at $z=4-7$ improves the $\fnl$ precision by almost a factor of 2; see the dark orange curves in \fig{fnl_LSST_S4}.
The reason for this is that the size of the scale-dependent $\fnl$ bias in \eqq{fnlbias} increases with increasing redshift: $\Delta b/b\propto (b-1)/[bD(z)]\propto (b-1)\sim z$, if the Gaussian bias evolves with redshift as $b(z)\propto D^{-1}(z)$ where $D(z)$ is the growth function.
Despite the improvement from including high redshifts, the more conservative LSST $i<25$ sample at $z\le 4$ is at most a factor of 2 to 2.5 worse than the more optimistic $i<27$, $z\le 7$ sample, and can still reach $\sigma(\fnl)\sim 1$ if very large scales can be measured.
In contrast, dropping LSST entirely and using only CMB-S4, SDSS and DESI degrades the $\fnl$ precision by a factor of 10, showing that it is crucial to include a deep galaxy survey like LSST to achieve $\sigma(\fnl)\sim 1$.

\begin{figure}[tbp]
\includegraphics[width=0.5\textwidth]{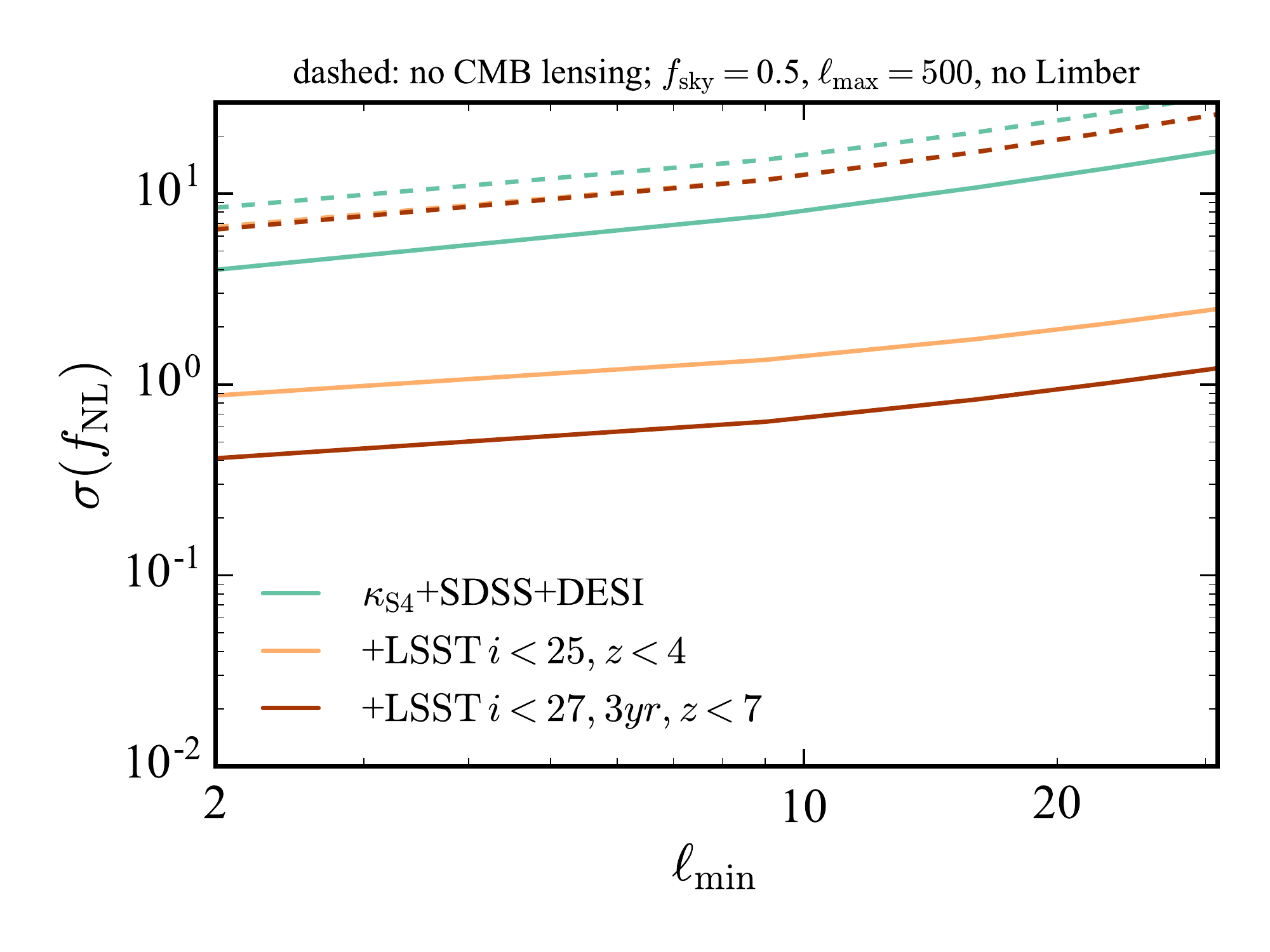}
\caption{$\fnl$ precision if only $gg$ spectra are used without any CMB lensing (dashed).
The precision is not competitive in that case, showing that CMB lensing is crucial.
Note that the yellow dashed line is barely visible because it overlaps with brown dashed.
}
\label{fig:fnl_lmin_CMBImpact}
\end{figure}

\fig{fnl_lmin_CMBImpact} shows that it is also critical to include CMB lensing measurements. 
The $\fnl$ precision with CMB lensing is 18 times stronger than without any CMB lensing measurement if we assume the most optimistic LSST scenario, and about 8 times stronger if we assume the least optimistic LSST scenario.
Without CMB lensing (dashed) we get $\sigma(\fnl)=6$ at best.
Thus, only the addition of CMB lensing makes it possible to achieve $\sigma(\fnl)<1$, by improving $\sigma(\fnl)$ by an order of magnitude.

\begin{figure}[tbp]
\includegraphics[width=0.5\textwidth]{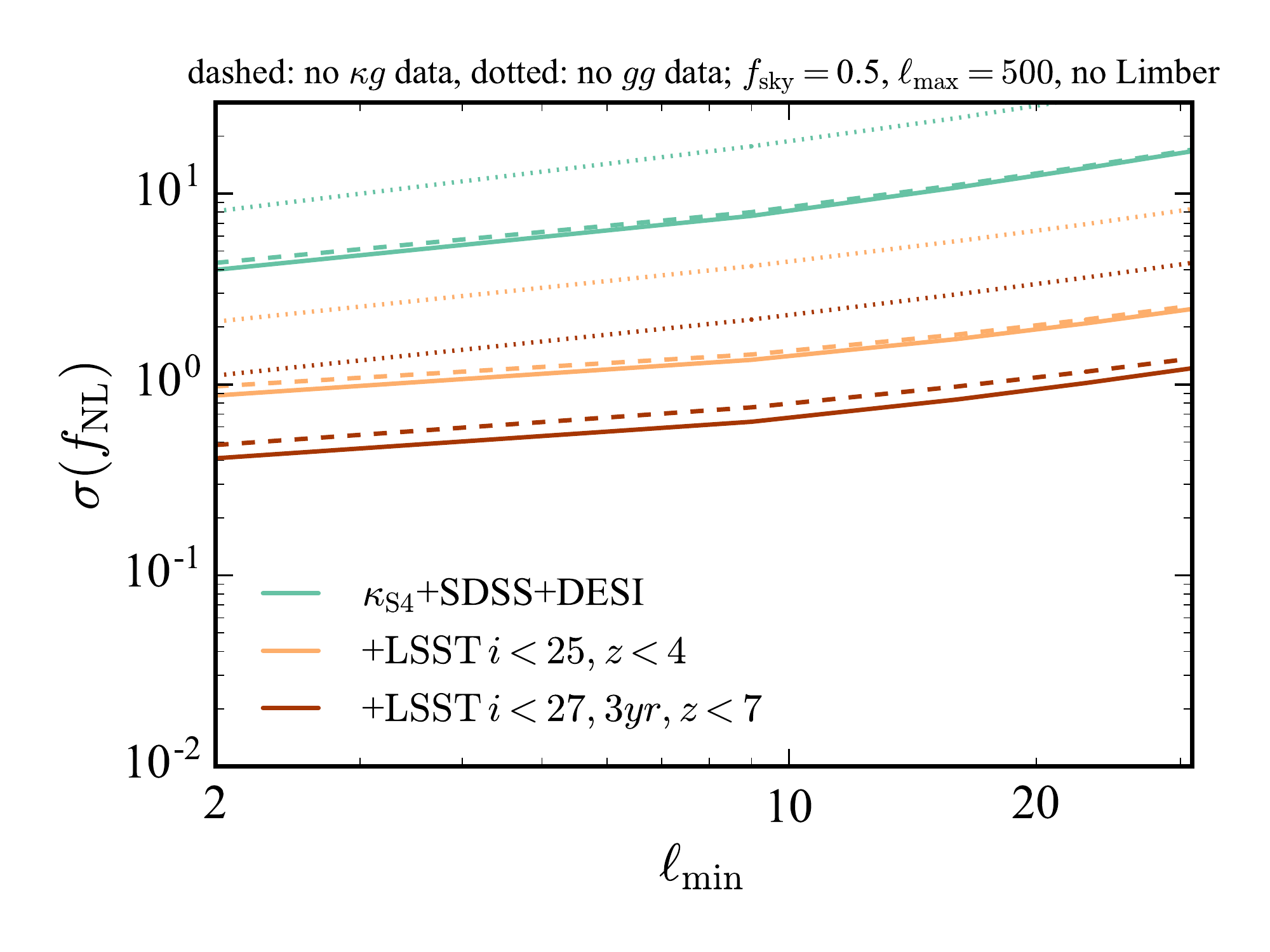}
\caption{Impact of different observed power spectra on the $\fnl$ precision. 
Solid assumes we observe all spectra, $\kappa\kappa$, $\kappa g$ and $gg$;
dashed assumes we only observe $\kappa\kappa$ and $gg$ but not $\kappa g$;
dotted assumes we only observe $\kappa\kappa$ and $\kappa g$ but not $gg$.
We only truncate the data vector in the Fisher analysis and do not modify any of the covariances, i.e.~all curves assume perfect sky overlap between all observed fields.
}
\label{fig:fnl_lmin_spectraImpact}
\end{figure}

\fig{fnl_lmin_spectraImpact} shows in more detail which power spectra are most important to observe, assuming all experiments observe the same patch of sky.
Observing no $\kappa g$ spectra and using only $\kappa\kappa$ and $gg$ degrades the $\fnl$ precision by at most $20\%$ relative to using all spectra (dashed vs solid). In many cases the degradation is smaller, i.e.~it is not important to measure $\kappa g$.
Notice however that the nonzero $\kappa g$ correlation is still exploited in the analysis because the covariance between measured $\kappa\kappa$ and $gg$ spectra involves $(C^{\kappa g})^2$ (if $C^{\kappa g}$ is nulled in the covariance, the precision degrades more; see dashed curves in \fig{fnl_LSST_S4}).

In contrast, $gg$ power spectra are very important for $\fnl$:
Without them, using only $\kappa\kappa$ and $\kappa g$ spectra, the $\fnl$ precision degrades by up to a factor 4 relative to using all spectra (dotted vs solid in \fig{fnl_lmin_spectraImpact}).
Part of this is caused by the fact that $gg\propto b^2$ is more sensitive to $\fnl$ than $\kappa g\propto b$, but the full improvement from $gg$ is somewhat larger than the expected factor of 2. 
Without measuring $gg$, using only $\kappa\kappa$ and $\kappa g$, we can reach $\sigma(\fnl)=1$ only if $\ell_\mathrm{min}=2$, which is rather challenging, especially from the ground.
It may thus be more promising to control $gg$ systematics like stellar contamination in the galactic plane, e.g.~using mode projection \cite{1992ApJ...398..169R,Slosar:2004fr,Leistedt1404,Leistedt1405,Kalus1607,Kalus1806}, and then use the large-scale $gg$ power spectra.

\begin{figure}[tbp]
\includegraphics[width=0.5\textwidth]{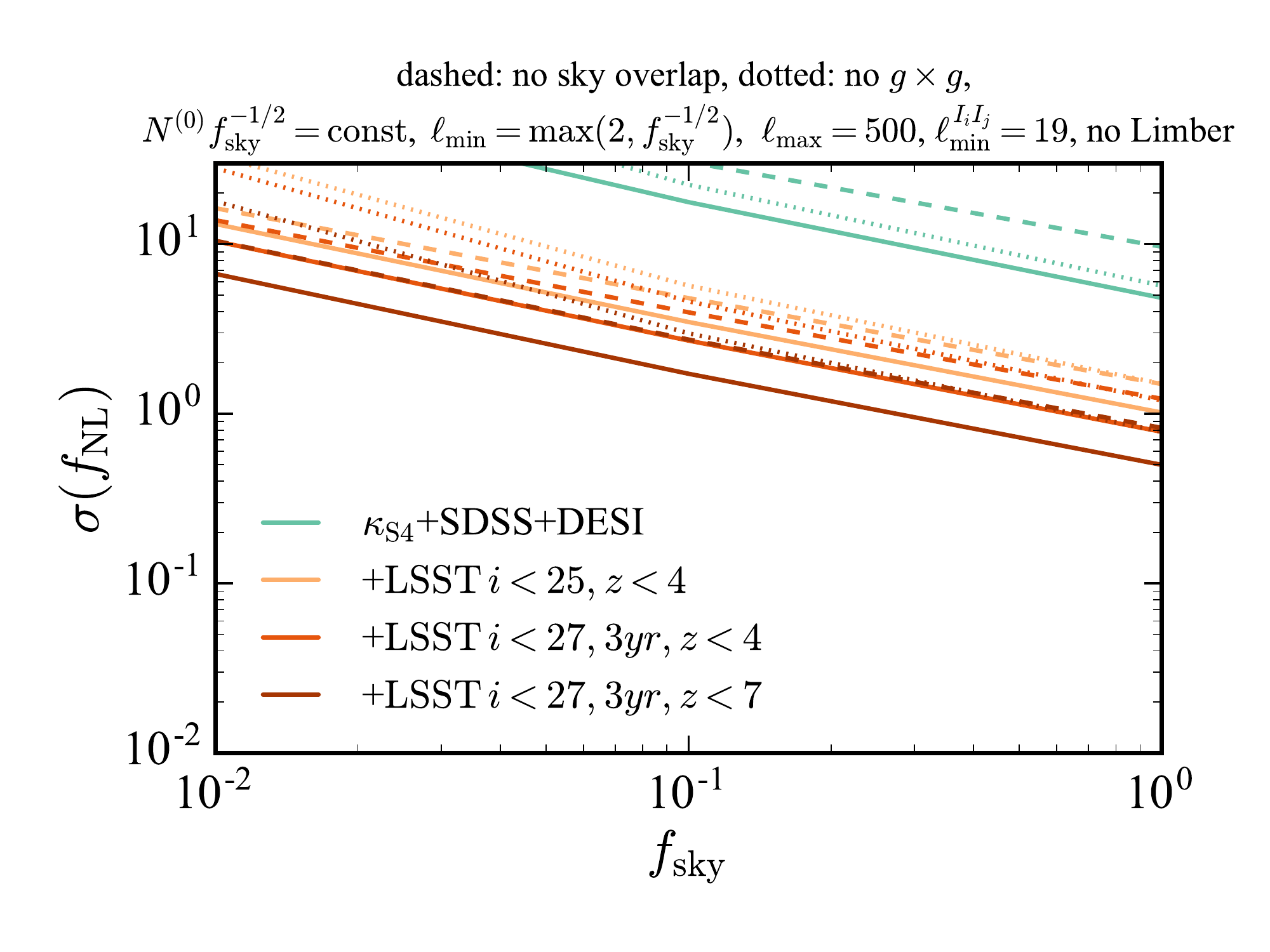}
\caption{Same as \fig{fnl_fsky}, but the solid curves exclude galaxy-galaxy spectra $C^{I_iI_j}_\ell$ at $\ell<18$.
}
\label{fig:fnl_fsky_lminIjIj20}
\end{figure}

Rather than hoping to project out large-scale $gg$ systematics, a more aggressive approach to avoid such systematics would be to drop all measured $gg$ spectra on large scales and use those measurements only on smaller scales where we believe systematics to be sufficiently small.
This is illustrated in \fig{fnl_fsky_lminIjIj20} where we drop all $gg$ auto- and cross-spectra at $\ell<18$ but include them at higher $\ell$.
In that case the $\fnl$ precision degrades roughly by a factor of 2 to 3 relative to using $gg$ spectra at all $\ell$.
Still, it is possible to reach $\sigma(\fnl)= 0.7$ for $\fsky=0.5$. 
This shows that very good $\fnl$ precision is possible even if large-scale $gg$ power spectra are excluded due to potential systematics.

\subsubsection{Optimizing experiments}

\begin{figure}[tbp]
\includegraphics[width=0.5\textwidth]{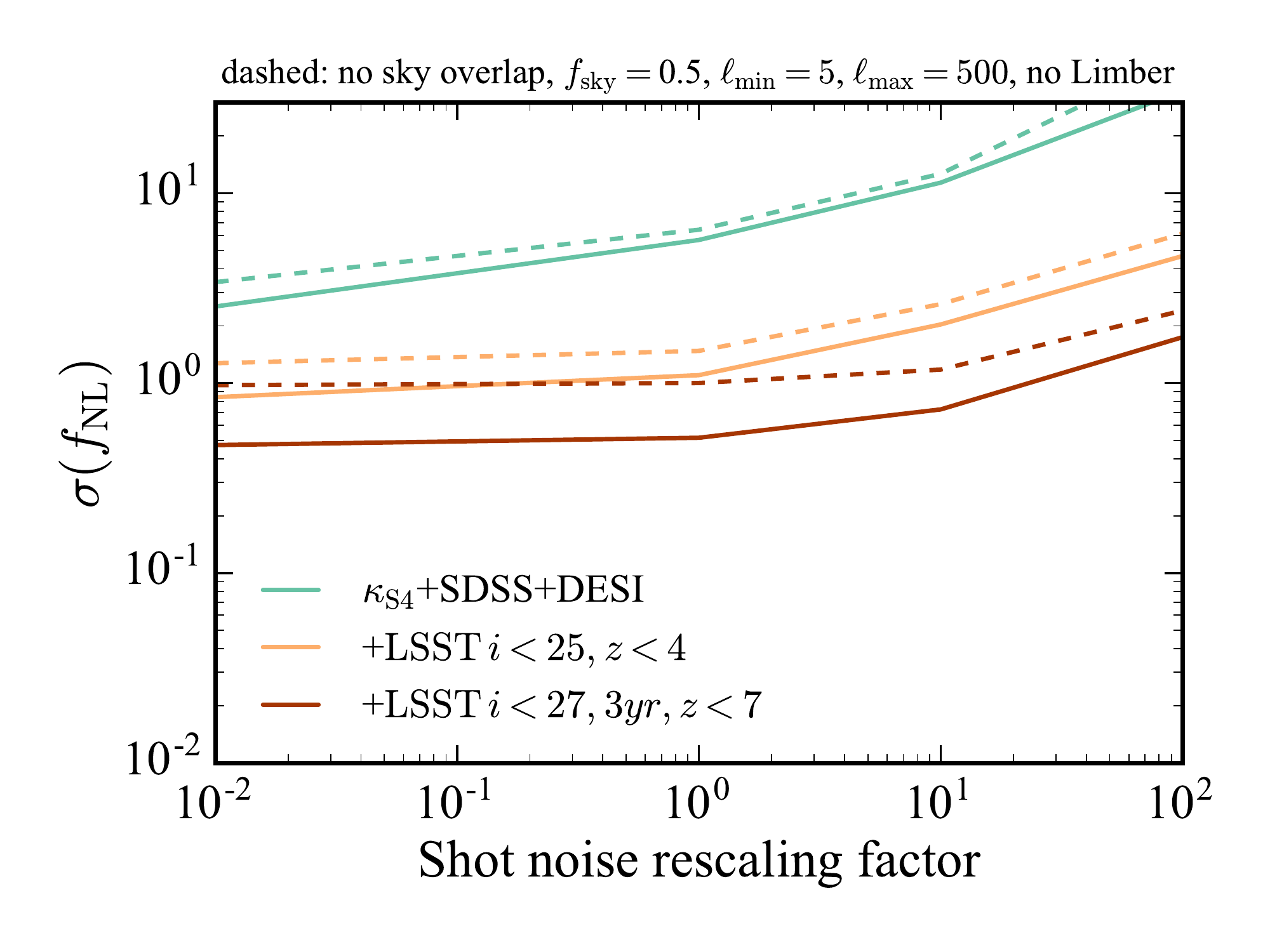}
\caption{$\fnl$ precision when multiplying the shot noise component of each power spectrum by a global rescaling factor that is varied on the horizontal axis, with 1 corresponding to the fiducial LSS survey specifications.
}
\label{fig:fnl_shotnoise}
\end{figure}

In \fig{fnl_shotnoise} we show how the $\fnl$ precision depends on the shot noise of the LSS surveys. If the shot noise is larger than in our default survey specifications, the $\fnl$ precision degrades in all cases.
Similarly, lower shot noise improves the precision for SDSS, DESI, and the conservative $i<25,z\le 4$ LSST sample.
But lower shot noise does not improve $\fnl$ for the optimistic $i<27,z\le 7$ LSST scenario. 
A possible reason for that saturation could be that tracers at redshift $z>7$ might be needed to increase the correlation coefficient with CMB lensing further, or CMB lensing noise might become the limiting factor if shot noise is very low.

How can the CMB lensing experiment be optimized for $\fnl$?
We find that reducing the CMB lensing noise by a factor of 5 relative to CMB-S4 does not visibly change the $\fnl$ precision if we assume the fiducial LSS survey specifications (not shown).
The $\fnl$ precision is thus not limited by CMB lensing noise and does not improve by improving CMB-S4 beyond our assumed 1 arcmin beam and $\Delta_{T}=1\,\mu$K arcmin noise, at least not before LSS surveys improve over LSST.
To optimize CMB-S4 for $\fnl$ thus means to push to as low lensing-$\ell$ as possible.
Note that this low lensing-$\ell$ comes from high CMB multipoles (for example lensing-$\ell=10$ can be obtained from the correlation of $\ell_\mathrm{CMB}=3000$ and $\ell_\mathrm{CMB}=3010$).
One foreseeable challenge of such measurements is that the lensing measurement would have to be consistent over a wide sky area, which poses challenges for example for accurate mean field characterization across the entire patch.

\begin{figure}[tbp]
\includegraphics[width=0.5\textwidth]{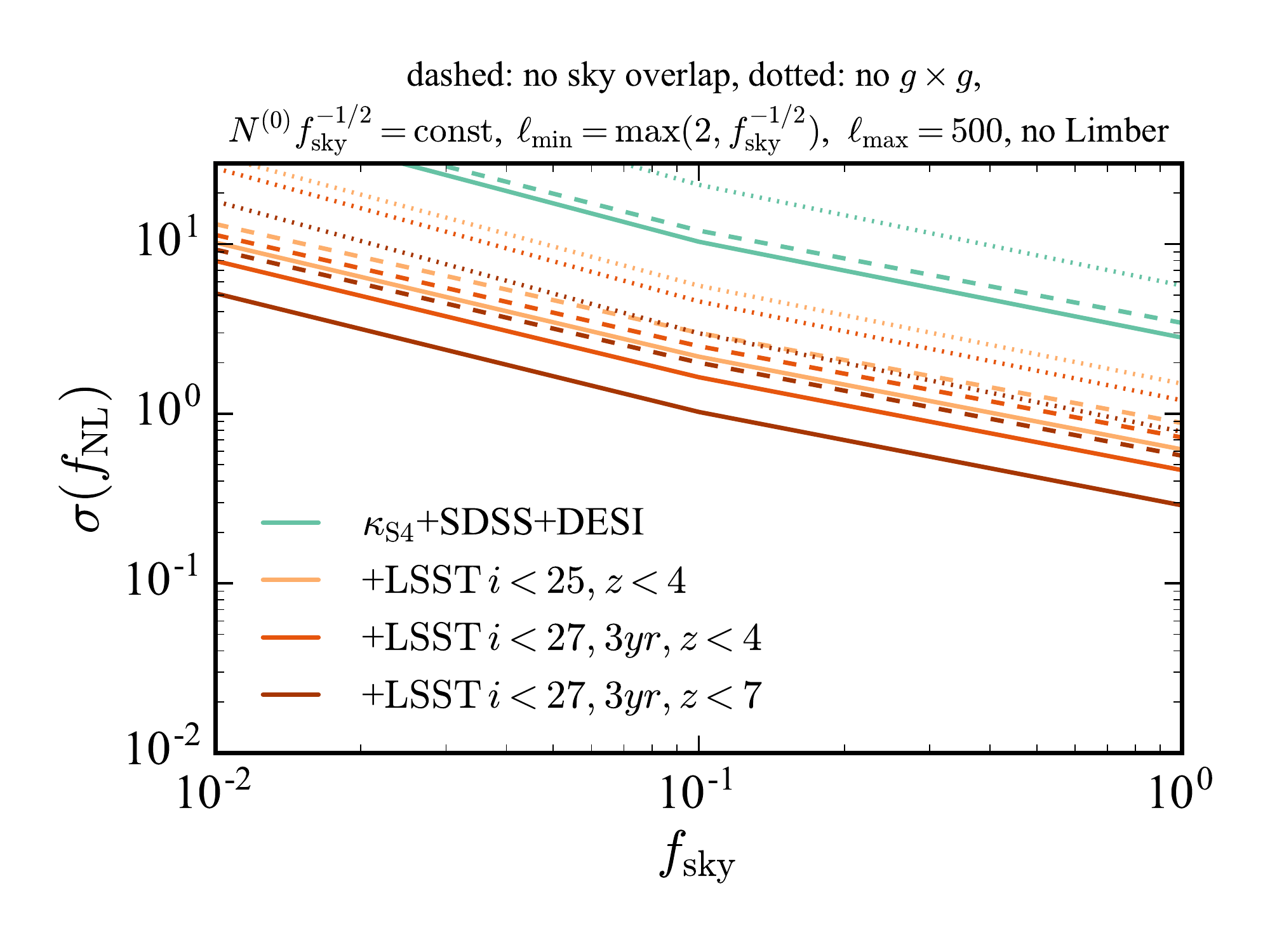}
\caption{
$\sigma(f_\mathrm{NL})$ precision as a function of sky fraction $f_\mathrm{sky}$ for approximately constant CMB observation time (keeping $N^{(0)}f_\mathrm{sky}^{-1/2}=\mathrm{const}$). We use $\ell_\mathrm{min}=\mathrm{max}(2,f_\mathrm{sky}^{-1/2})$).
Solid lines use $\kappa\kappa$, $\kappa g$ and $gg$ spectra assuming perfect sky overlap between all fields.
Dashed lines assume no sky overlap between any two fields. 
Dotted lines exclude $gg$ spectra from the data vector but assume perfect sky overlap. 
}
\label{fig:fnl_fsky}
\end{figure}

As in the previous section, we can also ask how the $\fnl$ precision depends on the sky fraction of the CMB lensing experiment if we assume approximately constant CMB observation time.
Since very low $\ell$ modes cannot be measured for too small $\fsky$, we also assume $\ell_\mathrm{min}=\mathrm{max}(2,f_\mathrm{sky}^{-1/2})$.
\fig{fnl_fsky} shows the resulting $\fnl$ precision as a function of $f_\mathrm{sky}$.
This shows that even for constant CMB observing time, the $\fnl$ precision strongly improves when increasing the sky fraction, preferring a wide shallow over a small deep CMB-S4 patch.
This makes sense because larger sky fraction reduces sample variance and enables measurements on larger scales where the $\fnl$ signal is maximal. 
The lower lensing noise that would be achievable with a small deep CMB patch does not improve the $\fnl$ precision, which is consistent with the finding above that lensing noise does not limit $\fnl$ in the regime studied here.

In conclusion, the above results suggest that LSS surveys can be optimized for $\fnl$ by  pushing to large scales and observing a high number density of tracers out to high redshift (like LSST), with maximal sky overlap with a wide shallow patch used to measure CMB lensing.

\subsubsection{Impact of marginalizations}
For the above $\fnl$ forecasts we marginalized over changes in the matter power spectrum parametrized by the $f_\mathrm{NL}^\mathrm{fake}$ parameter, and over linear galaxy bias. 
We briefly discuss the impact of these marginalizations.

\begin{figure}[tbp]
\includegraphics[width=0.5\textwidth]{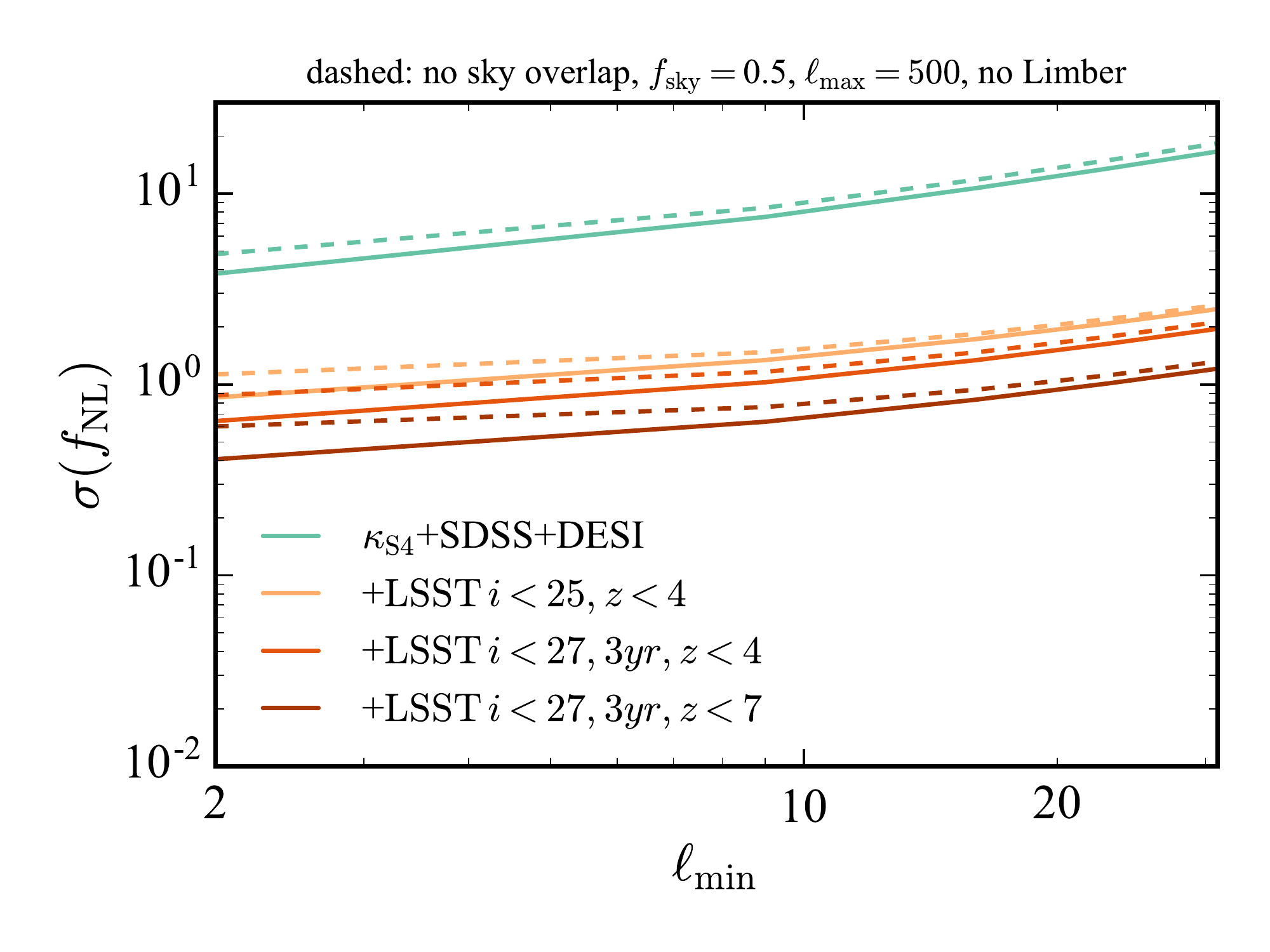}
\caption{$\fnl$ precision marginalized over galaxy bias but not over $f_\mathrm{NL}^\mathrm{fake}$.
The precision is the same as for marginalized $f_\mathrm{NL}^\mathrm{fake}$ if patches overlap perfectly on the sky (solid).
For independent, non-overlapping patches (dashed) the $\fnl$ precision is somewhat better than for the marginalized $f_\mathrm{NL}^\mathrm{fake}$ case which suffers from the degeneracy between $\fnl$ and $f_\mathrm{NL}^\mathrm{fake}$ that cannot be broken if patches do not overlap.
}
\label{fig:fnl_lmin_noFnlFake}
\end{figure}

First, to determine the relevance of the shape of the matter power spectrum, \fig{fnl_lmin_noFnlFake} shows the same forecast as \fig{fnl_LSST_S4} but without marginalizing over $f_\mathrm{NL}^\mathrm{fake}$. 
In that case, the improvement from having perfect sky overlap relative to having independent patches is about a factor $1.5$ for $\ell_\mathrm{min}=2$ and a factor $1.1$ for $\ell_\mathrm{min}=20$ (dashed vs solid in \fig{fnl_lmin_noFnlFake}).
Comparing with the corresponding improvement factors of 2 and 1.5 due to sky overlap in \fig{fnl_LSST_S4} that marginalized over $f_\mathrm{NL}^\mathrm{fake}$, this suggests that sample variance cancellation and breaking of degeneracies between $\fnl$ and $f_\mathrm{NL}^\mathrm{fake}$ using $\kappa g$ both contribute significantly for $\ell_\mathrm{min}=2$, but breaking of parameter degeneracies is the dominant effect for $\ell_\mathrm{min}=20$. 
The main point is still that a joint analysis of CMB-S4 and LSS clustering on the same patch can improve the $\fnl$ precision by a factor of 2 to 1.5.

\begin{figure}[tbp]
\includegraphics[width=0.5\textwidth]{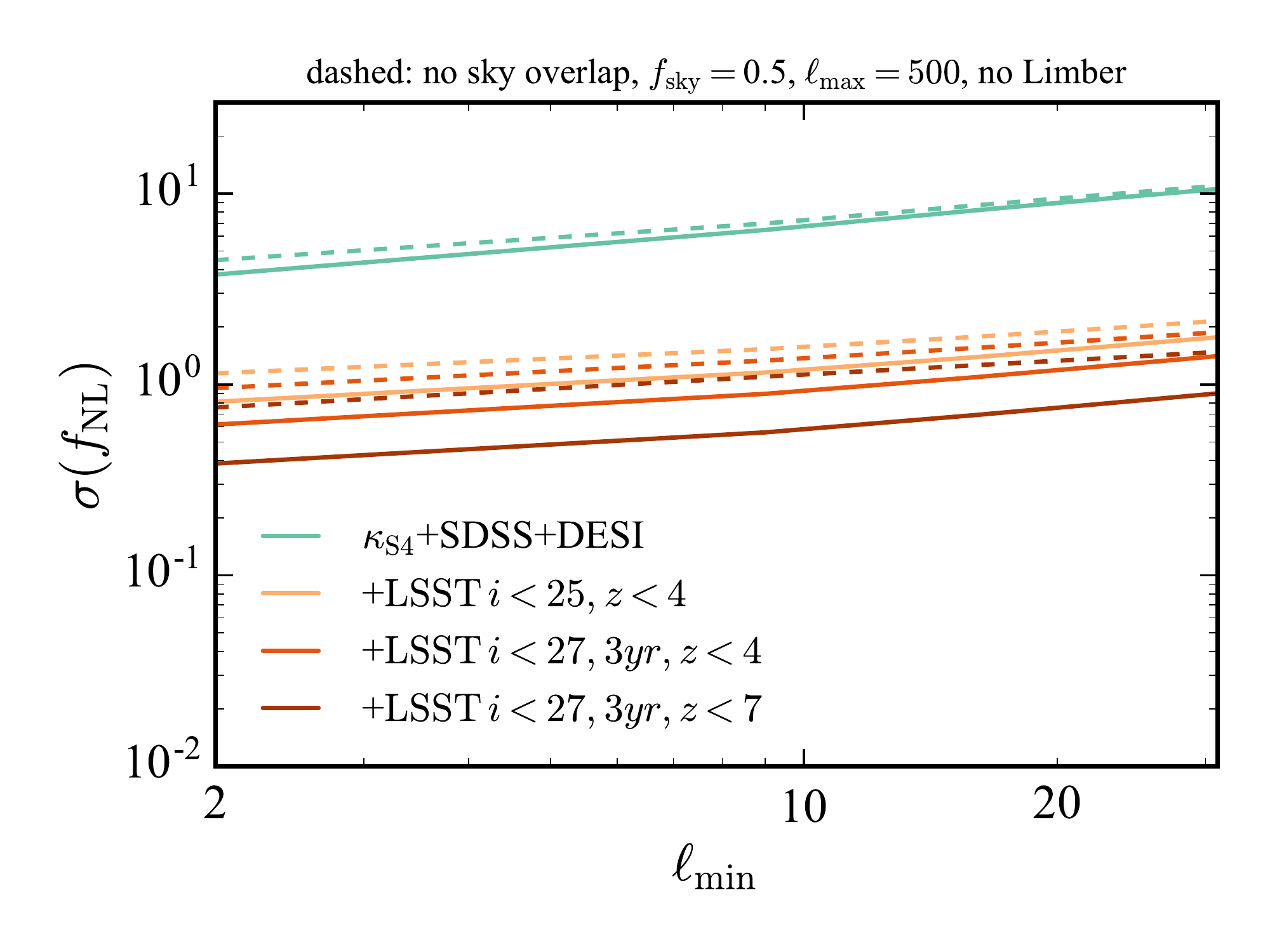}
\caption{$\fnl$ precision marginalized over $f_\mathrm{NL}^\mathrm{fake}$ but not over galaxy bias.
This has no effect for $\ell_\mathrm{min}=2$, but it improves the precision somewhat for $\ell_\mathrm{min}\gtrsim 10$.
}
\label{fig:fnl_lmin_noBias}
\end{figure}

Second, to determine the impact of marginalizing over the fiducial galaxy bias, \fig{fnl_lmin_noBias} shows the same forecasts as \fig{fnl_LSST_S4} but without marginalizing over galaxy bias.
This improves the $\fnl$ precision somewhat for $\ell_\mathrm{min}\gtrsim 10$. 
For example, we obtain $\sigma(\fnl)=0.9$ for $\ell_\mathrm{min}=20$ for the most optimistic LSST version.
If we can measure galaxy bias better, e.g.~using redshift space distortions and statistics beyond power spectra, we could achieve this somewhat higher precision. 
Note that for very low $\ell_\mathrm{min}=2$, knowing the galaxy bias does not improve over the $\sigma(\fnl)=0.4$ precision we obtained before.

\subsubsection{Exact integration vs Limber approximation}
\label{se:FnlLimberDiscussion}

\begin{figure}[tbp]
\includegraphics[width=0.5\textwidth]{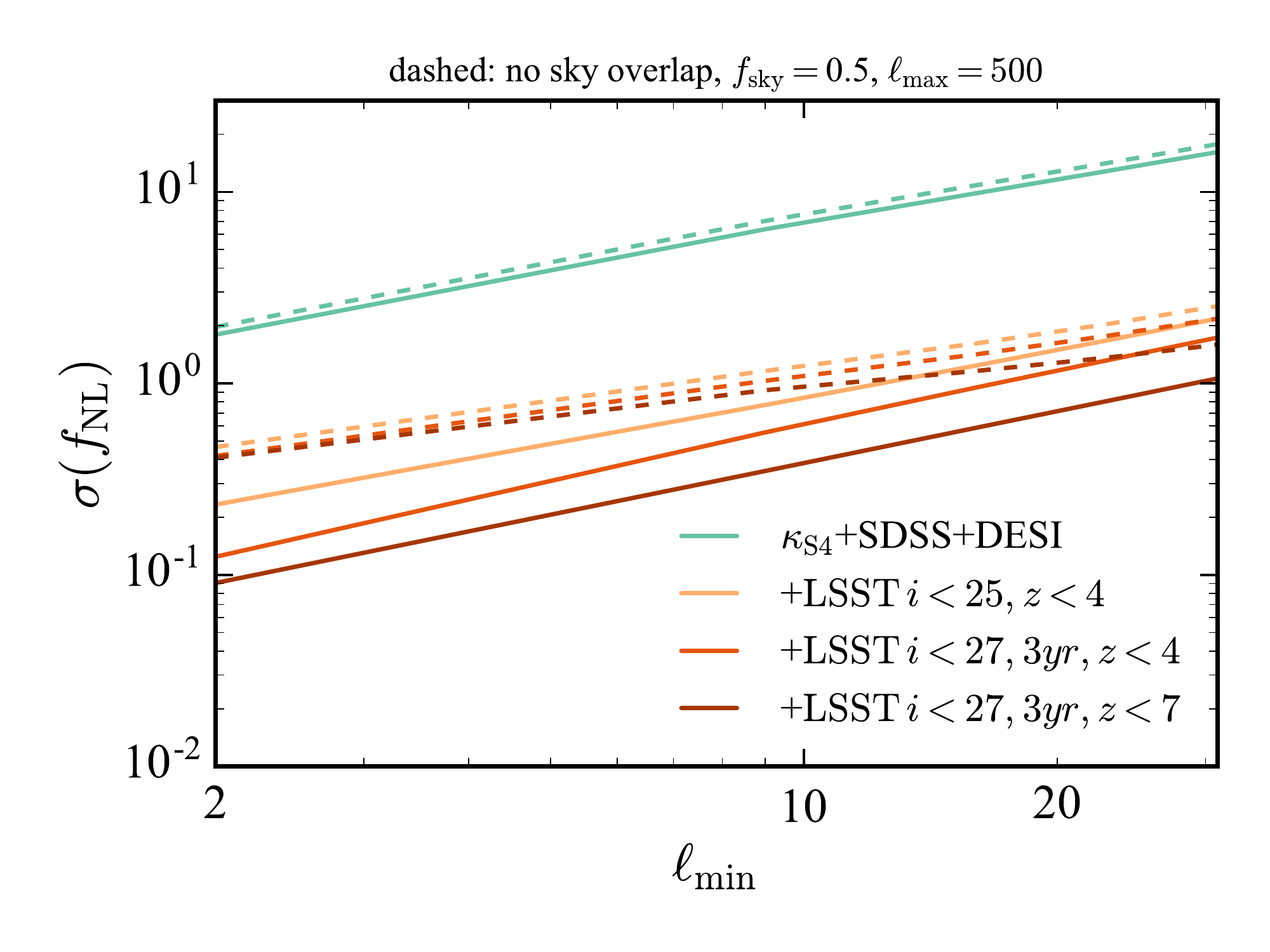}
\caption{
$\fnl$ precision as a function of $\ell_\mathrm{min}$ as in \fig{fnl_LSST_S4}, but using the Limber approximation at all wavenumbers $\ell$.
For low $\ell_\mathrm{min}$, the Limber approximation would wrongly suggest $\sigma(\fnl)$ to be a few times smaller than the exact result of \fig{fnl_LSST_S4}.
}
\label{fig:fnl_LSST_S4_limber}
\end{figure}

\fig{fnl_LSST_S4_limber} shows the $\fnl$ precision when wrongly assuming the Limber approximation on all scales. 
In that case the precision looks a few times better than for exact line-of-sight integrals \NEW{(see \cite{Jeong0910} for similar results)}.
The reason for this is that power spectra, especially galaxy-galaxy auto-spectra, are larger when computed exactly rather than with Limber, while derivatives with respect to $\fnl$ are smaller.
The fractional change of galaxy auto-spectra due to nonzero $\fnl$ is thus much smaller in the exact calculation than if the Limber approximation was assumed.
This reduction in $\fnl$ sensitivity relative to Limber-approximated power spectra is partially compensated because the exact integration adds a nonzero signal in cross-spectra between non-overlapping redshift bins, but these spectra are very small and their signal is much smaller than the sensitivity lost in the galaxy auto-power spectra.
This shows that it is important to use exact line-of-sight integrals at $\ell\le 50$ when forecasting $\fnl$ to avoid overly optimistic forecasts.

\subsection{Neutrino mass from scale-dependent bias}

\subsubsection{Setup}
As motivated above and discussed in more detail in \app{NeutrinoBasicsAppdx}, we can probe the sum of neutrino masses by trying to observe the small scale-dependent difference in the transfer functions relevant for CMB lensing and galaxy clustering.
To forecast how well this works with CMB-S4 lensing and LSST clustering, we marginalize over linear galaxy bias like in the last sections. 
Additionally we marginalize over a parameter $m_\nu^\mathrm{fake}$ that rescales the total matter power spectrum in a way that mimics the scale-dependence of the true neutrino mass bias as described in \eqq{ClNeutrinos}, analogeously to the $f_\mathrm{NL}^\mathrm{fake}$ parameter used in the last section.  This effectively marginalizes over uncertainties in the shape of the total power spectrum.
We do not include any information from the primary CMB.

\subsubsection{Results}

\begin{figure}[tbp]
\includegraphics[width=0.5\textwidth]{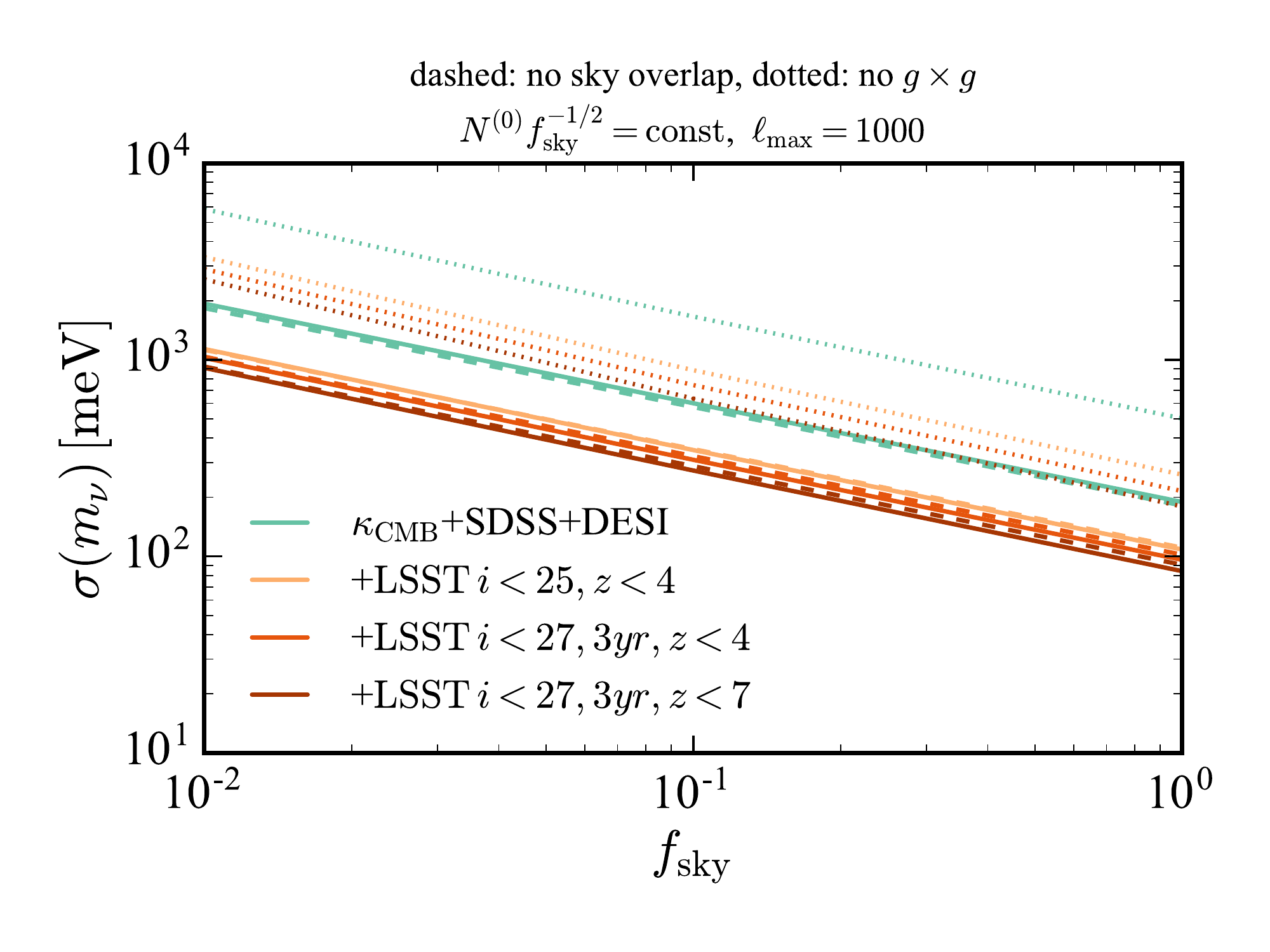}
\caption{
Expected precision for the sum of neutrino masses from scale-dependent bias between CMB-S4 lensing and galaxy clustering, as a function of sky fraction, assuming approximately constant CMB observation time (keeping $N^{(0)}f_\mathrm{sky}^{-1/2}$ fixed), and using multipoles $20\le \ell\le 1000$.
We marginalize over one scale-independent linear bias parameter per LSS redshift bin, and over  the $m_\nu^\mathrm{fake}$ parameter that rescales $P_\mathrm{tot}$ with the same signature as the scale-dependent difference between the total power spectrum (based on $T_\mathrm{cb\nu}$) and the matter-only power spectrum (based on $T_\mathrm{cb}$).
}
\label{fig:mnu_fsky}
\end{figure}

\fig{mnu_fsky} shows the expected neutrino mass precision from the scale-dependent bias effect.
In the most optimistic scenario we obtain $\sigma(m_\nu)\simeq 90\,\mathrm{meV}$.
Unfortunately, this is not competitive with the conventional method to measure neutrino mass from the suppression of small-scale clustering power at low redshift relative to the power of the primary CMB at high redshift:  
Using that method, a joint analysis of DESI galaxy power spectrum, DESI BAO, and Planck can achieve $\sigma(m_\nu)\simeq 20\,\mathrm{meV}$ (see Table 2.11 in \cite{DESIFDRDoc}),
with comparable precision also expected from CMB-S4 lensing, DESI BAO and an external $\tau$ prior \cite{CMBS4SciBook}.

We can ask what impact marginalizing over $m_\nu^\mathrm{fake}$ and galaxy bias has.
We find that not marginalizing over $m_\nu^\mathrm{fake}$ 
has virtually no impact on the precision  of $m_\nu$.
Uncertainties in the shape of the underlying total matter power spectrum do therefore not limit the neutrino forecast.

\begin{figure}[tbp]
\includegraphics[width=0.5\textwidth]{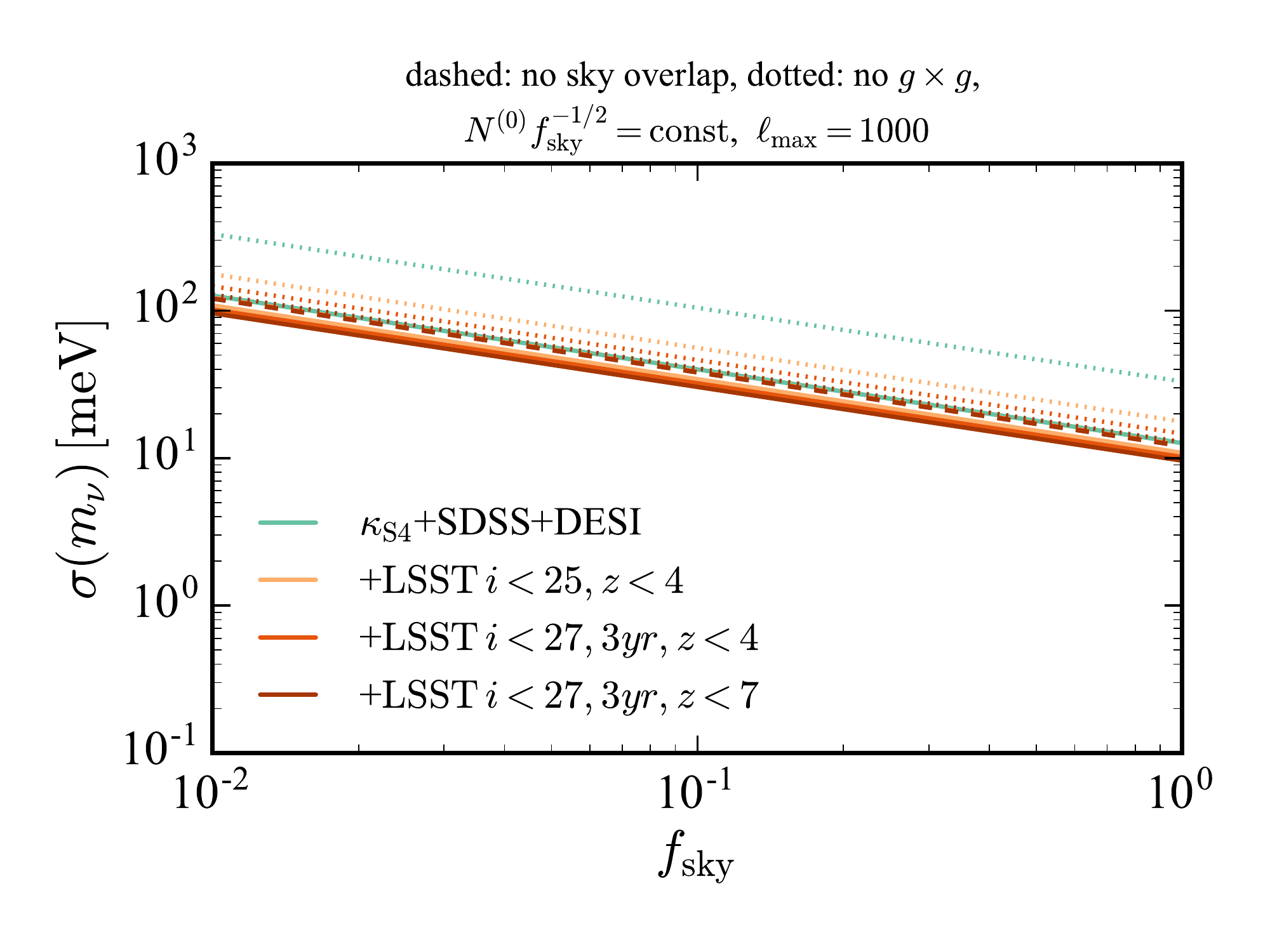}
\caption{Neutrino mass precision marginalized over $m_\nu^\mathrm{fake}$ but not over galaxy bias. In that case the precision is much better. Notice the different scale of the vertical axis.
}
\label{fig:mnu_fsky_not_marginalizing_bias}
\end{figure}

In contrast, \fig{mnu_fsky_not_marginalizing_bias} shows that if we do not marginalize over galaxy bias, the neutrino mass precision improves by a factor of 10, reaching $\sigma(m_\nu)=10\,\mathrm{meV}$ in the best case.
This shows that galaxy bias is the key limitation for the neutrino forecasts.
Improved bias measurements or modeling could thus lead to interesting neutrino constraints from the scale-depedent bias effect. 
2-point cross-correlations between CMB lensing and galaxy clustering alone are not able to provide such accurate bias measurements even if we only assume linear bias (otherwise this would have shown up in forecasts marginalizing over galaxy bias in \fig{mnu_fsky}).
But measurements of higher-order N-point functions or redshift space distortions should be able to determine galaxy bias more accurately.
This could push the neutrino mass precision closer to the case without marginalizing over galaxy bias, although this requires more detailed studies that include statistics beyond the power spectrum and nonlinear galaxy bias.

In conclusion, measuring the sum of neutrino masses using scale-dependent bias between lensing and clustering alone is likely not competitive with other methods.
This is consistent with similar findings for correlating galaxy weak lensing shear and galaxy clustering \cite{LoVerde1602}.
Nevertheless, if neutrino mass is sufficiently large to be detectable using this scale-dependent bias effect, it may serve as a useful cross-check that would be independent from other cosmological neutrino mass measurements and independent of $\tau$.
Future improvements may be possible if bias parameters can be measured better, e.g.~using redshift space distortions and higher-order N-point functions, or if we can improve models for the measured power spectra involving fewer bias parameters while reaching smaller scales.

\NEW{
\section{Redshift errors}
\label{se:zerrors}
\subsection{Types of errors}
The above forecasts ignored redshift errors throughout. 
This is an important caveat because dense imaging surveys rely on photometric redshifts that are subject to two types of redshift errors.

First, there are non-catastrophic errors that smear out the true redshifts.
They can be described by adding to the true redshifts Gaussian random noise, with a typical rms of $dz/(1+z)\sim 0.05$ for LSST.
This error is smaller than our broad tomographic redshift bins which satisfy $\Delta z/(1+z)>0.2$ in all cases. 
We therefore continue to ignore these errors.

Second, there are catastrophic redshift errors, where galaxies are attributed to completely wrong redshifts.
This can severely bias angular power spectra. For example, if low-redshift galaxies (say $z=0.1$) are wrongly attributed to high-redshift tomographic bins (say $z=3-4$), this adds spurious power to high-redshift tomographic bins.
This can then be confused with an $\fnl$ signal or high $\sigma_8$ at high redshift.
Projection effects make this systematic error scale-dependent.
It may be possible to identify and remove some of the catastrophic outliers by comparing with spectroscopic surveys like DESI, which would reduce the catastrophic error rate, but presumably not perfectly.

\subsection{A simple model for catastrophic redshift errors}
Due to their complicated nature a fully realistic treatment of catastrophic redshift errors would be beyond the scope of this paper.
To still get a sense of their impact, we introduce an idealized but simple analytical model:
We reshuffle galaxy redshifts such that some fraction of galaxies is assigned to the correct tomographic redshift bin, while the remaining galaxies are outliers that are assigned to other redshift bins.
Each tomographic redshift bin will then consist of galaxies with correctly assigned redshifts and outlier galaxies whose true redshift is outside the redshift bin.

To compute the overdensity $\delta_g(\vtheta)$ of the $i$th observed tomographic redshift bin, we therefore integrate over the modified number density
\begin{align}
  \label{eq:fcatastr}
  \frac{\d n}{\d z}\bigg|_{i,\mathrm{obs}}(z) \;=\;
  \begin{cases}
 (1-f^i_\mathrm{out})\,\frac{\d n}{\d z}(z) & \mbox{if $z\in i$th bin}, \\
 \frac{n_i}{n_\mathrm{tot}-n_i}\,f^i_\mathrm{out}\,\frac{\d n}{\d z}(z)  & \mbox{else},
  \end{cases}
\end{align}
where the first line comes from galaxies with correct redshifts and the second line is due to outliers.
We introduced the outlier fraction $f_\mathrm{out}^i$ as the probability that a galaxy assigned to the $i$th tomographic redshift bin (e.g.,~$z=3-4$) actually resides at a redshift outside that bin (e.g., $z=0.1$).
$\frac{\d n}{\d z}(z)$ is the fiducial angular number density of the survey, i.e.~our best estimate of the true redshift distribution that would be obtained if the survey had no redshift errors; 
$n_i\equiv\int_{z\in\mathrm{bin}\,i}\d z\frac{\d n}{\d z}(z)$ is the number of objects per steradian in the $i$th tomographic bin if there were no outliers; and $n_\mathrm{tot}\equiv\int \d z \frac{\d n}{\d z}(z)$ is the total number of observed objects per steradian if we integrate over all redshifts where $\frac{\d n}{\d z}(z)$ is nonzero ($0\le z\le 7$ in our case).
The normalization in \eqq{fcatastr} ensures that the total number of galaxies per tomographic bin is conserved when changing the outlier fraction $f^i_\mathrm{out}$.

The angular cross-power spectra between CMB lensing and observed galaxy redshift bins are then 
\begin{align}
  \label{eq:kg_outliers}
  C^{\kappa g}_\ell = (1-f_\mathrm{out})C^{\kappa\mathrm{c}}_\ell + f_\mathrm{out}C^{\kappa\mathrm{o}}_\ell,
\end{align}
where $C^{\kappa\mathrm{c}}$ is due to galaxies assigned to the correct tomographic redshift bin and $C^{\kappa\mathrm{o}}$ is due to redshift outliers.
Similarly, the auto-power spectra of clustering in observed redshift bins are
\begin{align}
  \label{eq:gg_outliers}
  C^{gg}_\ell = (1-f_\mathrm{out})^2C^{\mathrm{c}\mathrm{c}}_\ell
+2(1-f_\mathrm{out})f_\mathrm{out}C^{\mathrm{c}\mathrm{o}}_\ell
+f_\mathrm{out}^2C^{\mathrm{o}\mathrm{o}}_\ell,
\end{align}
which have contributions from the auto-correlation $C^{\mathrm{c}\mathrm{c}}$ of correctly assigned redshifts, from the cross-correlation  $C^{\mathrm{c}\mathrm{o}}$ between correct and outlier redshifts, which is only nonzero if beyond-Limber corrections are included, and from the auto-correlation $C^{\mathrm{o}\mathrm{o}}$ of outliers.\footnote{A similar expression holds for the cross-correlation between two different redshift bins if beyond-Limber corrections are included.  Also notice that the $C_\ell$ on the right hand side of \eq{kg_outliers} and \eq{gg_outliers} are independent of the fiducial value of $f_\mathrm{out}$ and are determined by the fiducial global number density $\d n/\d z$ of the survey.
To compute the shot noise of auto-power spectra with \eqq{NellShotNoise} we integrate over \eqq{fcatastr}.
Since outliers are just re-distributed between redshift bins and each galaxy still contributes only to a single redshift bin, cross-spectra between different redshift bins still have no shot noise.
If the fiducial $f_\mathrm{out}$ is nonzero, each fiducial angular $\kappa g$ and $gg$ power spectrum depends on galaxy bias amplitudes $B_i$ at all redshifts, which we will include in forecasts, but the dependence on bias parameters outside the nominal redshift bin is suppressed for small outlier fractions.
}

To study the impact of redshift outliers on forecasts, we will marginalize over the outlier rate. 
The fractional response of $C^{\kappa g}$ to a fractional change in the outlier rate is
\begin{align}
  \label{eq:kg_fout_response}
  \frac{f_\mathrm{out}}{C^{\kappa g}_\ell}\frac{\partial C^{\kappa g}_\ell}{\partial f_\mathrm{out}} = 
-\eta \frac{C_\ell^{\kappa \mathrm{c}}-C_\ell^{\kappa \mathrm{o}}}{C_\ell^{\kappa\mathrm{c}}+\eta C_\ell^{\kappa\mathrm{o}}},
\end{align}
where
\begin{align}
  \label{eq:eta_def}
  \eta \equiv \frac{f_\mathrm{out}}{1-f_\mathrm{out}}
\end{align}
 is small for small outlier fractions.
The response \eq{kg_fout_response} is shown in \fig{kg_fout_response}.
A $10\%$ ($100\%$) change in the outlier rate changes $C^{\kappa g}$ by at most $6\%$ ($60\%$).
The fractional response is largest at $\ell\lesssim 100$ and for high-redshift bins, because low-redshift galaxies  with large clustering power are wrongly assigned to high-redshift tomographic bins where the true clustering power is small.

\begin{figure}[tbp]
\includegraphics[width=0.48\textwidth]{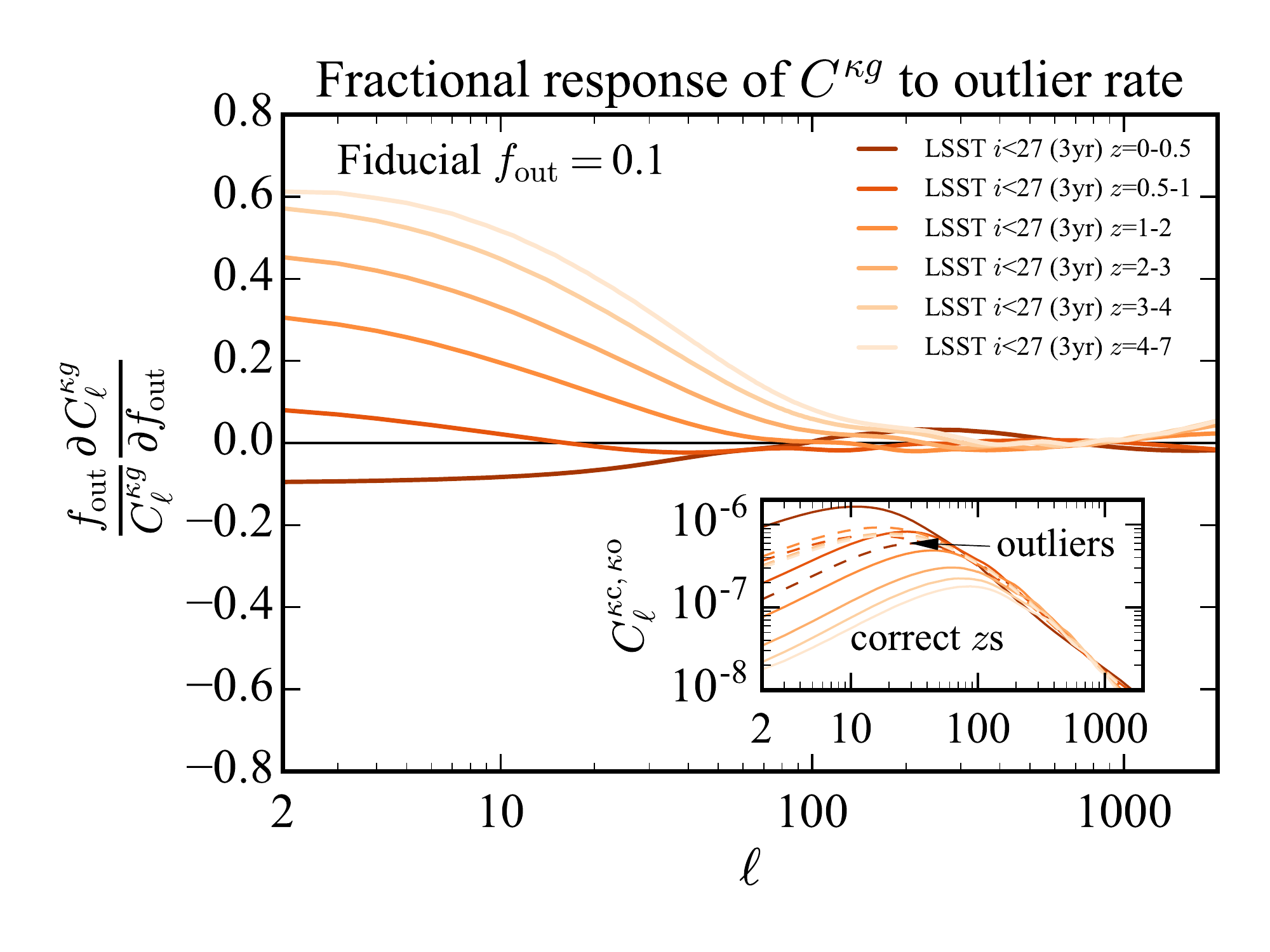}
\caption{Fractional response \eq{kg_fout_response} of CMB-lensing galaxy-clustering cross-spectra $C^{\kappa g}$ to a fractional change in the redshift outlier rate $f_\mathrm{out}$.
At low $\ell$ and for the lowest redshift bin, the contribution from correct redshifts $C^{\kappa\mathrm{c}}$ dominates over the contribution $C^{\kappa\mathrm{o}}$ from outliers so that the response \eq{kg_fout_response} becomes $-\eta=-0.11$.
For tomographic bins at higher redshift, the outlier term $C^{\kappa\mathrm{o}}$ dominates over the correct-redshift term $C^{\kappa\mathrm{c}}$, and low-$z$ galaxies that are incorrectly assigned to high-$z$ bins contaminate the high-$z$ bins.
This leads to a large low-$\ell$ response of high-redshift bins to the outlier fraction.
At $\ell\gtrsim 100$ the response \eq{kg_fout_response} vanishes because 
$C^{\kappa\mathrm{c}}\approx C^{\kappa\mathrm{o}}$.
}
\label{fig:kg_fout_response}
\end{figure}

The fractional response of galaxy auto-power spectra to a fractional change in the outlier rate is
\begin{align}
  \label{eq:gg_fout_response}
    \frac{f_\mathrm{out}}{C^{gg}_\ell}\frac{\partial C^{gg}_\ell}{\partial f_\mathrm{out}} = 
-2\eta \frac{C^\mathrm{cc}_\ell - (1-\eta)C^\mathrm{co}_\ell - \eta C^\mathrm{oo}_\ell}{C^\mathrm{cc}+2\eta C^\mathrm{co}+\eta^2C^\mathrm{oo}}
\approx -2\eta,
\end{align}
where the approximation in the last step is valid for small fiducial outlier fractions, $f_\mathrm{out}\lesssim 0.1$, because in that case $\eta\lesssim 0.1$ and $C^\mathrm{co}\ll C^\mathrm{cc}$.
 \fig{gg_fout_response} shows the response \eq{gg_fout_response}.
Indeed, it is close to $-2\eta$ for all $\ell$ and redshift bins.

\begin{figure}[tbp]
\includegraphics[width=0.48\textwidth]{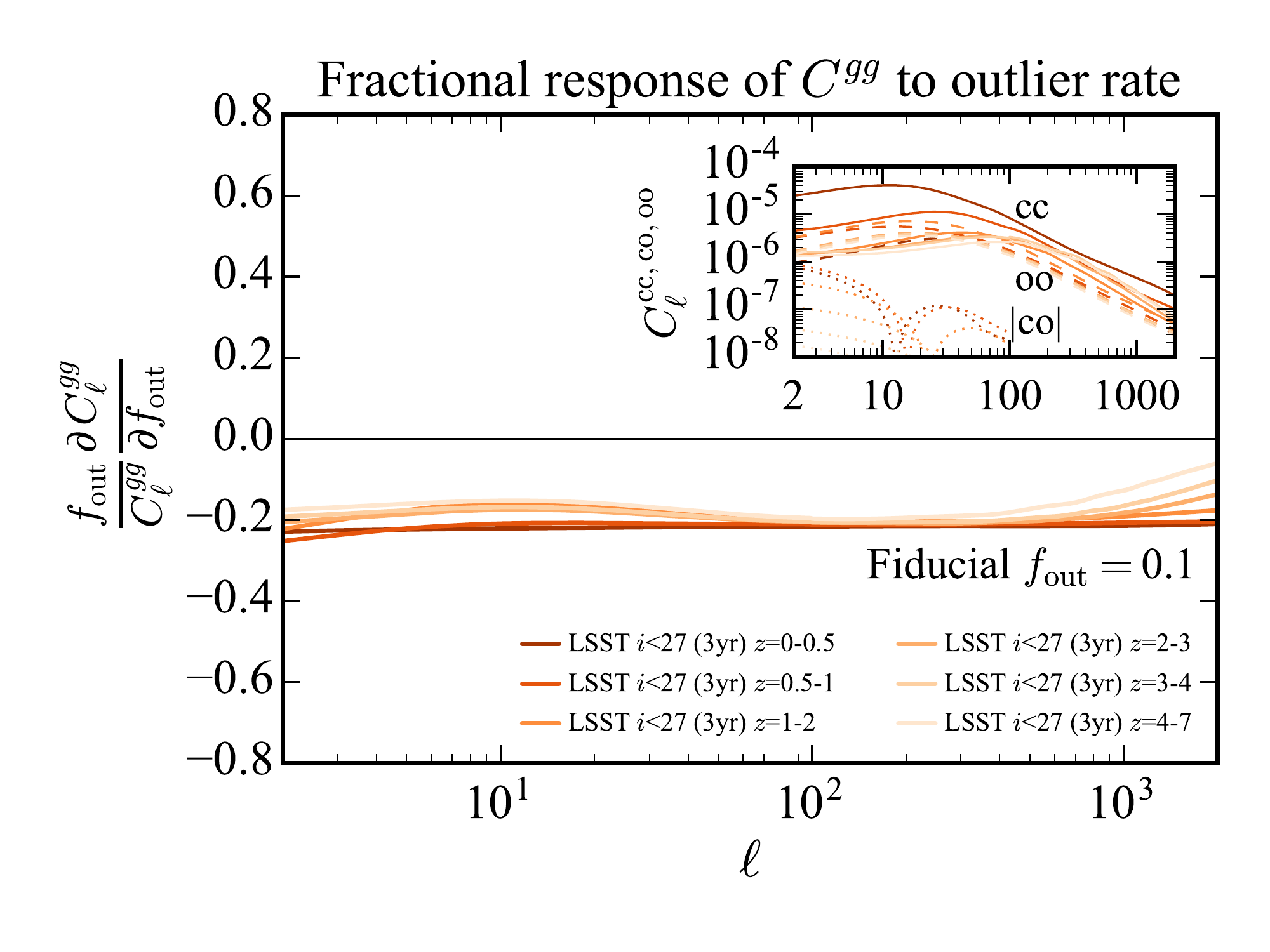}
\caption{Fractional response \eq{gg_fout_response} of galaxy auto-power spectra to a fractional change in the redshift outlier rate. 
The response is approximately $-2\eta=-0.22$, because in \eqq{gg_fout_response} the contribution $C^{cc}$ from the auto-correlation of correctly assigned redshifts dominates over the outlier contributions $(1-\eta)C^\mathrm{co}$ and $\eta C^\mathrm{oo}$ for the assumed fiducial outlier rate $f_\mathrm{out}=0.1$.
}
\label{fig:gg_fout_response}
\end{figure}

Thus, $\kappa g$ spectra respond to the outlier rate on large scales $\ell\lesssim 100$ but not on smaller scales, whereas the response of $gg$ spectra is approximately independent of scale and redshift.
The outlier rate can therefore be determined by measuring both $\kappa g$ and $gg$ spectra, as we discuss next.

}

\NEW{

\subsection{Impact on forecasts}

We perform a Fisher analysis that includes one outlier rate parameter $f_\mathrm{out}^i$ for each of our six tomographic redshift bins, assuming a fiducial outlier rate of $f_\mathrm{out}^i=0.1$ without any priors. We marginalize over one linear bias parameter per redshift bin, $f_\mathrm{NL}$, and $f_\mathrm{NL}^\mathrm{fake}$, and use all power spectra of CMB-S4 lensing and LSST clustering up to $\ell_\mathrm{max}=500$ (including beyond-Limber corrections at $\ell\le 50$).
We find that the uncertainty of $f_\mathrm{NL}$ degrades by only $3\%$ if we marginalize over the outlier fraction compared to assuming perfectly known redshifts.
The reason for this is that the outlier fraction can be measured from its distinct imprint on observable $\kappa g$ and $gg$ power spectra described above. 
Indeed, the outlier rate in each tomographic redshift bin can be determined with uncertainty $\sigma(f_\mathrm{out}^i)\sim 0.004$, i.e.~with subpercent level precision, similarly to the precision of bias parameters in this forecast.
For a fiducial outlier rate of $f_\mathrm{out}^i=0.3$, the degradation of $\sigma(f_\mathrm{NL})$ from marginalizing over $f_\mathrm{out}$ is $6\%$, which is still negligible; for a fiducial outlier rate of $f_\mathrm{out}^i=0.8$, which is unrealistically large, the degradation becomes a factor of a few.

As long as the fraction of catastrophic errors is less than $30\%$, catastrophic redshift errors modeled by the above idealized model have therefore almost no impact on measuring $f_\mathrm{NL}$.

An important caveat is that we assumed a simple model of catastrophic redshift errors and it is not clear how well it describes actual catastrophic redshift errors.  It would be interesting to check if the above results are also valid for more realistic models of catastrophic redshift errors, for example derived from simulated galaxy spectra. 
We also emphasize that our simple model assumes the true global number density $\d n/\d z$ as a function of redshift to be perfectly known. 
While one could calibrate this global $\d n/\d z$ by correlating with spectroscopic data or using clustering redshifts \cite{2008ApJ...684...88N,McQuinn:2013ib,Menard:2013aaa}, such a calibration would never be perfect in practice, but exploring this is beyond the scope of this paper.
Another potential worry is that outlier galaxies might correspond to a different galaxy popoluation than galaxies whose redshift is determined correctly, so that they might require independent bias parameters. 
Still, the small degradation of $\fnl$ constraining power for the idealized catastrophic redshift errors above gives us hope that more realistic catastrophic redshift errors can also be handled as long as their imprint on angular power spectra can be modeled.
}

\section{Conclusions}
\label{se:conclusions}

Cross-correlating future CMB lensing measurements from surveys like CMB-S4 with future clustering measurements from deep photometric redshift surveys like LSST promises great potential. 
The significant redshift overlap, low lensing noise, and high galaxy number density enable 150$\sigma$ to 260$\sigma$ measurements of  cross-spectra between CMB-S4 lensing and individual tomographic LSST redshift bins of width $\Delta z=0.5$ at $z<1$ and $\Delta z=1$ at $z>1$, assuming the experiments observe the same half of the sky.
Combining these tomographic LSST redshift bins with weights that match the CMB lensing kernel results in a combined tracer map that is more than $94\%$ correlated with CMB-S4 lensing on large scales. 
On smaller scales the correlation drops but remains greater than $60\%$ up to $\ell=1000$.

This CMB-lensing--clustering cross-correlation signal can be used to break parameter degeneracies and partially cancel cosmic variance when combining with auto-power spectrum measurements.
We study this using Fisher forecasts that combine all auto- and cross-power spectra of CMB-S4 CMB lensing and SDSS, DESI and LSST clustering measurements, focusing on three applications.

First, we find that the matter amplitude $\sigma_8(z)$ in redshift bins $z=0-0.5,0.5-1,1-2,2-3,3-4,4-7$, and $7-100$ can be determined to $1\%$ for $\ell_\mathrm{max}=100$ and to $0.2\%$ for $\ell_\mathrm{max}=1000$. 
This assumes a sky fraction of $\fsky=0.5$ and marginalizes over linear galaxy bias, assuming all other cosmological parameters to be fixed (more generally, the constraints should be interpreted as constraints on parameter combinations like $\Omega_m\sigma_8$ that are probed by the measured spectra).
Such high precision measurements of $\sigma_8(z)$ out to high redshift probe the growth of structure and the expansion of the Universe in a redshift range where the standard $\Lambda$CDM model has not been tested well observationally, offering significant discovery potential.
It would be interesting to project these $\sigma_8(z)$ forecasts forward to concrete models of accelerating expansion that differ from the standard cosmological constant, noting that one may want to define a new figure of merit to capture potential discovery potential at high redshift better than the conventionally used figure of merit and $w_0-w_a$ parametrization.
The $\sigma_8(z)$ measurements also carry information on the sum of neutrino masses when comparing with the amplitude of the primary CMB.

The error bars of $\sigma_8(z)$ are limited by the number of modes and therefore scale approximately as $f_\mathrm{sky}^{-1/2}\ell_\mathrm{max}^{-1}$, improving with larger sky area and with the smallest scale included in the analysis, which is limited by our ability to model nonlinear galaxy bias.
Without sky overlap between CMB-S4 and LSST, the $\sigma_8$ precision degrades by more than a factor of 20 because of the bias-$\sigma_8$ degeneracy.
Without any CMB lensing measurements, our forecast would not give any constraint on $\sigma_8$ because of its degeneracy with $b_1$.
A joint analysis of CMB-lensing and LSS clustering on a large overlapping patch of sky is therefore critical for high-precision $\sigma_8(z)$ measurements.

Similarly, we find that the linear galaxy bias of tomographic LSST redshift bins can be measured with subpercent-level precision even if we marginalize over $\sigma_8(z)$.
This could be used to obtain a 3-D map of the dark matter in the Universe by dividing the galaxy density by its bias in each redshift bin \cite{Pen0402008}.

A second application is to measure the amplitude of local primordial non-Gaussianity $\fnl$ using the scale-dependent bias it induces between lensing and clustering on large scales.
A joint analysis of CMB-S4 CMB lensing and LSST clustering can reach $\sigma(\fnl)=0.4$ if they observe the same half of the sky and if CMB lensing and clustering power spectra can be measured down to $\ell_\mathrm{min}=2$.
Measuring such large scales is observationally challenging but not impossible.
More conservatively, a minimum multipole of $\ell_\mathrm{min}=20$ gives $\sigma(\fnl)=1$, which is still five times more precise than the best current Planck measurement \cite{Planck15fnl}.
This improves slightly to $\sigma(\fnl)= 0.7$ if we exclude only large-scale galaxy-galaxy power spectra at $\ell\le 18$ (where systematics should be most problematic) but include CMB-lensing--clustering cross-spectra and CMB lensing auto-spectra down to $\ell=2$, assuming again $\fsky=0.5$.
If more detailed forecasts confirm the sensitivity to $\fnl\lesssim 1$, this would open the exciting possibility to rule out single-field inflation in a model-independent way with CMB-S4 and LSST.
More precise and robust constraints may be possible when combining with the proposed SPHEREx experiment, which can reach $\sigma(\fnl)<1$ using clustering measurements alone \cite{Spherex1412}. 

Without CMB lensing the $\fnl$ precision degrades by an order of magnitude, showing that it is critical to include CMB lensing.
The $\fnl$ forecast also benefits from observing as large scales as possible, which requires CMB-S4 and LSST to observe a wide patch of sky. 
If CMB-S4 and LSST observe on the same patch rather than independent patches, this improves the $\fnl$ precision by a factor of 1.5 to 2 due to partially cancelling cosmic variance and breaking degeneracies with the shape of the matter power spectrum.
Including LSST galaxies at high redshift $z=4-7$ improves the precision by another factor of 2.

The third application is to measure the sum of neutrino masses from the small scale-dependent bias they induce between lensing and clustering.
This is a rather clean signature that only involves linear theory and is not limited by the optical depth $\tau$ to the CMB that is a limiting factor for conventional cosmological neutrino mass measuremements from the suppression of power relative to the primary CMB \cite{Allison1509}.
In the most optimistic scenario we find $\sigma(m_\nu)\simeq 90\,\mathrm{meV}$ using only the scale-dependent bias effect, where we marginalized over linear galaxy bias.
Unfortunately, this is not competitive with other future experiments.
The precision could potentially be improved by improving the precision of galaxy bias parameters; for perfectly known bias one may reach $\sigma(m_\nu)\sim 10\,\mathrm{meV}$.
It would be interesting to study this further using redshift space distortions and statistics beyond power spectra.
\NEW{Another method to measure neutrino mass without $\tau$ information would be to use the precise measurements of $\sigma_8(z)$ and look for the small change in the cosmic growth function induced by nonzero neutrino mass (see Yu~\emph{et al.~in prep.} \cite{ByeongheeInPrep}).}

For our LSST forecasts we always included earlier observations from SDSS and DESI, but this is not actually required because LSST has the highest number density.
Indeed, if we observe CMB-S4 lensing and LSST clustering on the same patch, dropping SDSS and DESI degrades low-$z$ $\sigma_8$ and bias constraints by less than $50\%$\footnote{This degradation is mostly caused by SDSS photometric redshifts whose number density we assumed to be optimistically high.}, and has a negligible effect on $\sigma_8$ and bias at $z\ge 1$ as well as $\fnl$ and $m_\nu$.
Even with their lower number densities, SDSS and DESI spectroscopy should be very useful to calibrate LSST redshifts and reduce catastrophic redshift errors.

\NEW{Catastrophic redshift errors represent a possible source of confusion for $\fnl$ measurements because clustering of low-redshift galaxies may be confused with an $\fnl$ signal at high redshift.
Under the simplifying assumption that catastrophic redshift errors occur uniformly across the fiducial global galaxy number density, we found that this potential concern does not affect our $\fnl$ forecasts (marginalizing over one catastrophic error rate in each tomographic redshift bin has a negligible effect on $\sigma(\fnl)$ as long as the fiducial catastrophic redshift error rate is not greater than $\sim 30\%$; see \secref{zerrors}).
We argue that this is the case because there are enough observable power spectra to disentangle the effect of catastrophic errors from the $\fnl$ signature so that we can measure both at the same time.
This conclusion holds only for the simple toy model that we assumed about the nature of catastrophic redshift errors, warranting a more detailed study of the impact of more realistic, non-uniformly occuring catastrophic redshift errors.
}

There are a number of ways how one could improve the robustness and accuracy of our forecasts.
For example, one should properly marginalize over cosmological parameters within some priors rather than using the simplified parameterizations of the matter power spectrum that we marginalized over.
A more complete analysis would also include uncertainties in the true number density $\d n/\d z$ of observed galaxies, and account for photometric redshift errors that we ignored on the basis of using broad redshift bins and cross-correlating only with CMB lensing which has a broad redshift kernel.
Improving the modeling of nonlinear galaxy bias at high redshift should become a major priority if we want to measure $\sigma_8(z)$ at the subpercent level using cross-correlations (see \cite{Modi1706} and main text).
For $\fnl$, it is critical to understand better how well we can hope to deal with large-scale systematics and what minimum multipole $\ell_\mathrm{min}$ can be reached with future experiments.
Other corrections that may affect the forecasts are non-Gaussian covariance contributions, redshift space distortions, magnification bias, general relativistic corrections, higher-order CMB lensing biases, and differences between Monte-Carlo forecasts and Fisher forecasts.  
Given the promise of our idealized forecasts it is important to scrutinize their robustness against such corrections in the future.

There are also a number of ways to extend the forecasts by including additional measurements.
For example, it would be interesting to add shear measurements from galaxy weak lensing, which probe lower redshift than CMB lensing and might thus have better redshift overlap with clustering measurements (e.g., see  \cite{schaan1607} for such forecasts to calibrate multiplicative shear bias).
There are also a number of other LSS experiments that could lead to significant improvements, for example Hyper Suprime-Cam \cite{HSCwebsite}, HETDEX \cite{hetdexWebsite}, Euclid \cite{EuclidWhitePaper,euclidECWebsite,euclidESAWebsite}, WFIRST \cite{WFIRSTCDT,WFIRSTweb1,WFIRSTweb2}, and SPHEREx \cite{Spherex1412,spherexWebsite}.
Intensity mapping surveys may also be helpful because they can add high redshift information (e.g.~see \cite{Fonseca:2016xvi} for $\fnl$).
On the CMB side, it would be interesting to study how close the Simons Observatory \cite{SimonsObservatory} in combination with pre-LSST LSS surveys can get to the CMB-S4/LSST forecasts presented here.
An additional route to add information would be to include 3-point statistics or other summary statistics beyond the power spectrum.
A lot of such statistics are possible, e.g.~$\kappa\kappa\kappa$, $\kappa\kappa g$, $\kappa gg$, and $ggg$ bispectra, all of which should improve the forecasts.
For example, we know that for surveys with high number density these statistics constrain nonlinear galaxy bias rather well, which should improve the precision of $\sigma_8(z)$.
In the context of $\fnl$, galaxy bispectra alone are already rather promising \cite{Spherex1412,Yamauchi1611}, which should improve when adding CMB lensing.
These points deserve detailed future investigation given the promise of the forecasts presented here.

In summary, we find that cross-correlations of future CMB lensing surveys like CMB-S4 and photometric redshift surveys like LSST promise to be an exciting opportunity to measure the growth of cosmic structure and primordial non-Gaussianity with unprecedented precision, improving by an order of magnitude over the precision that can be obtained with any one of these surveys alone.

\section*{Acknowledgements}

We especially thank Pat McDonald, Blake Sherwin, Marko Simonovic, and Martin White for discussions and feedback, and Arka Banerjee, Bhuv Jain and the anonymous referee for discussions about catastrophic redshift errors.
We also thank many more members of the cosmology community for useful feedback during the preparation of this paper.
MS thanks the organizers of the Cosmic Visions meeting in Chicago, neutrino workshops at CITA and CCA, a CMB-S4 workshop at Harvard, and a CMB lensing workshop at Stanford for the opportunity to present and discuss parts of this work.
We acknowledge use of the publicly available \texttt{CAMB} Python wrapper \cite{cambpython,camb} and \texttt{quicklens} \cite{quicklens,PlanckLensing2015}.
This research used resources of the National Energy Research Scientific Computing Center (NERSC), a DOE Office of Science User Facility supported by the Office of Science of the U.S.~Department of Energy under Contract No.~DE-AC02-05CH11231.
It also used the COSMOS Shared Memory system at DAMTP, University of Cambridge, which is operated on behalf of the STFC DiRAC HPC Facility and funded by BIS National E-infrastructure capital grant ST/J005673/1 and STFC grants ST/H008586/1, ST/K00333X/1.
We acknowledge support of NASA grant NNX15AL17G.
MS gratefully acknowledges support from the Bezos Fund.

\appendix

\section{3-D Fourier to 2-D angular projection}
\label{app:3dto2d}

In this Appendix we summarize how we compute the 2-D angular power spectra used in the main text.

\subsection{Fields}

Our observables are the CMB lensing convergence $X(\vtheta)=\kappa(\vtheta)$ and galaxy number density contrast $X(\vtheta)=g(\vtheta)$ on the 2-D sky.  They are
line-of-sight projections of the 3-D density contrast $\delta_X$,
\begin{align}
  \label{eq:12}
  X(\vtheta) = \int_0^\infty \d z\,W_X(z)\,b_X(z)\,\delta_X(\chi(z)\vtheta,z),
\end{align}
where $\vtheta$ denotes angular position on the sky, $\chi(z)$ is the comoving distance to redshift $z$, $W_X(z)$ is a redshift kernel, and $b_X(z)$ is the fiducial linear tracer bias.\footnote{Equivalently, the integral over redshift can be written as an integral over comoving distance using $\d\chi=-\d z/H(z)$, which follows from $\chi(a)=\int_a^1\d \tilde a/[\tilde a^2H(\tilde a)]$ and $a=(1+z)^{-1}$.}

The CMB lensing convergence $\kappa_\mathrm{CMB}$ is an unbiased tracer of the total matter density contrast $\delta_{cb\nu}$, including cold dark matter '$c$', baryons '$b$', and neutrinos '$\nu$', because gravitational lensing is sensitive to all matter. 
We thus have
\begin{align}
  \label{eq:1}
  b_{\kappa_\mathrm{CMB}}&=1,\non\\
\delta_{\kappa_\mathrm{CMB}}(\vx) &= \delta_{cb\nu}(\vx),\non\\
  W_{\kappa_\mathrm{cmb}}(z) &= \frac{3}{2}\Omega_{m,0}H_0^2 \frac{(1+z)}{H(z)}\chi(z)\frac{\chi(z_s)-\chi(z)}{\chi(z_s)},
\end{align}
where $\Omega_{m,0}$ is the fractional matter density today, and a spatially flat universe is assumed.
The lensing kernel is evaluated for source photons emitted at the CMB last scattering surface at $z_s\simeq 1090$.
This kernel peaks roughly half way to the source plane, which is $6\!-\!7\,$Gpc away from us, corresponding to $z\simeq 2$, but it is extended over a wide range of redshifts.

Biased LSS tracers like galaxies are expected to form where dark matter and baryons gravitationally collapse, without being sensitive to the neutrino overdensity.  
The fractional number density contrast $g(\vtheta)$ on the sky thus follows from
\begin{align}
  \label{eq:16}
  b_g(z)&: \text{tracer-dependent},\non\\
  \delta_g(\vx) &= \delta_{cb}(\vx),\non\\
  W_g(z) & = \frac{1}{n_\mathrm{tot}} \frac{\d N}{\d z\d\theta^2},
\end{align}
where $\delta_{cb}$ is the CDM-baryon density contrast.
The redshift kernel is determined by the redshift distribution $\d N/(\d z\d\theta^2)$ of the observed objects, and by the total number density of objects per steradian (e.g.~\cite{FontRibera1308})
\begin{align}
  \label{eq:11}
n_\mathrm{tot} = \int_0^\infty\d z\,\frac{\d N}{\d z\d\theta^2}.
\end{align}
We assume that the linear tracer bias $b_g$ depends on redshift but not on scale.
This ignores higher-order scale-dependent bias corrections as discussed in the main text.

\subsection{Angular power spectra}

To compute angular power spectra of the above fields, we first expand in spherical harmonics, $X(\vtheta)=X_{\ell m}Y_{\ell m}(\vtheta)$, with
\begin{align}
  \label{eq:35}
  X_{\ell m} =\,& 4\pi i^\ell \int_0^\infty \d z\,W_X(z)b_X(z)\non\\
&\times\int\frac{\d^3\vk}{(2\pi)^3}j_\ell(k\chi(z))Y^*_{\ell m}(\hat\vk)\delta_X(\vk,z).
\end{align}
This follows by expanding the plane wave $e^{i\vk\cdot\vx}$ in spherical harmonics.
The angular power spectrum $\la X_{\ell m}(X'_{\ell'm'})^*\ra = \delta_{\ell\ell'}\delta_{mm'}C_\ell^{XX'}$ is then
\begin{align}
  \label{eq:FullLOSInt}
&  C_\ell^{XX'}  = 
\frac{2}{\pi}\int_0^\infty\d z\,W_{X}(z)b_X(z)
\int_0^\infty \d z'\,
W_{X'}(z')b_{X'}(z')\non\\
&\;\times
\int_0^\infty\frac{\d k}{k} j_\ell(k\chi(z))j_\ell(k\chi(z'))\,k^3 P_{\delta_X\delta_{X'}}(k,z,z').
\end{align}
Direct numerical evaluation of this integral is challenging because the spherical Bessel functions are highly oscillatory. 
We therefore use different evaluation techniques on small and large angular scales.

\subsubsection{Small scales: Limber approximation}
On small angular scales, $\ell\gtrsim 50$, we use the Limber approximation \cite{1953ApJ...117..134L,1992ApJ...388..272K},
\begin{align}
  \label{eq:33}
  \int_0^\infty \d k\,k^2j_\ell(k\chi)j_\ell(k\chi') f(k) \simeq \frac{\pi}{2\chi^2}\delta_D(\chi-\chi')f(\ell/\chi),
\end{align}
which gives the simple result
\begin{align}
  \label{eq:CXY}
 C_\ell^{XX'}  = \,
\int_z &
P_{\delta_X\delta_{X'}}(k=\ell/\chi(z),z)\non\\
&\times
W_{X}(z)b_X(z)
W_{X'}(z)b_{X'}(z).
\end{align}
The integration
\begin{align}
  \label{eq:RedshiftMeasure}
\int_z\equiv \int_0^\infty \d z\,\frac{H(z)}{\chi^2(z)}
\end{align}
includes a factor that converts volumes from $\mathrm{Mpc}^{3}$ to steradian times $\d z$.
We include nonlinear halofit corrections \cite{Takahashi:2012em,Mead:2015yca,Mead:2016zqy,Smith:2002dz} for the power spectrum $P_{\delta_X\delta_{X'}}(k,z)$, which we compute as a 2-D spline in $k$ and $z$ using CAMB Python \cite{camb,cambwebsite,cambpython}. 

We implement the Limber-approximated line-of-sight integral of \eqq{CXY} using matrix multiplication, $C_{\ell_i}=M_{ij}v_j$, where $M_{ij}\equiv P(\ell_i/\chi(z_j),z_j)$ and $v_j\sim W^2(z_j)b^2(z_j)\Delta z_jH(z_j)/\chi^2(z_j)$.
This enables fast on-the-fly computation of line-of-sight integrals in high-level languages such as Python.

\subsubsection{Large scales: Exact integration}
\label{app:FullLOS}

On large angular scales, $\ell\lesssim 50$, we compute exact line-of-sight integrals \eq{FullLOSInt} because the Limber approximation fails. 
We assume linear growth, i.e.~$P(k,z,z')=\bar D(z)\bar D(z')P(k,z=0)$ where $\bar D(z)\equiv D(z)/D(z=0)$ is normalized to unity at $z=0$.
Then,
\begin{align}
  \label{eq:32}
&  C_\ell^{XX'}  = 
\frac{2}{\pi}\int_0^\infty\d \chi\,\bar W_{X}(\chi)
\int_0^\infty \d \chi'\,
\bar W_{X'}(\chi')\non\\
&\;\times
\int_0^\infty\frac{\d k}{k} j_\ell(k\chi)j_\ell(k\chi')\,k^3P_{\delta_X\delta_{X'}}(k,z=0),
\end{align}
where we changed integration variables from $z$ to $\chi$ and absorbed all time-dependent factors in the kernel 
\begin{align}
  \label{eq:44}
\bar W_X(\chi)\equiv H(z)\bar D(z)W_X(z)b_X(z),  
\end{align}
where $\chi=\chi(z)$.

The conventional way to evaluate this is to first integrate over $\chi$ and $\chi'$, and then over $k$, i.e.
\begin{align}
  \label{eq:42}
&  C_\ell^{XX'}  = 
\frac{2}{\pi}
\int_0^\infty\frac{\d k}{k} \Delta_{X,\ell}(k)\Delta_{X',\ell}(k)k^3P_{\delta_X\delta_{X'}}(k,z=0),
\end{align}
where
\begin{align}
  \label{eq:43}
  \Delta_{X,\ell}(k) \equiv \int_0^\infty\d\chi\,\bar W_X(\chi)j_\ell(k\chi).
\end{align}
This transfer function scales as $\Delta_{X,\ell}(k)\sim k^\ell$ at low $k$, peaks at $k\simeq \ell/\chi_\mathrm{peak}$ where $\chi_\mathrm{peak}$ is the peak of the kernel $\bar W_X(\chi)$, and falls off at high $k$ because $j_\ell(k\chi)= (k\chi)^{-1}\sin(k\chi-\ell\pi/2)$ for $k\chi\rightarrow\infty$.

For fast evaluation of multiple line-of-sight integrals we tabulate $j_\ell(k\chi)$ at the discrete sampling points of the $k$ and $\chi$ integrations and use the large-argument limit of spherical Bessel functions for $k\chi>2000$.
The computational cost could be reduced further by using a generalized form of the FFTLog algorithm \cite{hamiltonfftlog} to evaluate the projection integrals \cite{Assassi:2017lea}.

To include scale-dependent bias from local primordial non-Gaussianity, we replace
$\Delta_{X,\ell}(k)$ by
\begin{align}
  \label{eq:45}
\Delta_{X,\ell}(k) 
+ 
\fnl\frac{3\Omega_{m,0}\delta_cH_0^2}{k^2T(k)c^2}
\int_0^\infty\d\chi\,\bar W_{X}^{\fnl}(\chi)j_\ell(k\chi),
\end{align}
where from \eqq{fnlbias}
\begin{align}
  \label{eq:46}
\bar W_{X}^{\fnl}(\chi) = \frac{H(z)W_X(z)}{D(z=0)}[b_X(z)-1].
\end{align}

\subsubsection{Noise}
Observable power spectra include noise, $\hat C_\ell=C_\ell+N_\ell$.
The noise power $N$ denotes either CMB lensing reconstruction noise $N^{\kappa_\mathrm{CMB}\kappa_\mathrm{CMB}}$ shown in \fig{N0}, or shot noise, which is given by (e.g.~\cite{FontRibera1308})
\begin{align}
  \label{eq:NellShotNoise}
  N^{gg}_\ell = \int_z \frac{W_g^2(z)}{n_\mathrm{com}(z)},
\end{align}
where the comoving number density is
\begin{align}
  \label{eq:15}
  n_\mathrm{com}(z) 
= \frac{\d N}{\d z\d\theta^2}\,
\frac{H(z)}{\chi^2(z)}.
\end{align}

\section{Origin of scale-dependent bias}

\label{app:FnlAndMnuBasicsAppdx}

In this appendix we provide some background that explains the origin of the scale-dependent bias from primordial non-Gaussianity $\fnl$ and neutrino mass. We also summarize the motivation to measure this.

\subsection{Primordial non-Gaussianity}

\label{app:fnlBasicsAppdx}

\subsubsection{Motivation to measure $f_\mathrm{NL}$}

LSS density perturbations are sourced by primordial density fluctuations generated in the early Universe.
Measuring statistical properties of the LSS can therefore give us clues about the physics that generated the primordial fluctuations.
In particular, within the paradigm of inflation, a primordial probability distribution function (pdf) that is not a Gaussian can only be produced by certain inflation models, involving for example multiple fields. 
Here we focus on the local type of primordial non-Gaussianity, where the primordial potential is the sum of a random Gaussian field and its square, $\phi(\vx)+\fnl(\phi^2(\vx)-\la\phi^2\ra)$, which has a non-Gaussian pdf. 
If we observe this with a large nonlinear amplitude, $f_\mathrm{NL}\gtrsim 1$, it will rule out single-field models of the inflationary expansion of the early Universe in a robust way \cite{Maldacena2003,Creminelli2004}.
This is one of few known observational means to rule out a whole class of currently viable early-universe models.

In practice the measurement is challenging because the threshold signal $f_\mathrm{NL}\simeq 1$ separating between single-field and multi-field models has a very small effect on observables.
The best upper limit, $f_\mathrm{NL}=0.8\pm 5.0$, comes from Planck CMB temperature and polarization measurements \cite{Planck15fnl}.

Observations of late-time LSS can improve the CMB limit on $f_\mathrm{NL}$ because they probe different Fourier modes, and because they can exploit the 
scale-dependent bias effect \cite{Dalal0710}. 
In brief, that effect is generated as follows. 
Inflation models with multiple fields can generate non-Gaussian correlations between long and short wavelength modes, $\langle\delta_l\delta_s\delta_s\rangle\neq 0$.
As a consequence, the small-scale power of fluctuations in a region depends on the realization of long wavelength modes in that region. 
Dark matter halos and galaxies thus form preferentially in regions where long-wavelength modes are high.
This leads to a scale-dependent bias between the matter and galaxy density that scales as $k^{-2}$ on large scales~\cite{Dalal0710}; see~\cite{Alvarez1412} for a recent review.
Observing such scale-dependent galaxy bias from local primordial non-Gaussianity would rule out single-field inflation because correlations between long and short modes are suppressed in all single-field inflation models \cite{Maldacena2003,Creminelli2004}.
Several forecasts have already demonstrated the high sensitivity of future LSS probes alone to $\fnl$, e.g.~\cite{Raccanelli1409,Camera1409,Spherex1412,dePutter1412,Alonso1507,Tucci1606}.  Many of these forecasts could potentially be improved by adding information from cross-correlations with CMB lensing.

\subsubsection{Scale-dependent bias}

Quantitatively, the non-Gaussian coupling between long and short wavelength modes imposed by local primordial non-Gaussianity rescales the bias $b_g$ between galaxies (forming in collapsed dark matter halos) and dark matter as
\begin{align}
  \label{eq:15}
  b_g(z) \,\rightarrow\, b_g(z)\left[1+f_\mathrm{NL}\beta(k,z)\right],
\end{align}
where the fractional bias change relative to Gaussian fluctuations is
\cite{Dalal0710,MatarreseVerde0801,Slosar0805} (also see \cite{Biagetti1611} and references therein)
\begin{align}
  \label{eq:fnlbias}
 \beta(k,z) =  \frac{\Delta b_g}{b_g} =  3\frac{(b_g-1)}{b_g}\frac{\Omega_{m,0}\delta_c}{k^2T(k)D(z)}\left(\frac{H_0}{c}\right)^2.
\end{align}
Here, $b_g(z)$ is the fiducial linear bias of the galaxy sample assuming Gaussian fluctuations, 
$\delta_c=1.686$ is the linear overdensity of spherical collapse,
$T(k)$ is the transfer function normalized to unity on large scales,
$D(z)$ is the linear growth function normalized to $(1+z)^{-1}$ in matter domination,
$\Omega_{m,0}$ is the matter density today,
and $H_0$ is the Hubble constant today.
\eqq{fnlbias} shows that the scale-dependent bias increases with higher redshift and with increasing fiducial galaxy bias.

Since the scale-dependent bias correction only applies to the galaxy overdensity but not to the lensing convergence, the galaxy-galaxy power spectrum scales like $(1+\fnl\beta)^2\approx 1+2\fnl\beta$, whereas the lensing-galaxy cross-spectrum scales like $1+f_\mathrm{NL}\beta$, and the lensing-lensing power spectrum is independent of $f_\mathrm{NL}$.
Comparing these power spectra therefore allows for a partial cancellation of cosmic variance as illustrated in \fig{fnlcartoon}.

\subsection{Neutrino mass scale-dependent bias from transfer functions}

\label{app:NeutrinoBasicsAppdx}

To describe the scale-dependent bias between lensing and clustering expected from neutrino mass, we define the fractional difference between the $cb\times cb\nu$ and $cb\nu\times cb\nu$ power spectra as 
\begin{align}
  \label{eq:23}
  \Delta(k,z) \equiv \frac{P_{cb,cb\nu}(k,z)}{P_{cb\nu,cb\nu}(k,z)}-1.
\end{align}
\fig{NeutrinoTransfer} shows $-\Delta(k,z)$ as a function of wavenumber $k$ for a few redshifts $z$.
On large scales, $k\lesssim 10^{-3}\,h\mathrm{Mpc}^{-1}$, the two transfer functions equal each other.
Over the range of scales $10^{-3}\,h\mathrm{Mpc}^{-1}\lesssim k\lesssim 10^{-1}\,h\mathrm{Mpc}^{-1}$, where neutrino free-streaming becomes relevant, the transfer functions smoothly separate from each other, reaching a maximal relative difference of $\Delta_\mathrm{max}=f_\nu$ at $k\gtrsim 0.1\,h\mathrm{Mpc}^{-1}$.
The transition is slightly redshift-dependent, with slightly larger scale-dependent bias at higher redshift for a given scale $k$.

In our forecasts, we focus on the neutrino mass information coming from the scale-dependent bias at $10^{-3}\,h\mathrm{Mpc}^{-1}\lesssim k\lesssim 10^{-1}\,h\mathrm{Mpc}^{-1}$, marginalizing over potential scale-dependent changes of the total matter power spectrum that could mimic a neutrino signature.
To implement this we write (with $k=\ell/\chi(z)$)
\begin{align}
  C^{\kappa\kappa}_\ell =\,& \int_z W_\kappa^2(z)\left[1+m_\nu^\mathrm{fake}\bar\Delta(k,z=1)\right]^2 P_{cb\nu,cb\nu}(k,z) \non\\
&\,\;+ N^{\kappa\kappa}_\ell,\\
C^{\kappa g_i}_\ell =\,& \int_z W_\kappa(z)W_{g_i}(z) b_{g_i}(z)
\left[1+m_\nu\bar\Delta(k,z)\right]\non\\
&\quad\;\times 
\left[1+m_\nu^\mathrm{fake}\bar\Delta(k,z=1)\right]^2
P_{cb\nu,cb\nu}(k,z),\\
C^{g_ig_j}_\ell =\,&
\int_z W_{g_i}(z)W_{g_j}(z) b_{g_i}(z)b_{g_j}(z)\left[1+m_\nu\bar\Delta(k,z)\right]^2\non\\
&\quad\;\times\left[1+m_\nu^\mathrm{fake}\bar\Delta(k,z=1)\right]^2 P_{cb\nu,cb\nu}(k,z)\non\\
&\,\; + \delta_{ij}^KN^{g_ig_i}_\ell,
\label{eq:ClNeutrinos}
\end{align}
where we marginalize over the `fake' parameter $m_\nu^\mathrm{fake}$ that rescales all power spectra in a way that resembles the shape of the scale-dependent transfer function difference $\Delta(k,z=1)$.
The `true' neutrino mass, whose precision we will forecast, enters only $\kappa g$ and $gg$ spectra.
We include it by linearly rescaling $\Delta$ in $m_\nu$, e.g.
\begin{align}
  \label{eq:8}
P_{cb,cb\nu}(k,z)=\left[1+m_\nu\bar\Delta(k,z)\right]P_{cb\nu,cb\nu}(k,z),
\end{align}
where we defined
\begin{align}
  \label{eq:4}
  \bar\Delta(k,z) \equiv \frac{\Delta(k,z)}{m_\nu^\mathrm{fid}}
\end{align}
and choose $m_\nu^\mathrm{fid}=60\,\mathrm{meV}$.
This approximation is sufficiently accurate for our purposes.

The above relations for angular power spectra show that if we express clustering and lensing power spectra in terms of the total matter power spectrum $P_{cb\nu,cb\nu}$, the scale-dependence from the different transfer functions can be treated as a scale-dependent bias correction
\begin{align}
  \label{eq:24}
  b_g(z) \,\rightarrow\, b_g(z)\left[1+m_\nu\bar\Delta(k,z)\right].
\end{align}

We do not include the redshift dependence of $\Delta(k,z)$ for the fake neutrino mass parameter because that parameter is intended to parameterize only the unknown scale dependence of the matter power spectrum. 
In practice, our results change very little if we include time dependence for the fake neutrino mass parameter.
As mentioned in the main text, we include redshift-independent bias amplitude parameters $B$ that rescale the fiducial redshift-dependent bias $b(z)$ in each tomographic redshift bin.

\section{Error of lensing-clustering cross-spectrum}
\label{se:AnalyticalEstimateNoSVC}
In this Appendix we discuss the sample variance error of the amplitude of the $\kappa g$ cross-spectrum, ignoring other power spectra and sample variance cancellation.
For a single mode,
\begin{align}
  \label{eq:tmp}
\mathrm{var}(C^{\kappa g})=(C^{\kappa g})^2+C^{gg}C^{\kappa\kappa}=(C^{\kappa g})^2(1+r_{\ell}^{-2}).
\end{align}
The fractional error per mode is thus $(1+r^{-2}_{\ell})^{1/2}$, which is $\sqrt{2}$ if lensing and clustering are perfectly correlated, $r_{\ell}=1$.
If they are not perfectly correlated, the fractional error increases and becomes $\sqrt{2}(1+\frac{\epsilon}{2})$ if $\epsilon\equiv 1-r_\ell$ and $|\epsilon| \ll 1$.

\begin{figure}[tbp]
\includegraphics[width=0.5\textwidth]{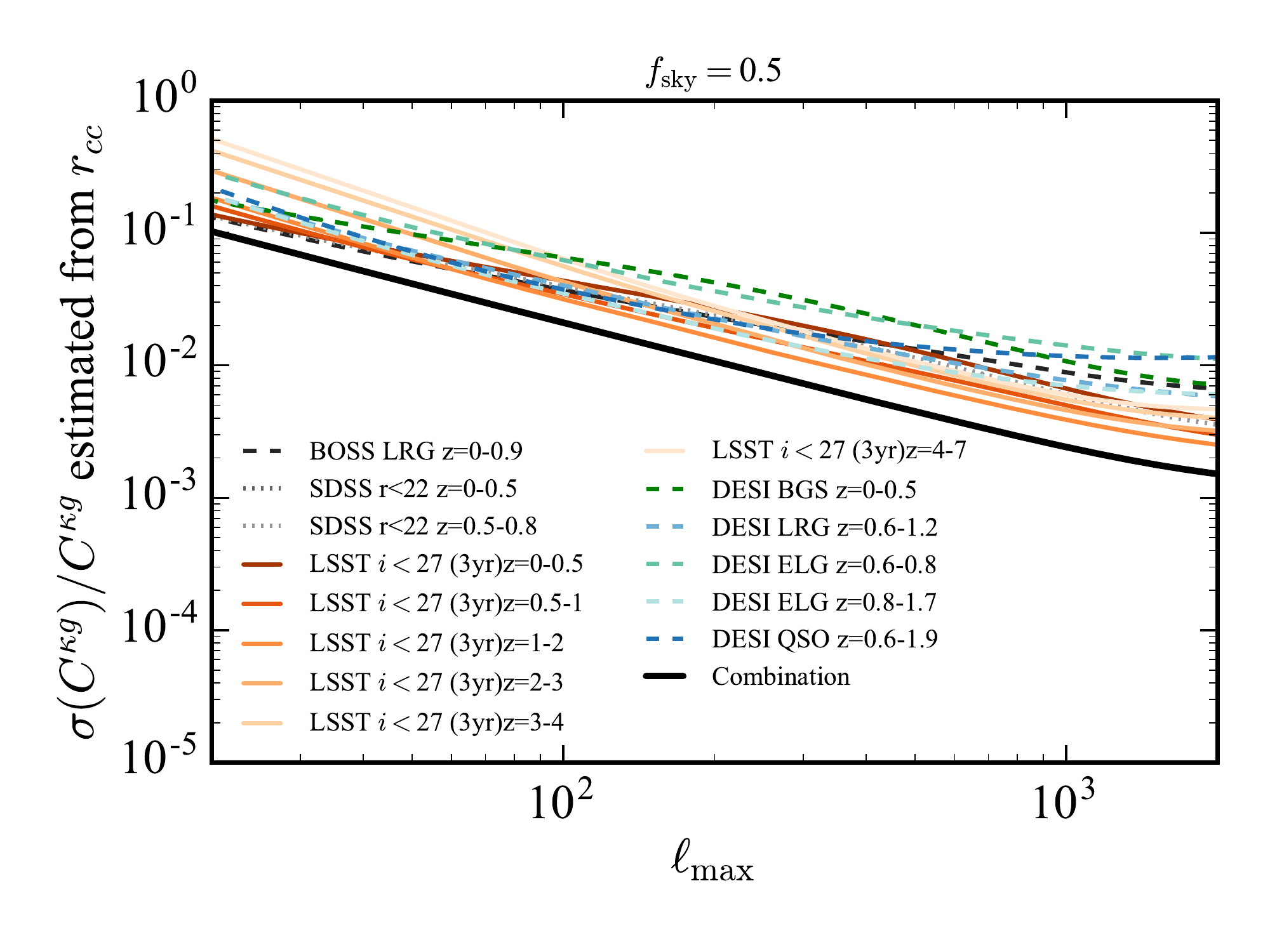}
\caption{Approximate estimate of the fractional error of the total $\kappa g$ spectrum based only on the correlation coefficients shown in \fig{rho}, computed using \eqq{CrossFractError}. 
}
\label{fig:KappaGFractErrorEstimate}
\end{figure}

Summing over all modes gives the fractional error of the cross-spectrum amplitude\footnote{Here we used $N_\mathrm{modes}=\fsky\sum_\ell(2\ell+1)\simeq \fsky\ell_\mathrm{max}^2$, and we introduced the mode-averaged cross-correlation coefficient $r_{cc}$,
\begin{align}
  \label{eq:30}
(1+r_{cc}^{-2})^{-1/2}=\frac{\sum_\ell (2\ell+1)(1+r_\ell^{-2})^{-1/2}}{\sum_{\ell'}(2\ell'+1)}.
\end{align}
In practice, $r_\mathrm{cc}$ as a function of $\ell_\mathrm{max}$ is similar to $r_\ell$ shown in \fig{rho} but slightly higher for large $\ell_\mathrm{max}$.  If $r_\ell=\mathrm{const}$ it would reduce to $r_{cc}=r_\ell$ exactly.
}
\begin{align}
  \label{eq:CrossFractError}
\frac{\sigma(C^{\kappa g})}{C^{\kappa g}}
=\left(\frac{1+r_{cc}^{-2}}{N_\mathrm{modes}}\right)^{1/2}
\simeq\left(\frac{1+r_{cc}^{-2}}{f_\mathrm{sky}\,\ell_\mathrm{max}^2}\right)^{1/2}.
\end{align}
The crucial point about \eqq{CrossFractError} is that the cross-correlation coefficient $r_{cc}$ that we need to insert is the one for the total combined sample. 
So if we are only looking at a single tracer at a single redshift, $r_{cc}$ may be low, but when we combine all tracers the total $r_{cc}$ is higher; see \fig{rho}.
This can also be seen in \fig{KappaGFractErrorEstimate} which evaluates \eqq{CrossFractError} using $r_{cc}$ from \fig{rho}.

The improvement factor of a multi-tracer analysis for measuring the $\kappa g$ amplitude relative to a single-tracer analysis is
\begin{align}
  \label{eq:MultiVsSingleImprov}
  \frac{\sigma_\mathrm{single}}{\sigma_\mathrm{multi}}\simeq
\left(\frac{1+r_{cc,\mathrm{single}}^{-2}}{1+r_{cc,\mathrm{multi}}^{-2}}\right)^{1/2},
\end{align}
if we only account for the sampling variance of $\kappa g$ as in \eqq{CrossFractError}.
\eqq{MultiVsSingleImprov} involves only the cross-correlation coefficient between CMB lensing and LSS tracers.
At $\ell=1000$ in \fig{rho} we have $r_{cc} \sim 0.22$ for a single tracer at $z=3-4$,
and $r_{cc} \sim 0.62$ for combined tracers, so the error improvement is $[(1+0.22^{-2})/(1+0.62^{-2})]^{1/2} = 2.4$.
At lower wavenumbers the improvement factor is similar, for example $[(1+0.3^{-2})/(1+0.88^{-2})]^{1/2}=2.3$ at $\ell=100$, and $[(1+0.22^{-2})/(1+0.94^{-2})]^{1/2}=3.2$ at $\ell=30$.
We thus expect to improve error bars by a factor of 2 to 3 in a multi-tracer analysis that combines all tomographic LSS redshift bins relative to working with just a single tracer.

By construction the improvement factor only captures the reduction in sample variance uncertainty of $\kappa g$ due to combining tracers and increasing $r_{cc}$, while ignoring improvements from other spectra and sample variance cancellation.
For example, this is the improvement one would expect when measuring $\sigma_8$ from $\kappa g$ without marginalizing over galaxy bias or any other parameters.
When we instead marginalize over galaxy bias, $\sigma_8$ cannot be determined from $\kappa g$ alone because $\sigma_8$ and galaxy bias are degenerate, and this can only partially be broken by including $gg$ spectra.

\begin{figure}[tbp]
\includegraphics[width=0.5\textwidth]{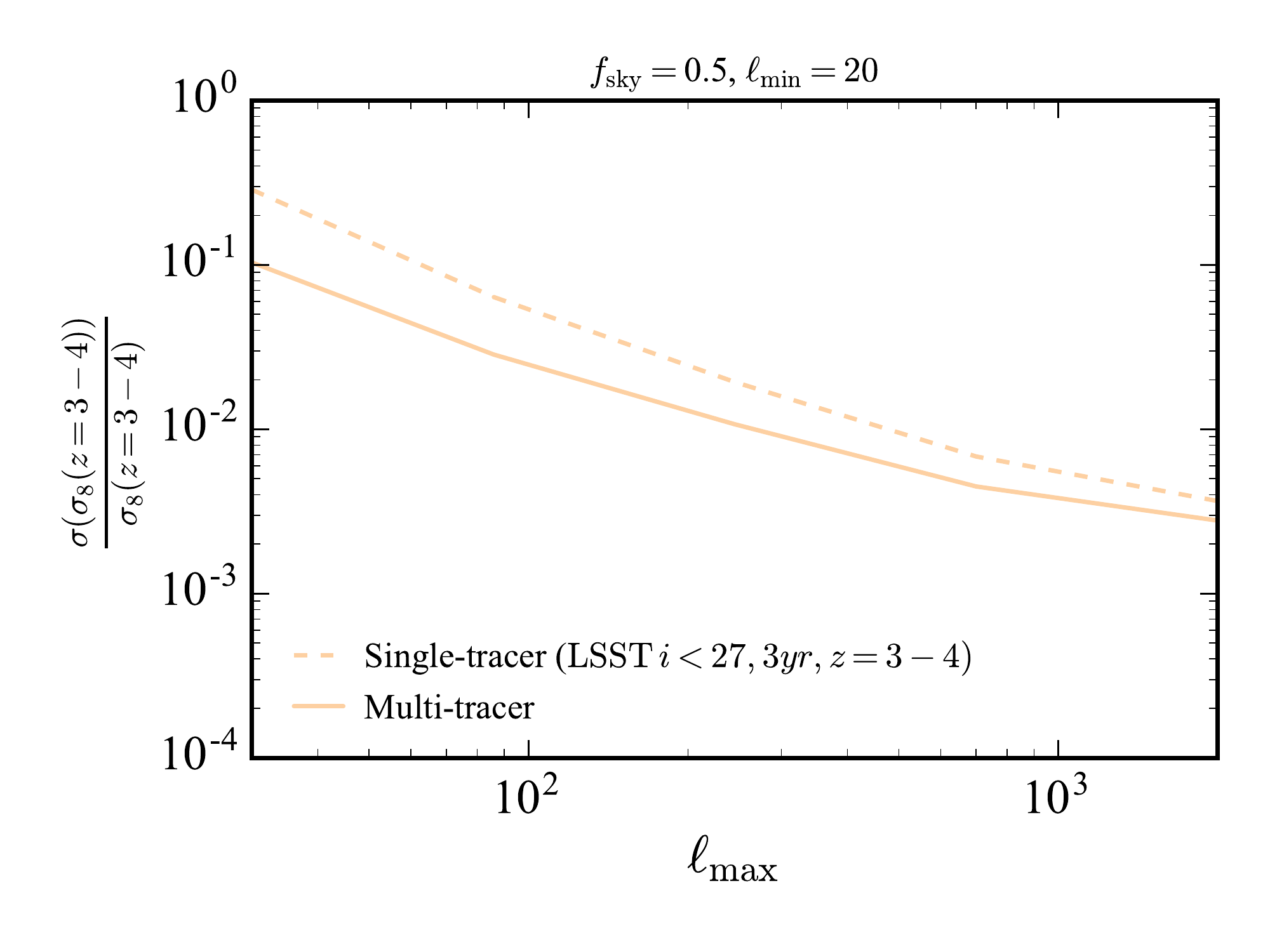}
\caption{Fractional statistical uncertainty of $\sigma_8(z=3-4)$ at a single redshift bin obtained from a multi-tracer analysis (solid) or a single-tracer analysis using only the LSST $z=3-4$ redshift bin (dashed). 
}
\label{fig:s8_lmax_multivssingletracer}
\end{figure}

We can consider a simple scenario where we restrict ourselves to using a single tomographic redshift bin of LSST at $z=3-4$ and constrain only the amplitude $\sigma_8(z=3-4)$ in the same redshift bin (i.e.~we use only $\kappa\kappa$, $\kappa g^\mathrm{LSST}_{z=3-4}$, and $g^\mathrm{LSST}_{z=3-4}g^\mathrm{LSST}_{z=3-4}$ power spectra, while still marginalizing over linear bias as usual).
The resulting single-tracer $\sigma_8$ precision is shown in dashed in \fig{s8_lmax_multivssingletracer}, where we also show the precision if all tracers are included (solid).
On large and intermediate scales the multi-tracer analysis yields 2 to 3 times better precision than the single-tracer analysis because of its increased cross-correlation coefficient with CMB lensing on large scales.
This is roughly consistent with the improvement factor estimated above.

\section{Combining observations}
\label{app:CombiningObs}

Using all auto- and cross-spectra of CMB lensing and tomographic redshift bins leads to a large data vector; in the most extreme case that we study it contains 120 power spectra of 15 fields.
While we make the idealized assumption of Gaussian covariances, real data often requires simulations to obtain accurate covariances.
Estimating the covariance of 120 power spectra, all of which should have $\mathcal{O}(10)$ or more bins in $\ell$, would require a rather large number of simulations, which may not be practical.

To address this potential issue, we explore two schemes to compress the data vector while trying to keep as much sensitivity to the signal of interest as possible.
The first compression scheme combines observed power spectra, while the second one combines  observations at the map level before computing power spectra.
\NEW{Related and more general compression schemes have been studied in more detail elsewhere, for example \cite{MOPED,2016PhRvD..93h3525Z,2018MNRAS.476L..60A}.}

\subsection{Combining power spectra}
\label{se:CombinedCls}
In the first compression scheme we combine the measured power spectra $\hat C^{XY}_\ell$ of all observed fields $X$, $Y$ to a single combined power spectrum $\hat D_\ell$ that retains full sensitivity to the parameter of interest.

Let us assume that the measured power spectra
\begin{align}
  \label{eq:7}
 \hat{\vec d}_\ell = 
(\hat C^{11}_{\ell}, \hat C^{12}_{\ell}, \dots, \hat C^{1N}_{\ell}, 
\hat C^{22}_{\ell}, \hat C^{23}_{\ell},  
  \dots,
\hat C^{NN}_{\ell})
\end{align}
follow a Gaussian likelihood\footnote{This is adequate for $\kappa\kappa$ assuming Planck-like noise levels \cite{Marcel1308}, but may be less accurate for lower lensing noise and for $\kappa g$ and $gg$ power spectra.
An additional term involving the determinant of the covariance is irrelevant under the assumption that the covariance is independent of cosmological and nuisance parameters $\theta_a$.  
}
\begin{align}
  \label{eq:12}
  -2 \ln \mathcal{L} = \sum_\ell (\hat\vd_\ell-\vd_\ell)\mathrm{cov}(\hat\vd_\ell,\hat\vd_\ell)^{-1}(\hat\vd_\ell-\vd_\ell)
\end{align}
with Gaussian covariance \eq{GaussianCov}.
The model $\vd_\ell = \la\hat\vd_\ell\ra$ depends on parameters $\theta_a$ which can be cosmological or nuisance parameters.
We approximate this dependence by a first-order Taylor expansion around the fiducial parameter values $\theta_a^\mathrm{fid}$,
\begin{align}
  \label{eq:13}
  \vd_\ell = \vd_\ell^\mathrm{fid} + \sum_a\frac{\partial \vd_\ell}{\partial\theta_a}(\theta_a-\theta_a^\mathrm{fid}),
\end{align}
i.e.~we assume that second derivatives of the model with respect to parameters are small.
If our goal is to measure a specific parameter $\theta_m$, its maximum-likelihood estimator follows by solving $\partial[-2\ln\mathcal{L}]/\partial\theta_m=0$ for $\theta_m$, which gives
\begin{align} 
  \label{eq:ThetaiEsti}
  \hat\theta_m = \theta_m^\mathrm{fid} + \sum_\ell 
\vw_\ell
(\hat\vd_\ell-\vd_\ell^\mathrm{fid}).
\end{align}
The weighting vector $\vw$ is (no sum over $\ell$)
\begin{align}
  \label{eq:optimalWeightClcombi}
\vw_\ell\equiv \sum_b (F^{-1})_{mb}\frac{\partial\vd_\ell}{\partial\theta_b}
\mathrm{cov}(\hat\vd_\ell,\hat\vd_\ell)^{-1},
\end{align}
or writing out all components of the data vector,
\begin{align}
  \label{eq:optimalWeightClcombi2}
(w_\ell)_i=\sum_{b} (F^{-1})_{mb}\sum_j\frac{\partial(d_\ell)_j}{\partial\theta_b}
[\mathrm{cov}(\hat\vd_\ell,\hat\vd_\ell)^{-1}]_{ji}.
\end{align}
Here $F^{-1}$ is the inverse of the Fisher matrix given by \eqq{FisherFast}.
The weight in Eqs.~\eq{optimalWeightClcombi} and \eq{optimalWeightClcombi2} has a simple interpretation: It first applies an inverse-covariance ('$C^{-1}$') operation on the data vector to down-weight noisy modes, and then projects on the expected signal from the parameter $\theta_m$ that we try to measure.
This is similar to a Wiener filter.

Guided by \eqq{ThetaiEsti} we can define a weighted combination $\hat D_\ell$ of all measured power spectra $\hat C^{XY}_\ell$ contained in $\hat\vd$ as
\begin{align}
  \label{eq:Dcompression}
  \hat{D}_\ell \equiv \mathbf{w}_\ell\hat\vd_\ell.
\end{align}
This is a compressed power spectrum, containing a single number for
every multipole $\ell$.
The estimator $\hat\theta_m$ then becomes
\begin{align}
  \label{eq:thetaEstiWithDl}
\hat\theta_m = \theta_m^\mathrm{fid} + \sum_\ell\big(\hat D_\ell - D_\ell^\mathrm{fid}\big).
\end{align}
Therefore, the maximum-likelihood estimate for $\theta_m$ can be obtained by fitting the measured compressed power spectrum $\hat D_\ell$ to the fiducial model.
It is straightforward to check that the Fisher information of \eqq{thetaEstiWithDl} agrees with the Fisher information \eq{FisherFast} if measuring all power spectra, i.e.~$\mathrm{cov}(\hat\theta_m,\hat\theta_m)=(F^{-1})_{mm}$.
The compression in \eqq{Dcompression} is therefore lossless if we aim to measure a single parameter $\theta_m$.

Generalizing the above, we can define one compressed power spectrum for each parameter of interest. 
This then gives $N_\mathrm{param}$ compressed power spectra if we are interested in $N_\mathrm{param}$ parameters.
Fitting these power spectra with a model retains full sensitivity to all parameters.

If the number of parameters $N_\mathrm{params}$ is smaller than the number of measured power spectra, the compression \eq{Dcompression} reduces the size of the data vector to $N_\mathrm{params}$ spectra at every $\ell$.
This is precisely what we were after: If 120 power spectra are measured but we are only interested in say 6 cosmological parameters, we can compress the measured power spectra to 6 combined power spectra $\hat D_\ell$ that retain full sensitivity to the parameters.
This is useful when estimating the covariance from a limited number of simulations.
It also has the nice property of down-weighting noisy modes and keeping only modes relevant for the signal of interest, similarly to a matched-filter estimator.

A subtlety of the above approach is that the inverse covariance of the full data vector $\hat \vd$ with all measured power spectra enters the weights $\vw$, so we still need to know the full covariance.
To address this, one could use an idealized theoretical covariance for the weights $\vw$ that would not require simulations.
If that covariance used for the weights deviates from the true covariance, the estimator becomes suboptimal and the compression is not perfectly lossless any more.
Importantly, however, one can then use a small number of simulations to characterize the true noise of the suboptimally compressed power spectra (i.e.~compute Monte-Carlo errors of $\hat D_\ell$).
This would account for corrections to the true covariance that are not captured by the idealized covariance model.
Final parameter estimates from the compressed power spectra can thus have larger error bars if the idealized covariance used for the weights is not accurate, but these larger error bars can still be estimated correctly using simulations.

A potential disadvantage of the compression is that the optimal weights \eq{optimalWeightClcombi} to compress the data vector depend on the parameters that are estimated and marginalized over, because the weights depend on the inverse Fisher matrix.
For example, the combined data vector that is optimal for $\fnl$ is not optimal for measuring neutrino mass, and vice versa.
If enough simulations are available for estimating the covariance, it may thus be simpler to work with the uncompressed data vector involving all observed spectra.
Checking results with  compressed power spectra may still provide useful cross-checks, for example if there is uncertainty about the accuracy of the covariance between all measured power spectra.

\subsection{Combining LSS tracer maps}

Rather than combining observations at the power spectrum level one may try to combine them already at the map level and then compute only few auto- and cross-spectra of combined maps.
For example, one could combine all biased LSS tracers to a combined map $I=\sum_i c_i \delta_{g_i}$ such that  it is maximally correlated with the CMB lensing convergence at the map level.\footnote{See Appendix A of \cite{BlakeMarcel1502} where this was used to optimize external CMB delensing whose efficiency depends only on the cross-correlation coefficient at the map level.}
However, such a weighting at the map level imposes relationships between the weights of individual LSS auto- and their cross-spectra with lensing.\footnote{For example, the auto-spectrum of the combined sample is
$C^{II} = \sum_{i,j}c_ic_jC^{g_ig_j}$
and its cross-spectrum with lensing is
$C^{\kappa I} = \sum_{i}c_iC^{\kappa g_i}$, 
where sums are over biased LSS tracers, so that the weight of $C^{\kappa g_i}$ is the square root of the weight of $C^{g_ig_i}$.}
We expect the resulting weights to be sub-optimal in general because the optimal weights of \eqq{optimalWeightClcombi} involve the response of power spectra with respect to cosmological and nuisance parameters included in the analysis and the inverse of a large covariance matrix. 
In general, combining biased LSS tracers at the map level is therefore expected to be sub-optimal, resulting in a data compression that is not lossless in general.

However, there are special cases in which lossless compression at the map level is possible.
One example is the situation where the only goal is to maximize the correlation coefficient with lensing to delens the CMB \cite{BlakeMarcel1502}.
Another example is the situation where we drop all observed LSS auto-spectra $C^{g_ig_j}$ from the observed data vector $\hat\vd$, which may be relevant if all LSS auto-spectra are dominated by systematics on the scales of interest.
In that case the reduced data vector $\hat\vd$ contains only $C^{\kappa\kappa}$ and $C^{\kappa g_i}$. 
The optimal weights $(w_\ell)_i$ for compressing these power spectra are then given by evaluating \eqq{optimalWeightClcombi} for the reduced data vector.
Since the sum of cross spectra is the same as the cross-spectrum of the sum, $\sum_i (w_\ell)_iC^{\kappa \delta_{g_i}}=C^{\kappa\sum_i(w_\ell)_i\delta_{g_i}}$, we can combine biased LSS tracers at the map level as
\begin{align}
  \label{eq:MapCompression}
I(\vl) \equiv \sum_i (w_\ell)_i \,\delta_{g_i}(\vl)
\end{align}
where the sum is over all biased LSS tracers.
The measured $C^{\kappa\kappa}$ and $C^{\kappa I}$ then contain the same Fisher information as the measured $C^{\kappa\kappa}$ and $C^{\kappa g_i}$.
\eqq{MapCompression} therefore represents a lossless data compression if and only if LSS auto-spectra $C^{g_ig_j}$ are excluded from the data analysis.

\bibliography{lsscross}

\label{lastpage}
\end{document}